\documentclass[12pt]{article}
\usepackage{amsmath} 
\usepackage{multirow} 
\usepackage{amsfonts}
\usepackage{amssymb}
\def\hybrid{\topmargin -20pt    \oddsidemargin 0pt
        \headheight 0pt \headsep 0pt
        \textwidth 6.25in       
        \textheight 9.5in       
        \marginparwidth .875in
        \parskip 5pt plus 1pt   \jot = 1.5ex}
\hybrid
\numberwithin{equation}{section}
\numberwithin{table}{section}
\newcommand{\beq}{\begin{equation}}
\newcommand{\eeq}{\end{equation}}
\newcommand{\bea}{\begin{eqnarray}}
\newcommand{\eea}{\end{eqnarray}}                  
\newcommand{\ba}{\begin{array}}
\newcommand{\ea}{\end{array}}
\newcommand{\bt}{\begin{tabular}}
\newcommand{\et}{\end{tabular}}
\newcommand{\bc}{\begin{center}}
\newcommand{\ec}{\end{center}}

\newcommand{\sx}{\sigma}

\newcommand{\Ox}{\Omega}

\newcommand{\Gx}{\Gamma}

\newcommand{\cO}{\mathcal{O}}

\newcommand{\cP}{\mathcal{P}}
\newcommand{\cC}{\mathcal{C}}
\newcommand{\cD}{\mathcal{D}}

\newcommand{\cK}{\mathcal{K}}

\newcommand{\cW}{\mathcal{W}}

\newcommand{\cH}{\mathcal{H}}

\newcommand{\cI}{\mathcal{I}}

\newcommand{\cM}{\mathcal M}

\newcommand{\OO}{\mathcal{O}}

\newcommand{\I}{\text{Im}}
\newcommand{\R}{\text{Re}}

\newcommand{\Kcs}{K^{\text{cs}}}

\newcommand{\Kh}{{\hat{K}}}
\newcommand{\Lh}{{\hat{L}}}
\newcommand{\Ah}{{\hat{A}}}

\DeclareMathOperator{\SU}{\mathit{SU}}

\newcommand{\rep}[1]{\mathbf{#1}}

\newcommand{\ii}{i}

\newcommand{\bbR}{\mathbb{R}}
\newcommand{\bbC}{\mathbb{C}}

\newcommand{\nn}{\nonumber}

\newcommand{\cref}{{\bf [check ref]}}

\newcommand{\WV}{\mathcal{W}}      

\newcommand{\simga}{\sigma}

\begin{document}

\begin{titlepage}
\begin{center}

\hfill hep-th/0602241\\
\hfill MAD-TH-06-2\\
\hfill DESY 06-017\\

\vspace*{2cm}

{\large \bf  Generalized $N=1$ Orientifold Compactifications \\[.2cm]
  and the Hitchin functionals}

\vskip 1.3cm

{\bf Iman Benmachiche$^{\, a,b}$ and Thomas W.\ Grimm$^{\, c}$}  \\
\vskip 0.3cm

{\em $^a$ DESY -- Theory Group \\
Notkestrasse 85, Lab. 2a \\
D-22603 Hamburg, Germany}\\[.6cm]

{\em $^b$ II. Institut f{\"u}r Theoretische Physik\\
Universit{\"a}t Hamburg\\
Luruper Chaussee 149\\
 D-22761 Hamburg, Germany}\\[.6cm]

{\em $^c$ Department of Physics \\
 University of Wisconsin \\
1150 University Avenue\\
Madison, WI 53706, USA}\\[.4cm]

{\tt iman.benmechiche@desy.de, grimm@physics.wisc.edu} \\

\end{center}

\vskip 0.5cm

\begin{center} {\bf ABSTRACT } \end{center}
\vspace{-2mm}
The four-dimensional $N=1$ supergravity theories arising in compactifications
of type IIA and type IIB on generalized orientifold backgrounds with background fluxes 
are discussed. The K\"ahler potentials are derived for reductions on $SU(3)$ structure 
orientifolds and shown to consist of the logarithm of the two Hitchin functionals. 
These are functions of even and odd forms parameterizing the 
geometry of the internal manifold, the B-field and the dilaton. 
The superpotentials induced by background fluxes and the non-Calabi-Yau 
geometry are determined by a reduction of the type IIA and type IIB fermionic actions
on $SU(3)$ and generalized $SU(3)\times SU(3)$ manifolds.
Mirror spaces of Calabi-Yau orientifolds with electric and part of the magnetic 
NS-NS fluxes are conjectured to be certain $SU(3)\times SU(3)$ structure manifolds. 
Evidence for this identification is provided by comparing the generalized type IIA and type IIB 
superpotentials.

\end{titlepage}

\section{Introduction}

The construction of semi-realistic type II string vacua for particle physics 
and cosmology attracted many efforts within the 
last years \cite{reviewPP,reviewcosmo}. Of particular interest are scenarios  
with space-time filling D-branes, which can provide 
for non-Abelian gauge groups on their world-volume. However, demanding 
the internal manifold to be compact, consistent setups also need to include 
orientifold planes carrying a negative tension. They arise 
in string theories modded out by a geometric symmetry of the background
in addition to the world-sheet parity operation \cite{JP,JPbook,AD}.  

From a phenomenological point of view orientifold compactifications
resulting in a four-dimensional $N=1$ supergravity theory are
of importance.  Prominent examples are type
II theories on Calabi-Yau orientifolds, since reductions 
on Calabi-Yau manifolds yield a four-dimensional 
 $N=2$ supergravity while the orientifold projection breaks 
the supersymmetry further down to $N=1$ \cite{Ori,GKP,AAHV,BBKL,BH,Granarev}.
In general these theories admit a large number of 
moduli fields which are flat directions of the potential and not fixed
in the vacuum. A possible mechanism to generate a 
non-trivial potential for these fields is the inclusion of background fluxes arising as
vacuum expectation values of field strengths 
in the supergravity theory \cite{Granarev}. This potential 
generically possesses supersymmetric vacua in which a part 
or all moduli are fixed 
\cite{Curio:2000sc,GKP,Kachru:2002he,BKL,Curio:2005ew,Denef:2005mm,Lust:2005dy,Derendinger:2004jn,VZ,DeWolfe:2005uu}. 
In order to study the properties of these 
vacua it is necessary to know the characteristic 
data of the corresponding four-dimensional $N=1$ supergravity theory.
Using a Kaluza-Klein reduction the four-dimensional $N=1$ theories of 
type II Calabi-Yau orientifolds were determined 
in refs.~\cite{GKP,Becker:2002nn,TGL1,TGL2,Soft} 
and reviewed, for example, in refs.~\cite{Frey:2003tf,Granarev,Jockers:2005pn,TGthesis}.

In this paper we determine the $N=1$ data for a more general 
class of compactifications arising if the internal manifold  $\cM_6$ 
is no longer restricted to be Calabi-Yau.
In order that the resulting 
four-dimensional theory still admits some supersymmetry $\cM_6$
cannot be chosen arbitrarily, but rather has  to 
admit at least one globally defined spinor. 
In case that $\cM_6$ has exactly one globally defined 
spinor the structure group of the manifold reduces to $SU(3)$ \cite{CS,salamonb,joyce}.
Equivalently, these manifolds are characterized by the existence 
of two globally defined forms, a real two-form $J$ and a complex three-form $\Omega$.
These forms are in general not closed, which indicates a deviation from the 
Calabi-Yau case. This difference can also be encoded by specifying a new connection 
on $\cM_6$ with torsion which replaces the ordinary Levi-Cevita connection. 
The torsion can be interpreted as a background flux of the metric connection.
Compactifications on $SU(3)$ structure manifolds were considered in the early works 
\cite{Rocek,StromingerT,hullT} and more recently extended in refs.~\cite{waldram}--\cite{Micu:2006ey}.
In specific settings these `metric fluxes'
arise as mirrors of Calabi-Yau compactifications with electric 
NS-NS fluxes \cite{GLMW,Fidanza:2003zi}.

Compactifying type II string theory on an $SU(3)$ structure manifold
leads to an effective four-dimensional $N=2$ supergravity theory
with a potential depending on the torsion of $\cM_6$. 
As we will discuss in more detail below,
one can still impose an appropriate orientifold projection which 
truncates this theory to an $N=1$ supergravity.  
For specific set-ups this was also argued in the recent works \cite{Camara:2005dc,Shelton:2005cf,
Aldazabal:2006up,Villadoro:2006ia}.
Supersymmetric orientifold projections yield setups with $O6$ planes
in type IIA while for type IIB reductions two setups with $O3$ and $O7$ 
as well as $O5$ or $O9$ planes are encountered. Our 
analysis focuses on the effective $N=1$ four-dimensional supergravity 
theory for the bulk fields of these configurations, while freezing 
all moduli arising from the D-brane sector.

In contrast to the standard Calabi-Yau compactifications 
the reduction on $SU(3)$ structure manifolds is more subtle.
This can be traced back to the fact that in these 
generalized compactifications the distinction 
between massless or light modes and the massive Kaluza-Klein 
modes is not anymore straightforward. Recall that in Calabi-Yau compactifications 
the massless modes are in one-to-one correspondence with the harmonic 
forms of $\cM_6$. Background fluxes generate a potential for these modes 
and can lift them to an intermediate mass scale. In reductions on $SU(3)$
structure manifolds a potential is induced by the non-trivial torsion of $\cM_6$.
However, the masses acquired by the four-dimensional fields need not 
be generated at an intermediate scale. The specification of a distinguished 
finite set of modes corresponding to the light degrees of freedom is missing so far.   
It is therefore desirable to avoid a truncation to light modes by working with 
general forms on the ten-dimensional background $M_{3,1} \times \cM_6$. 
Most of our calculations will be performed within this general approach. We will argue 
that it remains possible to determine the four-dimensional $N=1$ spectrum by  
imposing the orientifold projection. Only in a second step 
we specify a reduction to a finite set of modes in order to illustrate 
our results and to discuss mirror symmetry to Calabi-Yau orientifolds with 
background fluxes.
 
In this paper we will focus mainly on the chiral field space of the 
four-dimensional theory. We will determine the local metric on this 
space and show that it can be derived from a K\"ahler potential
as demanded by $N=1$ supersymmetry.
Since the orientifold projection induces 
a consistent reduction of a four-dimensional $N=2$ supergravity theory 
to $N=1$ this K\"ahler manifold is a subspace of the full $N=2$ scalar field
space \cite{ADAF,D'Auria:2005yg}. 
Locally it takes the form $\cM^{\rm K} \times \cM^{\rm Q}$ where 
$\cM^{\rm K}$ and $\cM^{\rm Q}$ are the subspaces 
of the $N=2$ special K\"ahler and quaternionic manifolds respectively.
$\cM^{\rm Q}$ has half the dimension of the quaternionic space. 
The K\"ahler potentials for both manifolds are shown to be the 
logarithms of the Hitchin functionals \cite{Hitchin:2000jd,Hitchin:2004ut,Hitchin:2005in} 
for specific even and odd forms on $\cM_6$. 

In compactifications on $SU(3)$ structure manifolds a scalar potential 
is induced by the torsion as well as possible background fluxes. 
Due to the $N=1$ supersymmetry it can be encoded by 
a holomorphic superpotential and D-terms arising due 
to non-trivial gaugings. In this work we derive the 
superpotentials for both type IIA and type IIB orientifold setups
by evaluating appropriate fermionic mass terms. 
This extends and confirms 
the results already present in the literature \cite{Derendinger:2004jn,VZ,Berglund, GLW, Camara:2005dc,Granarev}. 
The knowledge of the superpotential 
together with the K\"ahler potential is necessary to determine the 
conditions on four-dimensional supersymmetric vacua. It is readily checked that 
these conditions evaluated for the orientifold set-ups are in accord with the $N=1$ 
conditions on ten-dimensional backgrounds derived in ref.~\cite{Grana:2005sn,Jeschek:2005ek}.    

Recently, it was argued that more general four-dimensional supergravity theories can 
arise in compactifications of type II string theory on generalized manifolds with $SU(3) \times SU(3)$ 
structure \cite{Rocek,GMPT,JW,KL,Grana:2005sn, GLW}. The notion of a generalized (complex) manifold was
first introduced by Hitchin \cite{Hitchin:2004ut} and Gualtieri \cite{Gualtieri}. 
An intensive discussion of $SU(3) \times SU(3)$ structures and their application in 
$N=2$ compactifications can be found in the work of 
Witt \cite{Witt:2005sk} and Gra\~na, Waldram and Louis \cite{GLW}. We will make some first 
steps in exploring $N=1$ orientifold compactifications on manifolds 
with $SU(3) \times SU(3)$ structure by extending the orientifold projection to 
these spaces and deriving the induced superpotential due to the non-Calabi-Yau 
nature of the internal space. Our aim is to use these extended superpotentials 
to discuss possible mirror geometries of type II Calabi-Yau compactifications with fluxes.

The completion of mirror symmetry in the presence of NS-NS background 
fluxes is an area of intense current research 
\cite{GLMW,Fidanza:2003zi,Tomasiello,Mathai:2005fd,Dabholkar:2005ve,GLW,Shelton:2005cf,Chuang:2005qd,Aldazabal:2006up}. For compactifications 
with  electric NS-NS fluxes it was argued in refs.~\cite{GLMW,Fidanza:2003zi} that 
the mirror geometry is a set of specific $SU(3)$ structure 
manifolds known as half-flat manifolds. To extend this conjecture 
to the magnetic NS-NS fluxes various more drastic deviations from 
the standard compactifications are expected~\cite{Mathai:2005fd, Dabholkar:2005ve,GLW,Shelton:2005cf}. 
We will use our results on the orientifold superpotentials to 
conjecture a possible mirror geometry of compactifications
with part of the electric and magnetic background fluxes.
These mirrors are extensions of generalized 
manifolds with $SU(3) \times SU(3)$ structure.%
\footnote{%
A similar conjecture was mentioned in ref.~\cite{GLW} and we are grateful to Jan Louis for discussions
on that point.}
Note however, that in order 
to accommodate the mirror of the magnetic NS-NS fluxes the mirror 
metric on the internal space might no longer be well-defined. 
In our analysis it will be sufficient to characterize these generalized 
spaces by the existence of special even and odd forms not making use 
of an associated metric.

This paper is organized as follows. At the end of this introduction we give 
a short summary of our results.  
In section \ref{SU3} we briefly review some mathematical facts 
about $SU(3)$ structure manifolds and 
comment on the compactifications of the type II supergravity on these 
spaces. We immediately turn to the definition of the orientifold projections
of the type IIA/B theories in section \ref{projection}. This
allows us to determine the $N=1$ spectrum of the four-dimensional 
supergravity theories arising in the orientifold compactifications
in section \ref{OSpectrum}. The K\"ahler potentials and their relation to the Hitchin functionals
are discussed in section \ref{Kaehlerpot}. In section \ref{Superpot} we derive the superpotentials
of type IIA and type IIB orientifolds induced by the background fluxes and the torsion of $\cM_6$.
In order to fully identify these superpotentials under mirror symmetry the compactifications
need to be performed on a more general class of spaces $\cM_{\tilde Y}$. In section \ref{Mirror} we 
use our results on the K\"ahler and superpotentials to conjecture a possible identification 
of part of the magnetic NS-NS fluxes with properties of the mirror space $\cM_{\tilde Y}$.   
\\[.4cm]
\textit{Summary of results \\[.2cm]}
 For the convenience of the 
reader we will here briefly summarize our results.
In type IIA orientifolds with $O6$ planes, the globally defined three-form $\Omega$ 
is combined into a normalized three-form $\Pi^{\rm odd} = C \Omega$, 
where $C$ is proportional to the dilaton $e^{-\hat \phi}$. 
The real part of this form is complexified with the R-R three-form $\hat C_3$ 
with indices entirely on $\cM_6$ into the combination 
$ \Pi^{\rm odd}_c = \hat C_{3} + i\R(\Pi^{\rm odd})$. The globally defined two-form
$J$ is complexified with the NS-NS field $\hat B_2$ as $J_c = -\hat B_2 + i J$. The chiral fields of the
four-dimensional theory arise by expanding the complex forms $J_c$ and $ \Pi^{\rm odd}_c$ 
into an appropriate, not necessarily finite, set of real two- and three-forms of $\cM_6$. The complex scalar 
coefficients in this expansion are the bosonic fields in the chiral multiplets.
The K\"ahler potential on the chiral field space is given by 
 \beq \label{Kaehler_A}
    K \big[ J_c ,\Pi_c^{\rm odd} \big] = - \ln\big[ -i \int_{\cM_6}  \big< \Pi^{\rm ev}, \bar \Pi^{\rm ev} \big>\big] - 2 \ln\big[ i \int_{\cM_6}   \big< \Pi^{\rm odd}, \bar \Pi^{\rm odd} \big> \big] \ ,
 \eeq
where $\Pi^{\rm ev} = e^{J_c}$. The anti-symmetric pairing $\big<\cdot , \cdot \big>$ is defined in \eqref{def-Mukai} 
and replaces the wedge product.
The K\"ahler potential can be identified as the logarithm of the Hitchin functionals for
two- and three-forms on $\cM_6$ \cite{Hitchin:2000jd}. The superpotential for the chiral multiplets is 
given by \footnote{See also refs.~\cite{Derendinger:2004jn,VZ,Berglund,GLW,Camara:2005dc,Granarev}.}
\beq \label{sup_IIA}
   W \big[ J_c ,\Pi_c^{\rm odd} \big] = \int_{\cM_6} \big< F^{\rm ev} + d_H \Pi^{\rm odd}_c,  \Pi^{\rm ev}\big>
\eeq
where $F^{\rm ev}$ is the background flux of the even R-R field strengths. The NS-NS background flux 
$H_3$ of the NS-NS field strength $d\hat B_2$ arises through the exterior derivative $d_H = d - H_3 \wedge$.
The superpotential is readily shown to be holomorphic in the complex $N=1$ chiral multiplets.
In the expression \eqref{sup_IIA} both $d \Pi^{\rm odd}_c$ and $d \Pi^{\rm ev}$ are linear in the complex 
coordinates and indicate a deviation 
from the Calabi-Yau compactifications where $J$ and $\Omega$ are closed.

In type IIB orientifold compactifications the role of even and odd forms is interchanged.
One combines the globally defined two-form $J$ together with the B-fields $\hat B_2$ and
the dilaton into the complex even form $\Phi^{\rm ev} = e^{-\hat \phi} e^{-\hat B_2 + iJ}$.
In orientifolds with $O3/O7$ planes the real part of this form is complexified with 
the sum of even R-R potentials while it contains the imaginary part of $\Phi^{\rm ev}$ for $O5/O9$ orientifolds:
\bea
   O3/O7:\quad  \Phi^{\rm ev}_c &=& e^{-\hat B_2} \wedge \hat C^{\rm ev} + i \R(\Phi^{\rm ev})\ , \\
   O5/O9:\quad  \Phi^{\rm ev}_c &=& e^{-\hat B_2} \wedge \hat C^{\rm ev} + i \I(\Phi^{\rm ev})\ ,
\eea
where $e^{-\hat B_2} \wedge \hat C^{\rm ev}$ contains only forms with all indices on the internal manifold.
The complex forms $\Phi^{\rm ev}_c$ are expanded into real even forms on the manifold $\cM_6$ 
with complex scalar coefficients in four space-time dimensions. These complex fields 
are the bosonic components of a set of chiral multiplets. The expansion 
is chosen in accord with the orientifold projection which differs for $O3/O7$ and $O5/O9$ orientifolds. 
Additional chiral multiplets are complex scalars $z$ parameterizing independent degrees of freedom 
of the globally defined three-form $\Phi^{\rm odd} = \Omega$. 
The K\"ahler potential for all chiral multiplets is given by 
\beq \label{Kaehler_B}
    K \big[ z ,\Phi_c^{\rm ev} \big] = - \ln \big[- i \int_{\cM_6}  \big< \Phi^{\rm odd}, \bar \Phi^{\rm odd} \big> \big]
    - 2 \ln \big[i \int_{\cM_6}   \big< \Phi^{\rm ev}, \bar \Phi^{\rm ev} \big> \big] \ .
\eeq
Non-trivial NS-NS and R-R background fluxes $H_3$ and $F_3$ as well as the torsion of $\cM_6$ 
induce a superpotential for the chiral fields. It differs for the two type IIB setups 
and reads$\ ^2$
\beq \label{sup_IIB}
     W^{O3/O7}= \int_{\cM_6}  \big<F_3 + d_H \Phi^{\rm ev}_c,\Phi^{\rm odd}  \big>\ ,\qquad \quad 
     W^{O5/O9} = \int_{\cM_6}  \big<F_3  + d \Phi^{\rm ev}_c , \Phi^{\rm odd}  \big> \ ,     
\eeq
where $d_H = d -H_3\wedge$.
In addition several D-terms arise due to fluxes and torsion, which are more carefully discussed in a 
separate publication \cite{TGinprep}. In both type IIA and type IIB dual linear multiplets can become massive.
The scalar function encoding their kinetic terms are the Legendre transforms of the K\"ahler potentials
given above. 

Note that in the large volume and large complex structure limit the type 
II K\"ahler potentials are formally mirror symmetric under the
exchange $\Pi^{\rm ev/odd} \leftrightarrow \Phi^{\rm odd/ev}$. The complex forms
$\Pi^{\rm odd}_c$ and $\Phi^{\rm ev}_c$ are linear in the complex fields 
and identified under the mirror map. The type IIA and type IIB 
superpotentials cannot be identified under mirror symmetry. This 
is due to the fact that the dual of half of the NS-NS flux $H_3$ has 
no mirror parter. Choosing a symplectic basis of harmonic 
three-forms on $\cM_6$ electric and magnetic NS-NS fluxes 
can be distinguished.
We propose that the mirror for part of the 
electric and magnetic NS-NS fluxes $H_3^{Q}$ arises if one 
compactifies on a more general class of spaces.
In this conjecture one can allow for all but one magnetic and electric flux direction.\footnote{Interpreting mirror symmetry 
as three T-dualities \cite{Strominger:1996it}, the forbidden magnetic flux is the one having only indices 
in the T-dualized directions. Setting this flux quantum to zero, the dual space was termed 
the $Q$-space in ref.~\cite{Shelton:2005cf}. Hence, the index $Q$ on $H^Q_3$.}
The dual spaces are extensions of almost generalized complex manifolds with a more 
generic globally defined odd form. More precisely, in addition 
to the three-form $\Omega$ the globally defined odd forms 
$\Pi^{\rm odd}$ and $\Phi^{\rm odd}$ locally contain a one- and 
five-form $\Omega_1$ and $\Omega_5$ as \footnote{%
From a mathematical point of view, the forms $\Pi^{\rm odd}$ and $\Phi^{\rm odd}$ are expected to 
undergo type changes when moving along the internal manifold \cite{Gualtieri}.}
\beq
  \Pi^{\rm odd} = e^{-\hat B_2}\wedge (C\Omega_1 + C\Omega + C\Omega_5)\ , \qquad \qquad \Phi^{\rm odd} = e^{-\hat B_2}\wedge(\Omega_1 + \Omega + \Omega_5)\ .
\eeq
For the general odd forms $\Pi^{\rm odd}$ and $\Phi^{\rm odd}$ 
the K\"ahler potentials \eqref{Kaehler_A} and \eqref{Kaehler_B} are replaced by 
the extended Hitchin functionals introduced in refs.~\cite{Hitchin:2004ut,Hitchin:2005in}. Furthermore, 
using a fermionic reduction the superpotentials
\eqref{sup_IIA} and \eqref{sup_IIB} are shown to naturally generalize to 
the odd forms $\Pi^{\rm odd} $ and $ \Phi^{\rm odd} $. Also the complex form 
$\Pi^{\rm odd}_c$ including the 
R-R fields is generalized to
\beq
    \Pi^{\rm odd}_c = e^{-\hat B_2} \wedge \hat C^{\rm odd} + i  \R(\Pi^{\rm odd})\ ,
\eeq
where $e^{-\hat B_2} \wedge \hat C^{\rm odd}$ contains only forms with all indices on the internal manifold.

In a finite reduction the magnetic fluxes arise as the mirror of the torsion $d\Omega_1$ such that 
\beq \label{mirror_magn}
    d \R(\Omega_1+ \Omega + \Omega_5) \quad \leftrightarrow\quad  H_3^{Q}\ ,
\eeq
where the electric NS-NS fluxes are identified as the mirrors of $d\Omega$ as proposed 
in ref.~\cite{GLMW}. Hence, `generalized half-flat' manifolds
obeying $d \I(\Omega_1+ \Omega + \Omega_5) = 0$ and generically non-zero 
$d\R(\Omega_1+ \Omega + \Omega_5)$ are candidate mirrors of NS-NS flux compactifications.
We provide evidence for the identification \eqref{mirror_magn} by comparing 
the holomorphic superpotentials including the corrections due to $d\Omega_1$. 
For these generalized spaces the role of the tangent bundle $T \cM_6$ is taken
by the generalized tangent bundle $E$ locally given by $T\cM_6 \oplus T^* \cM_6$ \cite{Hitchin:2005in}.
Supersymmetry implies that $E$ has a structure
group $SU(3)\times SU(3)$ \cite{KL,GLW}. These generalized geometries might not necessarily 
descend to standard Riemannian manifolds with metric on $T\cM_6$. It is 
expected that they are more closely resemble the non-geometric compactifications 
introduced in refs.~\cite{Dabholkar:2005ve}.
The relation to the non-commutative background suggested in refs.~\cite{Mathai:2005fd} has to 
be clarified.

\section{Manifolds with $SU(3)$ structure}\label{SU3}
%
It is a well-known fact that type II supergravity compactified on 
a Calabi-Yau sixfold leads to an $N=2$ supergravity theory in four
space-time dimensions. In the absence of fluxes the effective four-dimensional 
theory contains no potential for the scalar fields and all vacua are Minkowski preserving 
the full supersymmetry. This changes as soon as 
we include background fluxes and localized sources such as D-branes and 
orientifold planes. In these situations it is a non-trivial task 
to perform consistent compactifications such that the four-dimensional effective theory remains supersymmetric. 
In particular, this is due 
to the fact, that the inclusion of sources forces 
the geometry to back-react. For example in 
orientifolds with $D3$ branes and fluxes the spacetime has to be non-trivially 
warped over an internal conformally Calabi-Yau manifold \cite{GKP}.   
In other situations the internal manifold is no longer directly related to 
a Calabi-Yau manifold and a more general class of compactification manifolds has to be 
taken into account \cite{Granarev}. 

In this section we discuss such a more general set 
of six-manifolds which yield upon compactification an $N=2$ supergravity 
theory in four space-time dimensions (see, for example, \cite{GLMW,Cardoso:2002hd,GLW}). 
To start with we specify the Kaluza-Klein Ansatz for the metric background. 
Topologically our ten-dimensional space-time is taken to be a product $M_{3,1}\times \cM_6$,  
where $M_{3,1}$ is a four-dimensional non-compact space and $\cM_6$
is a compact six-dimensional manifold.
The background metric is block-diagonal and reads
\beq\label{metricsplit}
     ds^2 = e^{2\Delta(y)} g_{\mu\nu}(x) dx^\mu dx^\nu + g_{mn}(y) dy^mdy^n\ ,
\eeq
where $x^\mu,\mu=0,\ldots, 3$ are the coordinates on $M_{3,1}$ while
$y^m, m=1,\ldots,6$ are the coordinates of $\cM_6$. Here $g_{\mu\nu}$ is 
the metric on $M_{3,1}$ and $g_{mn}$ is the metric on the internal 
manifold $\cM_6$. Note that the metric \eqref{metricsplit} generically includes a 
non-trivial warp factor $\Delta(y)$. However, in the following we
restrict our analysis to a large volume regime where $\Delta$ is approximately constant.\footnote{It would be 
desirable to extend our analysis to a general $\Delta(y)$ along the 
lines of \cite{Giddings:2005ff,Firouzjahi:2005qs}.}    

The amount of supersymmetry preserved by $\cM_6$ can 
be obtained by counting supercharges. 
Type II theories admit 32 supercharges in $D=10$ which can be represented 
by two (Majorana-Weyl) spinors $\epsilon^{(10)}_{1,2}$. 
In type IIA the two spinors have opposite chirality, while in type IIB
they are of the same chirality. Demanding
$N=2$ supersymmetry in four space-time dimensions the internal manifold has to admit 
one globally defined spinor $\eta$.\footnote{Note that in \cite{GLW} it was argued that $N=2$ supersymmetry can
be obtained by compactifying on a manifold with two globally defined spinors, which may coincide 
at points in $\cM_6$. We will come back to this generalization in section \ref{Mirror}.}
We decompose the ten-dimensional spinors as 
\bea\label{epsilon}
    \text{IIA:}\quad \epsilon^{(10)}_1&=& \epsilon_1\otimes\eta_+ +\bar \epsilon_1\otimes
      \eta_-
    \, \ \ \ \quad \text{IIB:}\quad  \epsilon^{(10)}_1= \epsilon_1\otimes\eta_+ + \bar \epsilon_1
    \otimes\eta_-\, ,\nn\\
    \epsilon^{(10)}_2&=& \epsilon_2\otimes\eta_- +\bar \epsilon_2\otimes
    \eta_+\qquad\qquad \quad \epsilon^{(10)}_2 = \epsilon_2\otimes\eta_+ + \bar \epsilon_2
    \otimes\eta_-\, ,
\eea
where $\epsilon_{1,2}$ and $\bar \epsilon_{1,2}$ are four-dimensional Weyl 
spinors which label the preserved $N=2$ supersymmetry. 
The spinors are chosen such that $\epsilon_{1,2}$ have positive 
four-dimensional chiralities and $\bar \epsilon_{1,2}$ have negative chiralities.
We indicate the six-dimensional chirality of the globally defined spinor $\eta$
by a subscript $\pm$. These spinors are related by complex conjugation 
$(\eta_\pm)^* = \eta_\mp$ and normalized as $\eta^\dagger_\pm \eta_\pm = \frac{1}{2}$.
We summarize our spinor conventions in appendix \ref{Spinors}.

The existence of one globally defined spinor $\eta$ reduces
the structure group of the internal manifold from $SO(6)$ to $SU(3)$ \cite{CS,salamonb,joyce}. 
If this spinor is also covariantly constant with respect to the 
Levi-Civita  connection the manifold has $SU(3)$ holonomy and hence satisfies 
the Calabi Yau conditions. For a general $SU(3)$ structure manifold the spinor $\eta$ is not any more 
covariantly constant. The failure of the Levi-Civita connection to 
annihilate the spinor $\eta$ is measured by the contorsion tensor $\tau$. Using $\tau$
one defines a new connection $D^{T}_m$ such that 
\beq \label{torsion_cond}
    D^{T}_m \eta = \big(D^{LC}_m- \tfrac{1}{4}\tau_{m np} \gamma^{np}\big) \eta = 0\ ,
\eeq
where $\gamma^{mn} = \frac{1}{2!} \gamma^{[m} \gamma^{n]}$ is the anti-symmetrized product 
of six-dimensional gamma matrices. The spinor $\eta$ is now covariantly constant with 
respect to the new connection $D^{T}_m$, which additionally contains the information about the torsion $\tau$.

Equivalently to the spinor language, $SU(3)$ structure manifolds can be characterized by the
existence of two no-where vanishing forms $J$ and $\rho_\eta$. 
The form $J$ is a real two-form while $\rho_\eta$ is a real three-form on $\cM_6$.
We denote the space of real $n$-forms on $\cM_6$ by 
\beq
   \Lambda^n T^* \equiv \Lambda^n ( T^* \cM_6)\ ,
\eeq
such that $J \in  \Lambda^2 T^*$ and $\rho_\eta \in  \Lambda^3 T^*$. The index $\eta$ 
indicates a specific normalization chosen as we define $J$ and $\rho_\eta$ in terms of the spinor $\eta$.
Here we first give a characterization independent of $\eta$ following the definition of Hitchin \cite{Hitchin:2000jd}.
In this case one demands that $J$ and $\rho_\eta$ are stable forms, i.e.~are elements of
open orbits under the action of general linear transformations $GL(6,\bbR)$ at every point of the 
tangent bundle $T\cM_6$. These forms define a reduction of the structure group from $GL(6,\bbR)$ to
$SU(3)$ if they furthermore satisfy
\beq \label{SU(3)-constr}
    J  \wedge J \wedge J = \tfrac{3}{2} \rho _\eta\wedge \hat \rho_\eta\ ,\qquad \qquad
    J \wedge \rho_\eta=0\ ,
\eeq
where $\hat\rho_\eta=*\rho_\eta$ is shown to be a function of $\rho_\eta$ only as we review in appendix \ref{GCG-HF}.

The spinor and the form descriptions of the $SU(3)$ structure are related by expressing the components of the
two-form $J$ and the complex three-form $\Ox_\eta=\rho_\eta+i\hat\rho_\eta$ in terms of the spinor $\eta$ as 
\beq\label{JOdef}
    J^{mn} \ =\ \mp 2\ii \eta^\dagger_\pm\gamma^{mn}\eta_\pm \ , \qquad \
    \Omega^{mnp}_\eta = 2\eta^\dagger_-\gamma^{mnp}\eta_+ \ , \qquad \ 
   \bar \Omega^{mnp}_\eta = 2\eta^\dagger_+\gamma^{mnp}\eta_-\ .
\eeq
Later on we will relate $\Omega_\eta$ to the three-form $\Omega$ used 
in the compactification by an appropriate rescaling. In the normalization \eqref{JOdef} one can 
apply Fierz identities to derive the $SU(3)$ structure 
constraints equivalent to \eqref{SU(3)-constr},
\beq \label{SU(3)-cond}
   J \wedge J \wedge J =\tfrac{3\ii}{4}\,\Omega_\eta \wedge \bar \Omega_\eta \ ,
   \qquad \qquad J \wedge \Omega_\eta = 0\ .
\eeq
Moreover, defining $I_m^{\ n} = J_{mp}g^{pn}$ by raising one of the indices on $J$ by the metric $g_{mn}$ one
shows that 
\beq
  I_p^{\ n} I_{m}^{\ p} = - \delta_m^n\ , \qquad \qquad I_n^{\ p} I_{m}^{\ q}\, g_{pq}= g_{mn} \ .
\eeq
This implies that $I_m^n$ is an almost complex structure with respect to 
which the metric $g_{mn}$ is hermitian. The almost complex structure 
can be used to define a $(p,q)$ grading of forms. Within this decomposition the form  
$J$ is of type $(1,1)$ while $\Omega_\eta$ is of type $(3,0)$.

The condition \eqref{torsion_cond} can be translated to the form language
implying that neither $J$ nor $\Ox_\eta$ are closed. The non-closedness is parameterized 
by the torsion $\tau$ which decomposes under $SU(3)$ into irreducible representations.  
The representations are conveniently encoded by five torsion classes $\cW_i$ defined as
\cite{CS,salamonb,Cardoso:2002hd},
\bea\label{dJ}
    dJ&=&-\tfrac{3}{2}\I (\WV_1\overline\Ox_\eta)+\WV_4\wedge J+\WV_3\nn\\
     d\Ox_\eta
   &=&\WV_1 J \wedge J+\WV_2\wedge J+\overline\WV_5\wedge\Ox_\eta \ ,
\eea
with constraints $J\wedge J\wedge\WV_2=J\wedge\WV_3=\Ox_\eta \wedge\WV_3=0$.
The pattern of vanishing torsion classes defines the properties of 
the manifold $\cM_6$. For example, $\cM_6$ is complex in case $\cW_1 = \cW_2 = 0$.
Of particular interest are half-flat manifolds, since they are believed to arise as mirrors of 
flux compactifications \cite{GLMW}. These are defined by $\cW_4= \cW_5=0$ and $\I \cW_1 = \I \cW_2 = 0$. 
Equivalently, by using \eqref{dJ}, half-flat manifolds are defined 
by the two conditions 
\beq \label{def-half}
   dJ\wedge J=0\, ,\quad d\I \Ox_\eta=0\ ,
\eeq 
while $dJ$ and $d\R \Ox_\eta$ are not necessarily vanishing.

As discussed in the beginning of this section, the compactification on $SU(3)$ structure 
manifolds leads to an $N=2$ supergravity theory. The supersymmetry is further reduced to $N=1$
by imposing an appropriate orientifold projection. The aim of the next section is to define this
projection and to determine the characteristic data of the four-dimensional supergravity theory
obtained by compactification on an $SU(3)$ structure orientifold.

%
\section{Type II $SU(3)$ structure orientifolds}

In this section we study compactifications of type IIA and type IIB supergravity 
on $SU(3)$ structure orientifolds. As reviewed in the previous section 
compactifications on $SU(3)$ structure manifolds lead to four-dimensional 
theories with $N=2$ supersymmetry. The inclusion of D-branes and orientifold planes
further reduced the amount of supersymmetry. In order that the four-dimensional effective 
theory possesses $N=1$ supersymmetry the D-branes and orientifold planes can not 
be chosen arbitrarily but rather have to fulfill certain supersymmetry 
conditions (BPS conditions).\footnote{In addition, the configurations of 
D-branes, orientifold planes and fluxes have to obey consistency conditions 
such as the cancellation of tadpoles 
\cite{VZ, DeWolfe:2005uu,Camara:2005dc,Shelton:2005cf,Aldazabal:2006up,Villadoro:2006ia}.} In this paper our main focus will be the bulk theory. 
In section \ref{projection} we specify 
the orientifold projections which yield supersymmetric orientifold planes preserving
half of the $N=2$ supersymmetry. We show in section \ref{OSpectrum} that the orientifold invariant  
spectrum arranges into $N=1$ supermultiplets. Performing a Kaluza-Klein reduction 
allows us to determine the K\"ahler potential 
of the four-dimensional theory in section \ref{Kaehlerpot}.
The discussion of the superpotential induced by the fluxes and torsion 
will be presented in section \ref{Superpot}.

\subsection{The orientifold projection \label{projection}}

In this section we specify the orientifold projections under consideration. 
We start from type II string theory and compactify on a $SU(3)$ structure manifold $\cM_6$. 
In addition we mod out by orientation reversal of the string world-sheet $\Omega_p$ together with an  
internal symmetry $\sigma$ which 
acts solely on $\cM_6$ but leaves the $D=4$ space-time untouched.
We will restrict ourselves to involutive symmetries ($\sigma^2 = 1$) of $\cM_6$. 
In a next step we have to specify additional properties of $\sigma$ in order 
that it provides a symmetry of the string theory under consideration. The type IIA and type IIB cases
are discussed in turn. 
\\[.4cm]
\textit{\underline{Type IIA orientifold projection}}\\[.4cm]
%
The orientifold projection for type IIA $SU(3)$ structure orientifolds can be 
obtained in close analogy to the Calabi-Yau case. Recall that for 
Calabi-Yau orientifolds the demand for $N=1$ supersymmetry implies 
that $\simga$ has to be an anti-holomorphic and isometric involution 
\cite{AAHV,BBKL,BH}.  This fixes the action of $\simga$ on the K\"ahler form $J$
as $\sigma^* J= -J$, where $\sigma^*$ denotes the pull-back of the map $\sigma$. 
Furthermore, supersymmetry implies that $\sigma$ acts 
non-trivially on the holomorphic three-form $\Omega$.
This naturally generalizes to the $SU(3)$ structure case, since we can still
assign a definite action of $\simga$ on the globally defined two-form 
$J$ and three-form $\Omega$ defined in \eqref{JOdef}.  
Together the orientifold constraints read
\beq \label{constrA}
  \sigma^* J\ =\ -J\ ,\qquad\qquad\sigma^* \Omega\ =\ e^{2i\theta}\bar\Omega\ ,
\eeq
where $e^{2i\theta}$ is a phase and we included a factor 2 for later 
convenience. Note that the second condition in \eqref{constrA} can be 
directly inferred from the compatibility of $\sigma$ with the 
$SU(3)$ structure condition $\Omega \wedge \bar \Omega \propto J \wedge J \wedge J$ given in \eqref{SU(3)-cond}.
In order that $\sigma$ is a symmetry of type IIA string theory 
it is demanded to be an isometry. Hence, the first condition in \eqref{constrA}
implies that $\sigma$ yields a minus sign when applied to the almost 
complex structure $I_n^m=J_{np}g^{pm}$ introduced in the previous section. 
This reduces to the anti-holomorphicity of $\sigma$ if $I_n^m$ is integrable as in the 
Calabi-Yau case.
The complete orientifold projection takes the form 
\footnote{The factor $(-1)^{F_L}$ is included in $\mathcal{O}$
                              to ensure that $\OO^2=1$ on all states.}
\beq \label{oproj}
  \mathcal{O} = (-1)^{F_L} \Omega_p \sigma\ ,
\eeq
where $\Omega_p$ is the world-sheet parity and $F_L$ is the space-time 
fermion number in the left-moving sector. 

The orientifold planes arise 
as the fix-points of $\sigma$. Just as in the Calabi-Yau case supersymmetric 
$SU(3)$ structure orientifolds generically contain $O6$ planes.  
This is due to the fact, that the fixed point set of $\sigma$ in 
$\cM_6$ are three-cycles $\Lambda_{O6}$ supporting the internal part of the 
orientifold planes. These are calibrated with respect to the real form 
$\R(e^{-i\theta}\Omega)$ such that 
\beq \label{calibr-O6}
  \rm{vol}(\Lambda_{O6})\propto \R(e^{-i\theta}\Omega)\ ,\qquad \quad \I(e^{-i\theta}\Omega)|_{\Lambda_{O6}} = J|_{\Lambda_{O6}} = 0
\eeq 
where $\rm{vol}(\Lambda_{O6})$ is the induced volume form on $\Lambda_{O6}$ and the overall normalization of $\Omega$ was left undetermined. 
The conditions \eqref{calibr-O6} also allow us to give a more explicit expression for the phase $e^{i\theta}$
as
\beq\label{theta}
   e^{-2i\theta} = \bar Z(\Lambda_{O6})/ Z(\Lambda_{O6})\ ,
\eeq
where $Z(\Lambda_{O6})$ is given by $Z(\Lambda_{O6}) = \int_{\Lambda_{O6}} \Omega$.
This expression determines the transformation 
behavior of $\theta$ under complex rescalings of $\Omega$. Later on we include $e^{-i\theta}$
to define a scale invariant three-form $C\Omega$. \\[.4cm]
%
\textit{\underline{Type IIB orientifold projection}}\\[.2cm]
%
Let us turn to type IIB $SU(3)$ structure orientifolds. 
Recall that for type IIB Calabi-Yau orientifolds consistency requires $\sigma$ to be a 
holomorphic and isometric involution of $\cM_6$ \cite{AAHV,BH}. A holomorphic isometry leaves both the metric
and the complex structure of the Calabi-Yau manifold invariant, such that $\sigma^* J = J$. 
We generalize this condition to the $SU(3)$ structure case by demanding that the 
globally defined two-form $J$ defined in \eqref{JOdef} transforms as
\beq \label{transJB}
  \sigma^* J = J \ .
\eeq
Once again we impose that $\sigma$ is an isometry of the manifold $\cM_6$, 
such that \eqref{transJB} translates to the invariance of the almost complex structure $I_m^n$.
Due to this fact the involution respects the $(p,q)$-decomposition of forms. Hence
the $(3,0)$ form $\Omega$ defined in \eqref{JOdef} will be mapped to a 
$(3,0)$ form. Demanding the resulting form to be globally defined we have two possible choices
\beq \label{Omegatransf}
  (1)\quad O3/O7: \quad \sigma^* \Omega = - \Omega\ ,\qquad \qquad 
  (2)\quad O5/O9: \quad  \sigma^* \Omega = + \Omega\ ,
\eeq
where the dimensionality of the orientifold planes is determined by the 
dimension of the fix-point set of $\simga$ \cite{AAHV}. 
Correspondingly, depending on the transformation properties of  $\Omega$ two different 
symmetry operations are possible \cite{Sen,DP,AAHV,BH} 
\beq \label{o3-projection}
\mathcal{O}_{(1)} = (-1)^{F_L} \Omega_p \, \sigma\ ,\qquad \qquad
\mathcal{O}_{(2)} = \Omega_p \, \sigma
\eeq 
where $\Omega_p$ is the world-sheet parity and $F_L$ is the space-time 
fermion number in the left-moving sector. The type IIB analog of the calibration condition 
\eqref{calibr-O6} involves a contribution from the NS-NS two-form $\hat B_2$. It 
states that the even cycles of the orientifold planes in $\cM_6$ are calibrated with 
respect to the real or imaginary parts of $e^{-\hat B_2 + i J}$. The explicit form of
this condition can be found, for example, in refs.~\cite{BBS,MMMS,D-branecalib}.

\subsection{The orientifold spectrum \label{OSpectrum}}
%
Having specified the orientifold projections \eqref{oproj} and \eqref{o3-projection} of the type
IIA and type IIB orientifolds we can examine the invariant spectrum. Recall that the
bosonic NS-NS fields of both type IIA and type IIB supergravity are the scalar dilaton $\hat \phi$, 
the ten-dimensional metric $\hat G_{MN}$ and the two-form $\hat B_2$.\footnote{The hat on 
the fields indicates ten-dimensional quantities.} 
Considering the theory on the product space $M_{3,1}\times \cM_6$ these fields decompose 
into $SU(3)$ representation as summarized in the table \ref{NS} \cite{GLW}. We denote the $SU(3)$ representation $\mathbf{R}$
with four-dimensional spin $\mathbf{s}$ by $\mathbf{R_{s}}$. For example, a triplet under $SU(3)$ yielding 
a vector in four-dimensions is denoted by $\mathbf{3_1}$. A four-dimensional tensor (or pseudo-scalar) is
indicated by an index $\mathbf{T}$. 
\begin{table}[h]
\begin{center}
\begin{tabular}{|c|c|l|} \hline
\rule[-0.3cm]{0cm}{0.8cm}
\multirow{3}{6mm}[-3.5mm]{$ \hat G$}&$
g_{\mu\nu}$& $ \rep{1}_\rep{2}$ \\ 
 \cline{2-3}
\rule[-0.3cm]{0cm}{0.8cm}
   & $ g_{\mu m}$ & $(\rep{3}+\rep{\bar{3}})_\rep{1}$ \\\cline{2-3}
\rule[-0.3cm]{0cm}{0.8cm} 
 &$ g_{mn}$ & $\rep{1}_\rep{0}+(\rep{6}+\rep{\bar{6}})_\rep{0}+\rep{8}_\rep{0}$\\
  \hline
\rule[-0.3cm]{0cm}{0.8cm}
\multirow{3}{6mm}[-3.5mm]{$ \hat B_2$}& $
B_{\mu\nu}$ & $ \rep{1}_\rep{T}$ \\ \cline{2-3}
\rule[-0.3cm]{0cm}{0.8cm}
   & $ B_{\mu m}$ & $(\rep{3}+\rep{\bar{3}})_\rep{1}$ \\ \cline{2-3}
\rule[-0.3cm]{0cm}{0.8cm} 
 &$ B_{mn}$ & $\rep{1}_\rep{0}+(\rep{3}+\rep{\bar{3}})_\rep{0}+\rep{8}_\rep{0}$\\ 
  \hline\rule[-0.3cm]{0cm}{0.8cm}
$ \hat \phi $&$\phi$& $\rep{1}_\rep{0}$\\
\hline
\end{tabular}
\caption{\small 
\textit{ Decomposition of the NS sector in $SU(3)$ representations}}
\label{NS}
\end{center}
\end{table}

In the R-R sector type IIA 
consists of odd forms $\hat C_{2n-1}$, while type IIB consists of even forms 
$\hat C_{2n}$. Their decomposition into 
$SU(3)$ representations is displayed in 
tables \ref{RRIIA} and \ref{RRIIB} \cite{GLW}. We list only the decompositions of 
the $\hat C_1$ and $\hat C_3$ in type IIA and $\hat C_0, \hat C_2, \hat C_4$ 
in type IIB. The higher forms are related to these fields via Hodge duality of 
their field strengths. The form $\hat C_4$ has a self-dual field strength and hence only half of its components are physical.
\begin{table}[htb]
\begin{center}
\begin{tabular}{|c|c|l|} \hline
\rule[-0.3cm]{0cm}{0.8cm}
\multirow{2}{6mm}[-3.5mm]{$\hat  C_1$}&$
 C_{\mu}$& $ \rep{1}_\rep{1}$ \\ 
 \cline{2-3}
\rule[-0.3cm]{0cm}{0.8cm}
   & $ C_{m}$ & $(\rep{3}+\rep{\bar{3}})_\rep{0}$ \\
  \hline
\rule[-0.3cm]{0cm}{0.8cm}
\multirow{3}{6mm}[-3.5mm]{$ \hat C_3$}& $
 C_{\mu\nu p}$ & $(\rep{3}+\rep{\bar{3}})_\rep{T}$ \\ \cline{2-3}
\rule[-0.3cm]{0cm}{0.8cm}
   & $ C_{\mu np}$ & $\rep{1}_\rep{1}
+(\rep{3}+\rep{\bar{3}})_\rep{1} +\rep{8}_\rep{1}$ \\ \cline{2-3}
\rule[-0.3cm]{0cm}{0.8cm} 
 &$ C_{mnp}$ & $(\rep{1} + \rep{1})_\rep{0} +(\rep{3}+\rep{\bar{3}})_\rep{0}
        +(\rep{6}+\rep{\bar{6}})_\rep{0} $\\
\hline
\end{tabular}
\caption{\small 
\textit{ Type IIA decomposition of the RR sector in $SU(3)$
  representations}}
\label{RRIIA}
\end{center}
\end{table}%
\begin{table}[htb]
\begin{center}
\begin{tabular}{|c|c|l|} \hline
\rule[-0.3cm]{0cm}{0.8cm}
$\hat C_{0}$&$C_0$&$\rep{1}_\rep{0}$\\ \hline\rule[-0.3cm]{0cm}{0.8cm}
\multirow{3}{5mm}[-3.5mm]{$ \hat C_2$}& $
 C_{\mu\nu}$ & $ \rep{1}_\rep{T}$ \\ \cline{2-3}
\rule[-0.3cm]{0cm}{0.8cm}
   & $ C_{\mu m}$ & $(\rep{3}+\rep{\bar{3}})_\rep{1}$ \\ \cline{2-3}
\rule[-0.3cm]{0cm}{0.8cm} 
 &$ C_{mn}$ & $\rep{1}_\rep{0}+(\rep{3}+\rep{\bar{3}})_\rep{0}+\rep{8}_\rep{0}$\\
  \hline\rule[-0.3cm]{0cm}{0.8cm}
\multirow{2}{5mm}[-3.5mm]{$ \hat C_4$}&$
 C_{\mu npq}$& $  \tfrac{1}{2}\left[ 
     (\rep{1}+ \rep{1})_\rep{1} +  
     (\rep{3}+ \rep{\bar{3}})_\rep{1} + 
     (\rep{6}+ \rep{\bar{6}})_\rep{1}\right] $ \\ 
 \cline{2-3}
\rule[-0.3cm]{0cm}{0.8cm}
   & $  C_{mnpq}/ C_{\mu\nu mn}$ & $\rep{1}_\rep{0}+(\rep{3}+\rep{\bar{3}})_\rep{0}+\rep{8}_\rep{0}$ \\
  \hline
\end{tabular}
\caption{\small 
\textit{ Type IIB decomposition of the RR sector in $SU(3)$
  representations}}
\label{RRIIB}
\end{center}
\end{table}

The fields arising in this decomposition can be arranged into one $N=8$ gravitational multiplet.
As discussed in ref.~\cite{GLW}, a possible reduction to standard  $N=2$ supergravity theory with 
a gravity multiplet as well as some vector, hyper and tensor multiplets is obtained by removing 
all the triplets from the spectrum. In particular, this amounts to discarding all four-dimensional fields which arise in 
the expansion of the ten-dimensional fields into one- and five-forms on $\cM_6$. 

In a second step we impose the orientifold projection to further reduce  
to an $N=1$ supergravity theory. Independent of the properties 
of the internal manifold we can give the transformation behavior of all supergravity 
fields under the world-sheet parity $\Omega_p$ and $(-1)^{F_L}$ \cite{JPbook, AD}. $\Omega_p$ acts on $\hat B_2$
with a minus sign, while leaving the dilaton $\hat \phi$ and the ten-dimensional metric $\hat G$
invariant. To display the transformation behavior of the R-R fields we introduce 
the parity operator $\lambda$ by
\beq \label{def-lambda}
   \lambda(\mathcal{C}_{2n}) = (-1)^{n} \mathcal{C}_{2n}\ , \qquad \lambda(\cC_{2n-1}) 
   = (-1)^{n} \cC_{2n-1}\ ,
\eeq
where $\cC_{2n}$ are even and $\cC_{2n-1}$ are odd forms. Evaluated on the R-R forms 
$\lambda$ is minus the world-sheet parity operator $\Omega_p$ such that 
\beq \label{parity-transf}
   \Omega_p \hat C_{k} = - \lambda(\hat C_{k}) \ ,
\eeq
where $k$ is odd for type IIA and even for type IIB. Finally, $(-1)^{F_L}$
acts on the R-R bosonic fields of the supergravity theories with a minus sign
while leaving the NS-NS fields invariant. 
\\[.4cm]
%
%
%
\textit{\underline{The type IIA orientifold spectrum}}\\[.4cm]
Let us now determine the invariant spectrum for type IIA 
orientifolds. It turns out to be convenient to combine 
the odd R-R forms $\hat C_{2n+1}$ as \cite{BKORvanP}
\beq \label{def-Codd}
  \hat C^{\rm odd} = \hat C_1 + \hat C_3 + \hat C_5 + \hat C_7 + \hat C_9\ .
\eeq
Note that only half of the degrees of freedom in $\hat C^{\rm odd}$ 
are physical, while the other half can be eliminated by a duality 
constraint \cite{BKORvanP}. Invariance under 
the orientifold projection $\mathcal{O}$ implies by using  
the transformation of the fields under $\Omega_p$ and $(-1)^{F_L}$
that the ten-dimensional fields have to transform as 
\beq \label{trans_BC_A}
   \sigma^*  \hat B_2 = -\hat B_2 \ , \qquad \qquad 
   \sigma^* \hat \phi = \hat \phi\ ,\qquad \qquad 
   \sigma^* \hat C^{\rm odd} = \lambda \big(\hat C^{\rm odd}\big)\ , 
\eeq
where the parity operator $\lambda$ is defined in \eqref{def-lambda} and we used \eqref{parity-transf}. 
It turns out to be convenient to combine the forms $\Omega$ and $J$ with
the ten-dimensional dilaton $\hat \phi$ and  $\hat B_2$ into new forms $\Pi^{\rm ev/odd}$ as
\beq \label{def-Pi}
   \Pi^{\rm ev} = e^{-\hat B_2 + iJ}\ , \qquad \qquad \Pi^{\rm odd} = C\Omega \ ,
\eeq 
where 
\beq \label{def-C}
   C = e^{-\hat \phi -i\theta} e^{(K^{cs} - K^{\rm K})/2}\ , \qquad   
   e^{-K^{cs}}=i \Omega \wedge \bar \Omega\ ,\qquad e^{-K^{\rm K}} = \tfrac{4}{3} J \wedge J \wedge J\ .
\eeq
In the expression for $C$ the form contributions precisely cancel such that $C$ is a complex 
scalar on $\cM_6$.
It depends on the ten-dimensional dilaton $\hat \phi$ 
and fixes the normalization of $\Omega$ such that the combination $C\Omega$ stays 
invariant under complex rescaling of $\Omega$.%
\footnote{Note that also
$\theta$ depends on the three-form $\Omega$ as given in \eqref{theta}. Hence, using the scaling behavior
of $\theta$ and $\Kcs$ one finds $C \rightarrow C e^{-f}$ as $\Omega \rightarrow e^f \Omega$ for every complex function $f$. } 
The four-dimensional dilaton is defined as 
\beq \label{def-D}
   e^{-2D} = \tfrac{4}{3}\int_{\cM_6} e^{-2\hat \phi} J \wedge J \wedge J\ ,
\eeq
and reduces to the definition $e^{- D} = e^{- \hat \phi} \sqrt{\text{Vol}(\cM_6)}$ in case 
$\hat \phi$ is constant along $\cM_6$.
Applied to the forms $\Pi^{\rm ev/odd}$ and $\hat C^{\rm odd}$ the orientifold conditions
\eqref{constrA} and \eqref{trans_BC_A}  are expressed as
\beq \label{const_PA_A}
   \sigma^* \Pi^{\rm ev} = \lambda\big( \Pi^{\rm ev} \big)\ ,  \qquad \qquad
  \sigma^* \Pi^{\rm odd} = \lambda\big(\bar \Pi^{\rm odd} \big)\ .
\eeq

In order to perform the Kaluza-Klein reduction one needs to 
specify the modes of the internal manifold $\cM_6$ used in the 
expansion of $\Pi^{\rm ev/odd}$ and $\hat C^{\rm odd}$. This implies that
one needs to specify a set of forms on $\cM_6$ which 
upon expansion yields the light fields in the spectrum of the four-dimensional 
theory. In general this issue is very hard to address and 
one can only hope to find an approximate answer in certain limits 
where the torsion is `small'. Most of the difficulty is due to the 
fact that a non-trivial torsion may not generate an additional 
scale below the Kaluza-Klein scale.\footnote{This is in contrast to standard 
RR and NS fluxes, which correspond to background values of the field
strengths of $\hat C^{\rm odd}$ and $\hat B_2$. The quantization condition 
implies that these fluxes can generate an intermediate scale. This 
allows to keep modes of the order of the flux scale, but discard all massive Kaluza-Klein 
modes.} Hence, discarding the Kaluza-Klein modes needs some justification.
Surprisingly, much of the analysis performed below does not explicitly depend
on the basis used in the expansion of $\Pi^{\rm ev/odd}$ and $\hat C^{\rm odd}$. 
We therefore only assume that the triplets in the $SU(3)$ decomposition 
are projected out while otherwise keeping the analysis general \cite{GLW}.
Later on  we restrict to a particular finite number of modes.

To implement the orientifold projection we note that the operator $\cP_6 = \lambda \sigma^*$ squares 
to the identity and thus splits the space of two- and three-forms $\Lambda^2 T^*$ and $\Lambda^3 T^*$ on $\cM_6$ into two eigenspaces as
\beq \label{decomp_2_3}
   \Lambda^2 T^* = \Lambda^2_+T^* \oplus \Lambda^2_-T^*\ ,\qquad\Lambda^3 T^* = \Lambda^3_+T^* \oplus \Lambda^3_-T^*\ ,
\eeq
where $ \Lambda^2_\pm T^*$ contains forms transforming with a $\pm$ sign under $\cP_6$.

In performing the Kaluza-Klein reduction one expands the 
forms $\Pi^{\rm ev/odd}$ and $\hat C^{\rm odd}$ into the 
appropriate subset of $\Lambda^2T^*$ and $\Lambda^3T^*$ 
consistent with the orientifold projection. The coefficients arising 
in these expansions correspond to the fields of the four-dimensional 
theory. In the case at hand the compactification has to result in 
an $N=1$ supergravity theory. The spectrum of this
theory consists of a gravity multiplet a number
of chiral multiplets and vector multiplets. 
Note that before the truncation to the light modes 
the number of multiplets is not necessarily finite, as the Kaluza-Klein tower consist of an 
infinite number of modes. These modes can acquire a mass via a generalized Higgs mechanism.
For example, a two-from can become massive by `eating' a vector \cite{mass_tensors}. In the 
following we will discuss the massless field content before such a Higgsing takes place.

Let us first concentrate on the $N=1$ chiral multiplets arising in the expansion 
of the forms $\Pi^{\rm ev}$. Due to supersymmetry the bosonic components of these 
multiplets span a complex K\"ahler manifold. Its complex structure can be determined
by specifying appropriate complex combinations of the forms $J$ and $\hat B_2$ which upon 
expansion into modes of the internal manifold yield the complex 
chiral coordinates. The globally defined two-form $J$ combines 
with the B-field into the complex combination \footnote{Note that the complex 
combination \eqref{def-J_c} precisely gives the correct coupling to the 
string world-sheet wrapped around supersymmetric two-cycles in $\cM_6$.}
\beq \label{def-J_c}
   J_c \equiv - \hat B_2 + i J \qquad \in  \quad \Lambda^2_+ T^*_\bbC \ ,
\eeq
where $J$ is given in the string frame. The field $\hat B_2$ is only 
extended along $\cM_6$, since due to \eqref{trans_BC_A} 
the four-dimensional two-form in $\hat B_2$ transforms with the 
wrong sign under the orientifold symmetry $\simga^*$ and hence
is projected out. In comparison to the general 
$SU(3)$ decomposition of $\hat B_2$ given in table \ref{NS} we only kept the
$\mathbf{1_0 + 8_0}$ representations while all other components left the spectrum.
The complex form $J_c$ is expanded
in real elements of  $\Lambda^2_+ T^*$ consistent with the orientifold projection 
\eqref{constrA}, \eqref{trans_BC_A} and the definition of $\lambda$ given in \eqref{def-lambda}.%
\footnote{Note that the eigenspaces $\Lambda^2_\pm T^*$
are obtained from the operator $\cP_6=\lambda \simga^*$ and hence differ by 
a minus sign from the eigenspaces of $\simga^*$.}
The coefficients of this expansion 
are complex scalar fields in four space-time dimensions parameterizing 
a manifold $\cM^{\rm K}$ and provide the 
bosonic components of chiral multiplets. 

Turning to the expansion of the R-R forms $\hat C^{\rm odd}$ 
we first note that $\hat C_1$ (and hence $\hat C_7$)
are completely projected out from the spectrum. The 
four-dimensional part of $\hat C_1$ is incompatible with the orientifold symmetry
as seen in \eqref{trans_BC_A}. On the other hand the internal 
part of $\hat C_1$ is a triplet under $SU(3)$ and hence discarded following the
assumptions made above. In contrast the expansion of $\hat C_3$ yields
four-dimensional scalars, vectors and three-forms. Therefore, we decompose 
\beq \label{decomp_C}
  \hat C_3 = C^{(0)}_3 + C^{(1)}_3 + C^{(3)}_3\ ,
\eeq 
where $C^{(n)}_3$ are $n$-forms in $M_{3,1}$ times $(3-n)$-forms in $\cM_6$.
More precisely, in order to fulfill the orientifold condition \eqref{trans_BC_A} 
the components $C^{(0)}_3,\, C_3^{(1)}$ and $C_3^{(3)}$ are 
expanded in forms $\Lambda^3_+T^*,\, \Lambda^2_-T^*$
and $\Lambda^0 T^*$ of $\cM_6$ respectively. The coefficients in 
this expansion correspond to four-dimensional real scalars, vectors 
and three-forms. Let us note that we project out fields which  
arise in  the expansion into one-forms on $\cM_6$ as well as all other triplets.
In summary the components kept, are the $\mathbf{1_1 + 8_1}$ and $\mathbf{(1+1)_0 + (6+\bar 6)_0}$
while all other representations in table \ref{RRIIA} have left the spectrum.  

The four-dimensional real scalars in $C_3^{(0)}$ need to combine with 
scalars arising in the expansion of $\Pi^{\rm odd}$ to form the 
components of chiral multiplets. The complex structure on the 
corresponding K\"ahler field space is defined through 
the complex form 
\footnote{Note that the complex 
combination \eqref{def-Pi_c} precisely gives the correct coupling to D-branes 
wrapped around supersymmetric cycles in $\cM_6$ \cite{BBS,MMMS,D-branecalib}.}
\beq \label{def-Pi_c}
   \Pi^{\rm odd}_c \equiv  C_3^{(0)} + i \R( \Pi^{\rm odd}) \qquad \in  \quad \Lambda^3_+ T^*_\bbC \ . 
\eeq
where we used that $\R( \Pi^{\rm odd})$ transforms with a plus sign as seen from eqn.~\eqref{const_PA_A}.
The complex  coefficients of $\Pi^{\rm odd}_c$ expanded 
in real forms $ \Lambda^3_+ T^*$ are the bosonic components of chiral multiplets.
Note that in the massless case theses chiral multiplets can be dualized to linear 
multiplets containing a scalar from $\R( \Pi^{\rm odd})$ and a two-form dual to the 
scalar in $ C_3^{(0)}$ \cite{BGG}. Due to the generality of our discussion 
both chiral and linear multiplets can become massive. 
The full $N=1$ spectrum for type IIA orientifold 
is summarized in table \ref{spectrumIIA}. 

\begin{table}[htb]
\begin{center}
\begin{tabular}{|c|c|c|} \hline
\rule[-0.3cm]{0cm}{0.8cm}
multiplet& bosonic fields&$\cM_6$-forms\\ \hline \hline
\rule[-0.3cm]{0cm}{0.8cm}
gravity multiplet& $g_{\mu\nu}$&\\ \hline\rule[-0.3cm]{0cm}{0.8cm}
\rule[-0.3cm]{0cm}{0.8cm}
chiral multiplets& 
 $J_c$ & $ \Lambda^2_+T^*$ \\ \hline
\rule[-0.3cm]{0cm}{0.8cm}
chiral/linear multiplets   & $ \Pi_c^{\rm odd}$ & $ \Lambda^3_+T^*$ \\ 
  \hline\rule[-0.3cm]{0cm}{0.8cm}
vector multiplets& $C_3^{(1)}$&$ \Lambda^2_-T^*$\\ \hline
\end{tabular}
\caption{\small 
\textit{$N=1$ spectrum of type IIA orientifolds}}
\label{spectrumIIA}
\end{center}
\end{table}

The analysis so far was not restricted to a finite set of fields. Even though 
most of the calculations can be performed in this more general setting 
we will also give a reduction to a four-dimensional theory with 
finite number of fields. This is particularly useful in the discussion of 
mirror symmetry between $SU(3)$ structure orientifolds and Calabi-Yau orientifolds
with background fluxes. A finite reduction is achieved  by picking  
a finite basis of forms $\Delta_{\rm finite}$ on the $SU(3)$ structure manifold 
slightly extending the Calabi-Yau reductions \cite{GLMW,GLW}. The 
explicit construction of such a finite set of forms is difficult, however, we
can specify its properties. Before turning to the orientifold constraints 
let us briefly recall the construction of ref.~\cite{GLW}.

To define the properties of $\Delta_{\rm finite}$ we first need to introduce
an additional structure on $\Lambda^{\rm ev,odd}T^*$
known as  Mukai pairings. These anti-symmetric forms are defined by 
\beq \label{def-Mukai}
    \big<\varphi,\psi \big> = \big[\lambda(\varphi)\wedge\psi \big]_6 
         = \begin{cases}\quad \varphi_0 \wedge \psi_6 - \varphi_2 \wedge 
          \psi_4 +\varphi_4 \wedge \psi_2 - \varphi_6 \wedge \psi_0\ ,\\
         \quad -\varphi_1 \wedge \psi_5 + \varphi_3 \wedge \psi_3 -\varphi_5 
          \wedge \psi_1\ ,
      \end{cases}
\eeq
where $\lambda$ is given in eqn.~\eqref{def-lambda}  and $[\ldots]_6$ denotes the forms of degree 6. Clearly 
$\big<\varphi,\psi \big>$ is proportional to a volume form on $\cM_6$ and can be integrated over the 
manifold $\cM_6$. Demanding this integrated Mukai pairings to be non-degenerate on 
$\Delta_{\rm finite}$ puts a first constraint on the possible set of forms. To make this more
precise, let us denote the finite set of forms in $\Lambda^n T^*$ by $\Delta^n$, with dimensions 
$\dim \Delta^n$.  As a first condition we demand that
$\dim \Delta^0=\dim \Delta^6=1$ and assume that $\Delta^0$ consists of the 
constant functions while $\Delta^6$ contains volume forms $\epsilon \propto J \wedge J\wedge J$.
Moreover, demanding non-degeneracy of the integrated Mukai pairings on $\Delta^{\rm ev}$ one 
defines a (canonical) symplectic basis on this space. 
Denoting a basis of $\Delta^{0}\oplus \Delta^{2}$ by $\omega_{\hat A} = (1,\omega_A)$ one defines
its dual  basis $\tilde \omega^{\hat A}=(\tilde \omega^A, \epsilon)$ of $\Delta^{4}\oplus \Delta^{6}$ 
by
\beq \label{int_omega}
  \int_{\cM_6} \big< \omega_{\hat A}, \tilde \omega^{\hat B} \big>
   = \delta_{\hat A}^{\hat B}\ , \qquad \hat A,\hat B = 0, \ldots , \dim \Delta^2\ ,
\eeq
with all other intersections vanishing. Turning to the odd forms $\Delta^{\rm odd}$ 
we follow a similar strategy to define a symplectic basis. However, in accord 
with our assumption above, we will set $\dim \Delta^1 = \dim \Delta^5 =0$ 
such that no one- or five-forms are used in the expansion of the fields.\footnote{In section \ref{Mirror} we discuss 
a possible way to weaken this condition.} Hence, 
non-degeneracy of the integrated Mukai pairings implies that 
a symplectic basis $(\alpha_\Kh,\beta^\Kh)$  of $\Delta^3$ can be defined as
\beq \label{int_alpha_beta}
  \int_{\cM_6} \big<\alpha_\Lh , \beta^\Kh \big> = \delta_\Lh^\Kh\ , \qquad \Kh,\Lh=1,\ldots, \tfrac{1}{2}\dim  \Delta^3\ ,
\eeq
with all other intersections vanishing. Note that the non-degeneracy 
of the integrated Mukai pairings implies that $\Delta^{\rm ev/odd}$ contains
the same number of exact and non-closed forms. We will come back to this issue 
later on when we introduce torsion fluxes.

After this brief review let us now specify how the orientifold symmetry acts on $\Delta_{\rm finite}$.
Under the operator $\cP_6=\lambda \simga^*$ 
the forms $\Delta^n$ decompose into eigenspaces as 
\beq \label{split-Delta}
   \Delta^n = \Delta^n_+ \oplus \Delta^n_-\ .
\eeq
Using the properties \eqref{constrA} and \eqref{def-lambda} one infers $\dim \Delta^0_- = \dim\Delta_-^6=0$. 
Furthermore, under the split \eqref{split-Delta} the basis $(\omega_{\hat A},\tilde \omega^{\hat A})$
introduced in \eqref{int_omega} decomposes as 
\beq \label{split_omega}
 (\omega_{\hat A},\tilde \omega^{\hat A}) \quad \rightarrow \quad 
 (1,\omega_a, \tilde \omega^\alpha,\epsilon) \in \Delta^{\rm ev}_+\ ,\quad
 (\omega_\alpha,\tilde \omega^a)  \in \Delta^{\rm ev}_-\ ,
\eeq
where $\alpha = 1,\ldots ,\dim \Delta^2_-$ while $a = 1, \ldots ,\dim \Delta^2_+$. Using the 
intersections \eqref{int_omega} one infers that $\dim \Delta^2_\pm = \dim \Delta^4_\mp$.
Turning to the odd forms consistency requires that
\beq \label{LagrSubset}
  \int_{\cM_6} \big< \Delta^{3}_\pm , \Delta^{3}_\pm \big> = 0 \ , \qquad *\Delta^{3}_\pm 
   = \Delta^{3}_\mp\ ,
\eeq
where in the second equality we used the fact that $\sigma$ is an 
orientation-reversing isometry. The first condition is a consequence of the fact that 
$ \Delta^{3}_\pm \wedge  \Delta^{3}_\pm$ transforms with a minus sign under $\cP_6$
and hence is a subset of $\Delta^6_-$ up to an exact form.
The equations \eqref{LagrSubset} imply that $\Delta^{3}_\pm$ are 
Lagrangian subspaces of $\Delta^{3}$ with respect to the integrated 
Mukai parings. Hence, also the symplectic basis $(\alpha_\Kh, \beta^\Kh)$
introduced in \eqref{int_alpha_beta} splits as 
\beq \label{split_alpha_beta}
  (\alpha_\Kh,\beta^\Kh)\quad \rightarrow\quad (\alpha_k, \beta^\lambda) \in 
   \Delta^{3}_+ \ , \qquad (\alpha_\lambda, \beta^k) \in \Delta^{3}_-\ ,
\eeq
where the numbers of $\alpha_k$ and $\beta^\lambda$ in $\Delta^3_+$ equal to the numbers of $\beta^k$
and $\alpha_\lambda$ in $\Delta^3_-$ respectively. This is in accord with equation \eqref{int_alpha_beta}.

We are now in the position to give an explicit expansion of the fields into 
the finite form basis of $\Delta_{\rm finite}$. As discussed in the 
general case above the four-dimensional complex chiral fields 
arise in the expansion of the forms $J_c$ and $\Pi^{\rm odd}_c$ 
introduced in eqn.~\eqref{def-J_c} and \eqref{def-Pi_c}. Restricted to $\Delta_+^2$, $\Delta_+^3$ and $\Delta_-^2$ one has
\beq \label{exp_JcPi}
   J_c = t^a \omega_a\ ,\qquad \Pi^{\rm odd}_c= N^k \alpha_k + T_\lambda 
   \beta^\lambda  \ , \qquad C^{(1)}_3 = A^\alpha \omega_\alpha\ ,
\eeq
where the basis decompositions \eqref{split_omega} and \eqref{split_alpha_beta} were used.
Hence, in the finite reduction the $N=1$ spectrum consists of $\dim \Delta^2_+$ chiral multiplets 
$t^a$ and $\frac{1}{2} \dim \Delta^3$ chiral multiplets $N^k,T_\lambda$. In addition one finds 
$\dim \Delta^{2}_-$ vector multiplets, which arise in the expansion of 
$\hat C_3$. Moreover, one four-dimensional massless three-form arises in the 
expansion of $C^{(3)}_3$ into the form $1 \in \Delta^0_+$. It carries no degrees 
of freedom and corresponds to an additional flux parameter.
\\[.4cm]
%
\textit{\underline{The type IIB orientifold spectrum}}\\[.4cm]
%
Let us next turn to the spectrum of type IIB $SU(3)$ structure orientifolds.
To identify the invariant spectrum we first analyze the transformation properties 
of the ten-dimensional fields. In contrast to type IIA supergravity the type IIB theory 
consists of even forms $\hat C_{2n}$ in the R-R sector, which we conveniently combine as 
\cite{BKORvanP}
\beq \label{def-Cev}
  \hat C^{\rm ev} = \hat C_0 + \hat C_2 + \hat C_4 + \hat C_6 + \hat C_8\ .
\eeq
Only half of the degrees of freedom in $\hat C^{\rm ev}$ are 
physical and related to the second half by a duality constraint 
\cite{BKORvanP}. Using the transformation properties of the 
fields under $\Omega_p$ and $(-1)^{F_L}$ the invariance under the 
orientifold projections $\mathcal{O}_{(i)}$ implies that the ten-dimensional 
fields have to transform as \footnote{The transformation behavior of the R-R 
forms under the world-sheet parity operator $\Omega_p$ was given in eqn.~\eqref{parity-transf}.}
\beq \label{trans_BC_B}
  \sigma^* \hat B_2 = - \hat B_2\ ,\qquad \qquad
    \sigma^* \hat  \phi = \hat \phi\ , \qquad \qquad
   \sigma^* \hat C^{\rm ev} = \pm \lambda \big(\hat C^{\rm ev}\big)\ ,
\eeq
where the plus sign holds for orientifolds with $O3/O7$ planes, while the
minus sign holds for $O5/O9$ orientifolds. The parity operator $\lambda$
was introduced in eqn.~\eqref{def-lambda}. We combine the globally defined
forms $J$ and $\Omega$ with the fields $\hat B_2$, $\hat \phi$ and $\hat C^{\rm ev}$
as
\beq \label{def-Phi}
   \Phi^{\rm odd}= \Omega\ , \qquad  \qquad \Phi^{\rm ev} = e^{-\hat \phi} e^{-\hat B_2 +i J}\ ,\qquad  \qquad
   \hat A^{\rm ev} = e^{-\hat B_2} \wedge \hat C^{\rm ev}\ .
\eeq
where in comparison to \eqref{def-Pi} one finds that $\Phi^{\rm odd}$ takes the role of $\Pi^{\rm ev}$ and
$\Phi^{\rm ev}$ replaces $\Pi^{\rm odd}$. 
Applied to these forms the orientifold conditions \eqref{transJB}, \eqref{Omegatransf} and  
\eqref{trans_BC_B} read
\beq \label{const_PA_B}
    \sigma^* \Phi^{\rm odd} = \mp \lambda (\Phi^{\rm odd})\ , \qquad \qquad \sigma^* \Phi^{\rm ev} = \lambda (\bar \Phi^{\rm ev}) \ , 
    \qquad \qquad 
    \sigma^* \hat A^{\rm ev} = \pm \lambda(\hat A^{\rm ev})\ ,
\eeq
where the upper sign corresponds to $O3/O7$ and the lower sign to $O5/O9$ orientifolds.

In a next step we have to specify the basis of forms on $\cM_6$ used in the Kaluza-Klein reduction. 
In doing so we will face similar problems like in the type IIA case. Following the strategy 
advanced above we first briefly discuss the general case and later simplify the reduction
to the finite set of forms $\Delta_{\rm finite}$. The decomposition of 
the ten-dimensional fields into $SU(3)$ representations is given in 
tables \ref{NS} and \ref{RRIIB}. Also in the type IIB case we will remove all
triplets of $SU(3)$ from the spectrum \cite{GLW}. 

In order to perform the reduction we first investigate the splitting of the 
spaces of forms on $\cM_6$ under the operator $\cP_6 = \lambda \sigma^*$.
Since $\cP_6$ squares to the identity operator it splits the forms as in 
eqn.~\eqref{decomp_2_3}. More generally, we will need the decomposition of all even forms as
\beq \label{decomp_even}
   \Lambda^{\rm ev} T^*=   \Lambda^{\rm ev}_+ T^* \oplus  \Lambda^{\rm ev}_- T^*\ .
\eeq
The four-dimensional fields arising as the coefficients of $\Phi^{\rm ev/odd}$ and 
$\hat A^{\rm ev}$ expanded on $ \Lambda^{\rm 3}_\pm T^*$ and $ \Lambda^{\rm ev}_\pm T^*$
fit into $N=1$ supermultiplets. 

Firstly, we decompose the odd form $\Phi^{\rm odd}$ into the eigenspaces of 
$\cP_6$. In accord with the orientifold constraint \eqref{const_PA_B} we find 
\beq \label{exp_Phi_odd}
    O3/O7:\quad \Phi^{\rm odd}\ \ \in \ \ \Lambda_-^{3}T^*_\bbC \ , \qquad  \qquad  
    O5/O9:\quad \Phi^{\rm odd}\ \ \in \ \ \Lambda_+^{3}T^*_\bbC \ .
\eeq
Note that the actual degrees of freedom of $\Phi^{\rm odd} = \Omega$ are reduced by 
several constraints. More precisely, one has to specify forms $\Phi^{\rm odd}$ which are
associated to different reductions of the $\cM_6$ structure group to $SU(3)$.
As already discussed in section \ref{SU3} those reductions can  be parameterized by real 
three-forms $\rho=\R(\Phi^{\rm odd})$ which are in addition stable. 
The imaginary part $\I(\Phi^{\rm odd})$ can be expressed as a function of $\R(\Phi^{\rm odd})$
such that only half of the degrees of freedom in $\Phi^{\rm odd}$ are independent \cite{Hitchin:2000jd}.  
Moreover, as in the case of a Calabi-Yau manifold, different complex normalizations of 
$\Phi^{\rm odd}$ correspond to the same $SU(3)$ structure of $\cM_6$. 
Therefore, one additional complex degree of freedom in $\Phi^{\rm odd}$ is unphysical and has to 
be removed from the $D=4$ spectrum. 

In the reduction also the ten-dimensional form $\hat A^{\rm ev}$ is expanded 
into a basis of forms on $\cM_6$ while additionally satisfying the orientifold condition \eqref{const_PA_B}. 
In analogy to \eqref{decomp_C} we decompose 
\beq \label{decomp_A}
  \hat A^{\rm ev} = A^{\rm ev}_{(0)} + A^{\rm ev}_{(1)} + A^{\rm ev}_{(2)} + A^{\rm ev}_{(3)}\ ,   
\eeq 
where the subscript $(n)$ indicates the form degree in four dimensions. Note that in a general
expansion of $ \hat A^{\rm ev}$ in odd and even forms of $\cM_6$ as $\hat A^{\rm ev} = \rm ev|_4 \times ev|_6 + odd|_4 \times odd|_6$ 
it would be impossible to assign a
four-dimensional form degree as done in eqn.~\eqref{decomp_A}. This is due to the 
fact that such a decomposition only allows to distinguish even and odd forms in four dimensions. 
However, the orientifold imposes the constraint \eqref{const_PA_B} 
which introduces an additional splitting within the even and odd four-dimensional forms. 
Let us first make this more precise in the case of $O3/O7$ orientifolds where 
$\hat A^{\rm ev}$ transforms as $\simga^* \hat A^{\rm ev} = \lambda(\hat A^{\rm ev})$. 
Using the properties of the parity operator $\lambda$ one 
finds that the scalars in $A^{\rm ev}_{(0)}$ arise as coefficients of forms in $\Lambda^{\rm ev}_+ T^*$
while the two-forms in $A^{\rm ev}_{(2)}$ arise as coefficients of forms in $\Lambda^{\rm ev}_- T^*$.
Similarly, one obtains the four-dimensional vectors in $A^{\rm ev}_{(1)}$ as coefficients of $\Lambda^{\rm 3}_+ T^*$
and the three-forms in $A^{\rm ev}_{(3)}$ as coefficients of $\Lambda^{\rm 3}_- T^*$. 
In the case of $O5/O9$ orientifolds the ten-dimensional form $\hat A^{\rm ev}$ transforms 
as $\simga^* \hat A^{\rm ev} =- \lambda(\hat A^{\rm ev})$ and all signs in the $O3/O7$ expansions  
above are exchanged.
For both cases the decomposition \eqref{decomp_A} is well defined and we can analyze the multiplet structure
of the four-dimensional theory. 

In four dimensions massless scalars are dual to massless two-forms, while massless vectors are dual to vectors. 
Using the duality condition on the field strengths of the even forms  
$\hat A^{\rm ev}$ one eliminates half of its degrees of freedom. Indeed it can be shown that 
in the massless case the scalars in  $A^{\rm ev}_{(0)}$ are dual to the two-forms 
$A^{\rm ev}_{(2)}$. However, due to the generality of our discussion also massive 
scalars, vectors, two-form and three-forms can arise in the expansion \eqref{decomp_A}. In these 
cases the duality constraint gives a complicated relation between these fields. 
In the following we will first restrict our attention to the massless case and eliminate the two-forms in
$A^{\rm ev}_{(2)}$ in favor of the scalars in $A^{\rm ev}_{(0)}$. 

Let us start with the chiral multiplets. As the bosonic components these 
multiplets contain the real scalars in $A^{\rm ev}_{(0)}$ which are complexified 
by the real scalars arising in the expansion of $\R(\Phi^{\rm ev})$
or $\I (\Phi^{\rm ev})$. From the orientifold constraint \eqref{const_PA_B}
on infers that $\R (\Phi^{\rm ev})$ is expanded in forms of $\Lambda^{\rm ev}_+ T^*$ while 
$\I (\Phi^{\rm ev})$ is expanded in forms of 
$\Lambda^{\rm ev}_- T^*$. Therefore, one finds the 
complex forms \footnote{Note that also in the type IIB cases the complex forms $ \Phi^{\rm ev}_c$
encode the correct couplings to D-branes fully wrapped on supersymmetric cycles in $\cM_6$ \cite{BBS,MMMS,D-branecalib}.}
\beq \label{def-Phi_c}
  O3/O7: \quad  \Phi^{\rm ev}_c = A^{\rm ev}_{(0)} + i \R( \Phi^{\rm ev}) \ , \qquad \quad
   O5/O9:\quad \Phi^{\rm ev}_c = A^{\rm ev}_{(0)} + i \I (\Phi^{\rm ev})\ .
\eeq
The complex scalars arising in the expansion of the forms $ \Phi^{\rm ev}_c$ span 
a complex manifold $\cM^{\rm Q}$. This manifold is K\"ahler as discussed in 
the next section. Here  we conclude our general analysis of the spectrum of type IIB $SU(3)$ structure orientifolds
by summarizing the four-dimensional multiplets in table \ref{spectrumIIB}.

\begin{table}[htb]
\begin{center}
\begin{tabular}{|c|c|c|c|} \hline
\rule[-0.3cm]{0cm}{0.8cm}
multiplet& bosonic fields& \multicolumn{2}{c|}{$\cM_6$-forms}
\\ \hline \hline
\rule[-0.3cm]{0cm}{0.8cm}
& &$O3/O7$& $O5/O9$  \\ \hline
\rule[-0.3cm]{0cm}{0.8cm}
gravity multiplet& $g_{\mu\nu}$& & \\ \hline\rule[-0.3cm]{0cm}{0.8cm}
\rule[-0.3cm]{0cm}{0.8cm}
chiral multiplets& 
 $\Phi^{\rm odd}$ & $ \Lambda^3_-T^*$ & $ \Lambda^3_+T^*$\\ \hline
\rule[-0.3cm]{0cm}{0.8cm}
chiral/linear multiplets   & $ \Phi_c^{\rm ev}$ & $ \Lambda^{\rm ev}_+ T^*$ &$ \Lambda^{\rm ev}_- T^*$ \\ 
  \hline\rule[-0.3cm]{0cm}{0.8cm}
vector multiplets& $A^{\rm ev}_{(1)}$&$ \Lambda^3_+T^*$ &$ \Lambda^3_-T^*$\\ \hline
\end{tabular}
\caption{\small 
\textit{$N=1$ spectrum of type IIB orientifolds}}
\label{spectrumIIB}
\end{center}
\end{table}

To end this section let us  give a truncation to a finite number of the four-dimensional fields.
As we have argued in the previous section this is achieved by expanding the ten-dimensional fields 
on the finite set  of forms on $\cM_6$ denoted by $\Delta_{\rm finite}$. This is 
done in accord with the orientifold constraints for $O3/O7$ and $O5/O9$ orientifolds. 
Once again, the $n$-forms $\Delta^n$ split as $\Delta^n = \Delta^n_+ \oplus \Delta^n_-$, where
$ \Delta^n_\pm$ are the eigenspaces of the operator $\cP_6= \lambda \simga^*$.
However, since $ \Delta^6$ contains forms proportional to $J \wedge J \wedge J$ 
one infers from condition \eqref{transJB} that $ \dim \Delta_+^6 = 0$. Clearly, one has $\dim \Delta_-^0 = 0$
since $\Delta^0$ contains constant scalars which are invariant under $\cP_6$.
A further investigation of the even forms in $\Delta^2$ and $\Delta^4$ shows that
the basis introduced in eqn.~\eqref{int_omega} decomposes as
\beq \label{decomp_omega_B}
  (\omega_{\hat A},\tilde \omega^{\hat A}) \quad \rightarrow \quad 
               (1,\omega_a, \tilde \omega^\alpha) \in \Delta^{\rm ev}_+\ ,\quad
               (\epsilon,\omega_\alpha,\tilde \omega^a)  \in \Delta^{\rm ev}_-\ ,
\eeq
where $\alpha = 1, \ldots, \dim \Delta^2_-$ and $a = 1, \ldots, \dim \Delta^2_+$. Using 
$J \wedge J \wedge J \in \Delta^6_-$ and eqn.~\eqref{int_omega} one finds that
$ \Delta^2_\pm = \Delta^4_\mp$. Together with the fact that $\int \big<\Delta^{\rm ev}_\pm,\Delta^{\rm ev}_\pm \big>=0$
one concludes that $\Delta^{\rm ev}_\pm$ are Lagrangian subspaces of $\Delta^{\rm ev}$. 
This is the analog of the Lagrangian condition \eqref{LagrSubset} found for the odd forms in type IIA.
Let us turn to the odd forms $\Delta^{3} = \Delta^3_+ \oplus \Delta^3_-$. 
Due to the condition \eqref{Omegatransf} the three-form $\Omega$ is an element of 
$\Delta^{3}_-$ for $O3/O7$ orientifolds, while it is an element of 
$\Delta^3_+$ for $O5/O9$ orientifolds. Note that in contrast to the even forms 
$\Delta^{\rm ev}_\pm$ the spaces $\Delta^{3}_-$ and $\Delta^{3}_+$ have generically different dimensions.
The basis of three-forms introduced in \eqref{int_alpha_beta} splits under the action of 
$\cP_6$ as
\beq \label{ab_split}
  (\alpha_\Kh,\beta^\Kh) \quad \rightarrow \quad (\alpha_\lambda,\beta^\lambda) 
                  \in \Delta^3_+\ ,\quad 
                  (\alpha_k,\beta^k) \in \Delta^3_-\ ,
\eeq
where $\lambda=1,...,\frac12\dim\Delta^3_+,\ k=1,...,\frac12\Delta^3_-$. 

Given the basis decompositions \eqref{decomp_omega_B} and \eqref{ab_split} we can explicitly 
determine the finite four-dimensional spectrum of the type IIB orientifold
theories. For orientifolds with $O3/O7$ planes one expands $\Phi^{\rm ev}_c$
and $A^{\rm ev}_{(1)}$ into $\Delta^{\rm ev}_+$ and $\Delta^{3}_+$ as
\beq \label{exp_Phi3}
  \Phi^{\rm ev}_c = \tau + G^a\omega_a + T_\alpha \tilde \omega^\alpha \ , \qquad A^{\rm ev}_{(1)} = A^\lambda \alpha_\lambda \ ,
\eeq
where $\tau,G^a,T_\alpha$ are complex scalars in four dimensions.
The vector coefficients of the forms $\beta^\lambda$ in 
the expansion of $A^{\rm ev}_{(1)}$ are eliminated by the duality constraint on the field strength 
of $\hat A^{\rm ev}$. In addition we find that $\Phi^{\rm odd}$ depends on $\frac12(\dim \Delta^3_- - 2)$
complex deformations $z^k$. Therefore the full $N=1$ spectrum consists of $\frac12(\dim \Delta^3_- - 2)$
chiral multiplets $z^k$ as well as $\dim \Delta^{2} + 1$ chiral multiplets $\tau,G^a,T_\alpha$.
Moreover, we find $\frac12 \dim \Delta^3_+$ vector multiplets $A^\lambda$. 

The story slightly changes for orientifolds with $O5/O9$ planes. In this case the 
chiral coordinates are obtained by expanding 
\beq \label{exp_Phi5}
  \Phi^{\rm ev} _c= t^\alpha \omega_\alpha + u_b\, 
   \tilde \omega^b + S\, \epsilon\ ,  \qquad A^{\rm ev}_{(1)} = A^k \alpha_k\ ,
\eeq
where $ t^\alpha, u_b,S$ are complex four-dimensional scalars and 
the volume form $\epsilon$ is normalized as $\int_{\cM_6} \epsilon =1$.
Moreover, the form $\Phi^{\rm odd}$ depends on $\frac12(\dim \Delta^3_+ - 2)$
complex deformations $z^\lambda$. In summary the complete $N=1$ spectrum consists of $\frac12(\dim \Delta^3_+ - 2)$
chiral multiplets $z^\lambda$ as well as $\dim \Delta^{2} + 1$ chiral multiplets $ t^\alpha, u_b,S$.
Finally, the expansion of $A^{\rm ev}_{(1)}$ yields $\frac12 \dim \Delta^3_-$ independent vector multiplets $A^k$. 
                
%
\subsection{The K\"ahler potential \label{Kaehlerpot}}
%
In this section we determine the K\"ahler potential encoding the kinetic 
terms of the chiral or dual linear multiplets. Recall that the standard 
bosonic action for chiral multiplets with bosonic components $M^I$ 
contains the kinetic terms \cite{WB}
\beq \label{kinetic_chiral}
  S_{\rm chiral} = \int_{M_{3,1}}\, G_{I \bar J}\, \mathbf{d} M^I \wedge *_4\, \mathbf{d} \bar M^J \ ,
\eeq
where $\mathbf{d}$ and $*_4$ are the exterior derivative and the Hodge-star on $M_{3,1}$.
The metric $G_{I \bar J} = \partial_{M^I} \partial_{\bar M^J }K$ is K\"ahler and locally given as the second 
derivative of a real K\"ahler potential $K(M,\bar M)$. In other words, the function $K$ determines
the dynamics of the system of chiral multiplets. Similarly, one can derive the kinetic terms for a 
set of linear multiplets from a real function, the kinetic potential $\tilde K$. Since in the 
massless case the linear multiplets are dual to chiral multiplets one can always translate
$\tilde K$ into an associated $K$ via a Legendre transformation \cite{BGG}.\footnote{For a brief 
review, see also section 4 of ref.~\cite{TGthesis}.} It therefore 
suffices to derive the K\"ahler potential. In the massive case the duality between 
chiral and linear multiplets is no longer valid, however, 
the function $\tilde K$ can still be formally related to a K\"ahler potential $K$.
In the following we will determine the K\"ahler potential $K$ for type IIA and type IIB
orientifolds in turn. 
\\[.4cm]
\textit{\underline{The IIA K\"ahler potential and the K\"ahler metric}}\\[.2cm]
%
Let us start by discussing the type IIA K\"ahler potential first. As in section \ref{OSpectrum} we 
will keep our analysis general and only later specify a finite reduction. 
We found in the previous section that the complex scalars in the chiral multiplets are obtained
by expanding the complex forms $\Pi^{\rm ev}$ and $\Pi^{\rm odd}_c$ into 
appropriate forms on $\cM_6$. Locally, the field space 
takes the form 
\beq \label{mod-space}
   \cM^{\rm K} \times \cM^{\rm Q}\ ,
\eeq
where $\cM^{\rm K}$ and $\cM^{\rm Q}$ are spanned by the complex scalars arising in the 
expansion of $\Pi^{\rm ev}$ and $\Pi^{\rm odd}_c$ respectively. $N=1$ supersymmetry demands 
that both manifolds in \eqref{mod-space} are K\"ahler with metrics locally encoded 
by K\"ahler potentials $K^{\rm K}$ and $K^{\rm Q}$.
From the point of view of an $N=2$ to $N=1$ reduction, the manifold $\cM^{\rm K}$ is a complex 
submanifold of the $N=2$ special K\"ahler manifold spanned by the complex scalars 
in the vector multiplets. As we will discuss momentarily 
the manifold $\cM^{\rm K}$ directly inherits its K\"ahler structure 
from the underlying $N=2$ theory. On the other hand, $\cM^{\rm Q}$ is
a submanifold of the quaternionic space spanned by the hyper multiplets
and has half its dimension. It is a non-trivial 
result that $\cM^{\rm Q}$ is a K\"ahler manifold since the underlying quaternionic 
manifold is not necessarily K\"ahler.

We analyze first the structure of the field space $\cM^{\rm K}$ spanned 
by the complex fields arising in the expansion of $J_c = -\hat B_2 + i J$ into forms $\Lambda^2_+ T^*$.
Note that as in the original $N=2$ theory not all forms $J$ are allowed and 
one restricts to the cases where $J$, $J\wedge J $ and $J \wedge J \wedge J$ measure
positive volumes of two-, four and six-cycles \cite{Candelas:1990pi}. We abbreviate this condition by writing 
$J \ge 0$. Hence, the coefficients of $J_c$ define the complex cone
\beq \label{def-MK_A}
   \cM^{\rm K} = \big\{ J_c \in \Lambda^2_+ T^*_\bbC: J \wedge J \wedge J \neq 0 \text{ and } J \ge 0 \big\}\ .
\eeq
This manifold has the same complex structure as the underlying $N=2$ special K\"ahler manifold.
It also inherits its K\"ahler structure with a K\"ahler potential given by \cite{Candelas:1990pi,GLW}
\beq \label{def-KK_A}
   K^{\rm K}(J_c) = - \ln\big[- i \int_{\cM_6} \big< \Pi^{\rm ev}, \bar \Pi^{\rm ev} \big> \big] 
                             = - \ln\big[\tfrac{4}{3} \int_{\cM_6} J \wedge J \wedge J \big]\ ,
\eeq
where $\Pi^{\rm ev} = e^{J_c}$ is introduced in \eqref{def-Pi} and the pairing $\big<\cdot,\cdot\big>$ is
defined in \eqref{def-Mukai}.\footnote{Note that in contrast to ref.~\cite{GLW} we included an integration 
in the definition of $K^{\rm K}$ such that it is independent of the coordinates on $\cM_6$. This implies that
four-dimensional supergravity theory takes the standard $N=1$ form. However, this also implies that 
we have to exclude modes which correspond the rescalings of $J$ by a function $\cM_6$ (see also appendix \ref{der_KQ}).
We will come back to this issue in a separate publication \cite{TGinprep}.}
The K\"ahler metric is obtained as the second derivative of $K^{\rm K}$ given in \eqref{def-KK_A} with respect to 
$J_c$ and $\bar J_c$. More precisely, one finds 
\beq \label{def-GK}
  G^{\rm K}( \omega,\omega') =\big[\partial_{J_c} \partial_{\bar J_c} K^{\rm K}\big]( \omega,\omega') = -2e^{K^{\rm K}} \int_{\cM_6} \big< \omega, *_6\, \omega' \big>\ , 
\eeq
where $*_6$ is the six-dimensional Hodge-star and $\omega,\, \omega'$ are two-forms in $\Lambda^2_+ T^*$\ . Note that 
in this general approach the derivatives are taken with respect to 
two-forms on $\cM_6$ containing the $D=4$ scalars such that the result needs to be 
evaluated on elements of $\Lambda^2_+ T^*$. The four-dimensional kinetic 
terms \eqref{kinetic_chiral}  read \footnote{The action \label{kin_Pi^ev} is given in the 
four-dimensional Einstein frame where the kinetic term for the metric takes the form $\frac12 R$.}
\beq \label{kin_Piev}
   S_{\Pi^{\rm ev}} = \int_{M_{3,1}} G^{\rm K}(\mathbf{d} J_c, *_4\ \mathbf{d} \bar J_c) 
\eeq
with four-dimensional derivative $\mathbf{d}$. From \eqref{def-GK} one concludes that the 
metric $G^{\rm K}$ only depends on $\Pi^{\rm ev}$. It is straight forward to evaluate 
\eqref{kin_Piev} in the finite basis $\omega_a\in\Delta_+^2$ introduced in equation \eqref{split_omega}. 
On this basis the complex form $J_c$ decomposes as $J_c = t^a \omega_a$ and one finds 
\beq
     S_{\Pi^{\rm ev}} = \int_{M_{3,1}} G^{\rm K}_{a\bar b}\ \mathbf{d}t^a\wedge *_4\, \mathbf{d}\bar t^{b}\ ,\qquad \qquad
     G^{\rm K}_{a \bar b} = 2e^{K^{\rm K}} \int  \omega_a 
   \wedge * \omega_b\ .
\eeq
In the finite basis the metric $G^K$ takes a form similar to the case where $\cM_6$ is a Calabi-Yau
orientifold \cite{TGL2}. However, since the forms $\omega_a$ are not necessarily harmonic
a potential for the fields $t^a$ is introduced as we will discuss in section \ref{Superpot}.

Let us now turn to the second factor in \eqref{mod-space} and investigate the K\"ahler 
structure of the manifold $\cM^{\rm Q}$. As introduced in section \ref{OSpectrum} the 
complex coordinates on this space are obtained by expanding the form 
$\Pi^{\rm odd}_c$ into elements of $\Lambda^3_+ T^*$. The metric 
on the field space $\cM^{\rm Q}$ is derived by inserting the expansion of 
$\Pi^{\rm odd}_c$ in the ten-dimensional effective action of type IIA supergravity.
For the form $C^{(0)}_3$ the reduction of the R-R sector yields the term
\beq \label{act_C_0}
   S_{C^{(0)}_3} =  \int_{M_{3,1}} G^{\rm Q} (\mathbf{d} C^{(0)}_3,*_4 \, \mathbf{d} C^{(0)}_3  )\ , 
\eeq
where the metric $G^{\rm Q} $ is defined as
\beq \label{def-GQ_A}
   G^{\rm Q}(\alpha, \alpha') = 2 e^{2D} \int_{\cM_6}  \big< \alpha, *_6\, \alpha' \big>\ ,
\eeq
with $ \alpha, \alpha' \in \Lambda^3_+ T^*$. The four-dimensional dilaton $D$ was defined in eqn.~\eqref{def-D}
and arises in \eqref{act_C_0} due to a Weyl rescaling to the four-dimensional Einstein frame.
The R-R field $C^{(0)}_3$ is complexified by $\R(\Pi^{\rm odd})$ as given in eqn.~\eqref{def-Pi_c}. Therefore, the full kinetic terms for the complex scalars in $\Pi^{\rm odd}_c$ are given by
\beq
     S_{\Pi^{\rm odd}_c} =  \int_{M_{3,1}} G^{\rm Q} (\mathbf{d}\Pi^{\rm odd}_c,*_4 \, \mathbf{d}\bar\Pi^{\rm odd}_c  )\ . 
\eeq
The metric $G^{\rm Q}$ is K\"ahler on the manifold $\cM^Q$ if we carefully specify the forms used in the expansion
of $\Pi^{\rm odd}_c$.
As already explained in section \ref{SU3}, the real three-forms $\rho = \R(\Pi^{\rm odd})$  
defining an $SU(3)$ structure manifold have to be `stable'.%
\footnote{The detailed definition of stable forms is given in 
appendix \ref{GCG-HF}.} 
We denote all stable forms in $\Lambda^3_+T^*$ by $U^3_+$. Using this definition the field space $\cM^{\rm Q}$ 
spanned by the complex scalars in $\Pi^{\rm odd}_c$ is locally of the 
form 
\beq \label{def-MQ_A}
   \cM^{\rm Q} = \big\{\R(\Pi^{\rm odd}) \in  U^3_+\big\} \times \Lambda^3_+T^*\ ,
\eeq
where $ \Lambda^3_+T^*$ is parameterized by the real scalars in the R-R field $C^{(0)}_3$.

In appendix \ref{der_KQ} we  show that the
metric $G^{\rm Q}$ can be obtained as the second derivative of a K\"ahler potential.
Note however, that we have to impose an additional constraint on the 
forms in  $\cM^{\rm Q}$ in order to obtain a K\"ahler potential independent 
of the coordinates on $\cM_6$. More precisely, we demand that all $(3,0)+(0,3)$ 
forms in $\cM^{\rm Q}$ are proportional to $\rho$ with a coefficient constant on $\cM_6$.\footnote{This condition 
can be weakened in case the K\"ahler potential is defined as a logarithm of a function varying along  
$\cM_6$ as we also discuss in appendix \ref{der_KQ}. 
In this case, the orientifold theory is an $N=1$ reformulation of the 
ten-dimensional supergravity theory \cite{TGinprep}.}   
On this the set of stable forms one shows that the metric $G^{\rm Q}$ is K\"ahler with a K\"ahler potential
given by
\beq \label{def-KQ_A}
     K^{\rm Q}(\Pi^{\rm odd}_c)=-2\ln\big[ i\int_{\cM_6} \big<\Pi^{\rm odd},\bar \Pi^{\rm odd} \big>\big] = - \ln \big[e^{-4D} \big]\ ,
\eeq 
where in the second equality we have used the definition of $\Pi^{\rm odd} = C\Omega$ given in equations \eqref{def-Pi}
and \eqref{def-C} to express $K^{\rm Q}$ in terms of the four-dimensional dilaton $e^{D}$ defined in eqn.~\eqref{def-D}. 
The functional appearing in the logarithm of the K\"ahler potential, 
\beq\label{Hitchinfunctinal}
   H \big[\R(\Pi^{\rm odd}) \big] = i \int \big<\Pi^{\rm odd},\bar \Pi^{\rm odd} \big> \ ,
\eeq
was first introduced by Hitchin in refs.~\cite{Hitchin:2000jd}.  A more explicit definition of $H$ as a functional of 
$\R(\Pi^{\rm odd})$ can be found in appendix \ref{GCG-HF}. 
The metric $G^{\rm Q}$ defined in \eqref{def-GQ_A} is obtained by the second derivative
\beq \label{second_der} 
   G^{\rm Q} (\alpha, \alpha') = \
   \big[ \partial_{\Pi^{\rm odd}_c }\, \partial_{\bar \Pi^{\rm odd}_c} K^{\rm Q} \big]  (\alpha, \alpha')\ .
\eeq
Note that $K^{\rm Q}$ is a function of $\R(\Pi^{\rm odd})$ and does not depend on the 
R-R fields $\R(\Pi^{\rm odd}_c) = C^{(0)}_3$. Hence, the metric $G^{\rm Q}$ possesses 
various shift symmetries and the second factor in \eqref{def-MQ_A} is a vector space.

Finally, we will restrict the results obtained for $\cM^{\rm Q}$ to the finite basis $\Delta_{\rm finite}$.
In order to do so, one expands the complex form $\Pi^{\rm odd}_c$ in the real
basis $\alpha_k , \beta^\lambda \in \Delta^3_+$ as given in eqn.~\eqref{exp_JcPi}. The 
coefficients of this expansion are complex scalars $N^k,T_\lambda$. 
The K\"ahler metric is the second derivative of $K^{\rm Q}$ given in eqn.~\eqref{def-KQ_A}
with respect to these complex fields. Explicitly, it takes the form
\bea
  \partial_{N^k} \partial_{\bar N^l} K &=& 2 e^{2D} \int_{\cM_6} \alpha_k \wedge *_6\,
  \alpha_l\ , \qquad 
  \partial_{N^k} \partial_{\bar T_\kappa} K = 2 e^{2D} \int_{\cM_6} \alpha_k 
  \wedge *_6\, \beta^\kappa\ ,\\
  \partial_{T_\kappa} \partial_{\bar T_\lambda} K &=& 2 e^{2D} \int_{\cM_6} 
  \beta^\kappa \wedge *_6\, \beta^\lambda\ . \nn
\eea

This ends our discussion of the K\"ahler  metric on the type IIA
field spaces $\cM^{\rm K} \times \cM^{\rm Q}$. We found that the K\"ahler potentials
are the two Hitchin functionals depending on real two- and 
three-forms on $\cM_6$. A similar result with odd and even forms exchanged 
is found for type IIB orientifolds to which we turn now.
\\[.4cm]
%
%
\textit{\underline{The IIB K\"ahler potential and the K\"ahler metric}}\\[.2cm]
%
In the following we investigate the K\"ahler structure of the scalar field space
in type IIB orientifolds.  The complex scalars in the chiral multiplets are obtained 
by expanding $\Phi^{\rm odd}$ and $\Phi^{\rm ev}_c$ into appropriate forms on $\cM_6$
as introduced in eqns.~\eqref{exp_Phi_odd} and \eqref{def-Phi_c}.
These complex scalars locally span the product manifold $\cM^{\rm K}\times\cM^{\rm Q}$,
where $\cM^{\rm K}$ contains the independent scalars in $\Phi^{\rm odd}$ while 
$\cM^{\rm Q}$ contains the scalars in $\Phi^{\rm ev}_c$. Note that we are now
dealing with two type IIB setups corresponding to two truncations of the 
original $N=2$ theory. 

As in the type IIA orientifolds the complex and K\"ahler structure of $\cM^{\rm K}$ is directly inherited 
from the underlying $N=2$ theory.  Independent reductions of the structure group of $\cM_6$ 
are parameterized by a set of real stable forms $\rho=\R(\Phi^{\rm odd})$ denoted by $U^3$. 
In order to satisfy the orientifold constraints \eqref{const_PA_B} this field space is reduced to $U^3_-$ for $O3/O7$
orientifolds and to $U^3_+$ for $O5/O9$ orientifolds. Furthermore, complex rescalings of the 
complex three-form $\Phi^{\rm odd}$ are unphysical.  
Hence, the moduli space encoded by $\Phi^{\rm odd}$ is obtained by dividing $U^3_\pm$
by reparameterizations $\Phi^{\rm odd} \rightarrow c\, \Phi^{\rm odd}$ for complex non-zero $c \in \bbC^*$.
The field space $\cM^{\rm K}$ is then defined as
\beq
    \cM^{\rm K}=\big\{ \R(\Phi^{\rm odd})\in U^{3}_\mp\big\} / \bbC^* \ ,
\eeq
where the minus sign stands for $O3/O7$ and the plus sign for $O5/O9$ orientifolds. 
The field space $\cM^{\rm K}$ is a complex K\"ahler manifold. This is shown in 
analogy to the $N=2$ case discussed in refs.~\cite{Hitchin:2000jd,GLW},
since the orientifold projections preserve the complex structure 
and only reduce the dimension of $\cM^{\rm K}$. We denote the 
complex scalars parameterizing $\cM^{\rm K}$ by $z$'s.  
The K\"ahler 
potential as a function of these fields and their complex conjugates 
is given by 
\beq \label{def-KK_B}
   K^{\rm K}(z,\bar z) =  - \ln\big[ -i \int_{\cM_6} \big<\Phi^{\rm odd}, \bar \Phi^{\rm odd}\big>\big] 
                    =  - \ln\big[ -i \int_{\cM_6} \Omega \wedge \bar \Omega\big] \ ,
\eeq
where in the second equality we used the definitions \eqref{def-Phi} and \eqref{def-Mukai} 
of $\Phi^{\rm odd}$ and the pairings $\big<\cdot,\cdot \big>$. 
The manifold $\cM^{\rm K}$ possesses a special geometry completely 
analogous to the $N=2$ case, such that in particular the three-form $\Omega(z)$ is a holomorphic 
function in the complex coordinates $z$ on $\cM^{\rm K}$. This special 
geometry was used in ref.~\cite{GLW} to derive the 
K\"ahler metric corresponding to $K^{\rm K}$. We will not review the result here, but rather immediately 
turn to the field space $\cM^{\rm Q}$ which is a K\"ahler field space in the 
$N=1$ theory.

Let us now determine the K\"ahler potential encoding the metric on the field space 
$\cM^{\rm Q}$. As discussed in section \ref{OSpectrum} the complex coordinates spanning $\cM^{\rm Q}$
are obtained by expanding $\Phi^{\rm ev}_c$ into elements of $\Lambda^{\rm ev}_\pm$ 
depending on whether we are dealing with $O3/O7$ or $O5/O9$ orientifolds. The 
precise definition of $\Phi^{\rm ev}_c$ was given in eqn.~\eqref{def-Phi_c}. Note 
that not every form in   $\Lambda^{\rm ev}_\pm$ corresponds to a reduction of the 
structure group of $\cM_6$ to $SU(3)$ and we have additionally to impose constraints on $\I(\Phi^{\rm ev}_c)$
analog to the stability condition discussed above. Recall that in the  $O3/O7$ 
case $\I(\Phi^{\rm ev}_c) = \R(e^{-\hat \phi} e^{-\hat B_2 + iJ})$ and in the $O5/O9$ case 
$\I(\Phi^{\rm ev}_c) = \I(e^{-\hat \phi} e^{-\hat B_2 + iJ})$ as given in eqns.~\eqref{def-Phi} and \eqref{def-Phi_c}.
In these definitions the real two-from $J$ has to satisfy $J\wedge J \wedge J \neq 0$ and $J \ge 0$ 
as in \eqref{def-MK_A}. Altogether the field space
$\cM^{\rm Q}$ locally takes the form
\beq \label{def-MQ_B}
   \cM^{\rm Q} = \big\{\I(\Phi^{\rm ev}_c) \in \Lambda_\pm^{\rm ev}: J \wedge J \wedge J \neq 0 \text{ and } J \ge 0 \big\} \times \Lambda^{\rm ev}_\pm T^* \ ,
\eeq
where the vector space $ \Lambda^{\rm ev}_\pm T^*$ is spanned by the fields $A^{\rm ev}_{(0)}$.
The plus sign in the expression \eqref{def-MQ_B} corresponds to 
orientifolds with $O3/O7$ planes while the minus sign stands for the $O5/O9$ orientifolds. The 
metric on the manifold $\cM^{\rm Q}$ is obtained by inserting the expansion of the R-R form $A^{\rm ev}_{(0)}$
into the ten-dimensional action of type IIB supergravity. Performing a Weyl rescaling to the 
four-dimensional Einstein frame one finds 
\beq\label{SA_B}
   S_{A^{\rm ev}_{(0)}} = \int_{M_{3,1}} G^{\rm Q}(\mathbf{d}A^{\rm ev}_{(0)}, *_4\,\mathbf{d}A^{\rm ev}_{(0)})\ ,
\eeq
where $G^{\rm Q}$ is defined on forms $\nu,\nu' \in \Lambda^{\rm ev}_\pm T^*$ as  
\beq\label{def-GQ_B}
   G^{\rm Q}(\nu,\nu')=2 e^{2D} \int_{\cM_6} \big<\nu, *_B \, \nu' \big>\ .
\eeq
The reason for this simple form is that we have replaced the ordinary Hodge-star 
by the B-twisted Hodge star $*_B$ acting on an even form $\nu$ as (see, for example, ref.~\cite{Witt:2005sk})
\beq
   *_B\, \nu = e^{\hat B_2} \wedge * \lambda(e^{-\hat B_2} \wedge \nu)\ ,   
\eeq
where $\lambda$ is the parity operator introduced in \eqref{def-lambda}.
In equation \eqref{def-GQ_B} the four-dimensional dilaton $D$
is defined as in the type IIA case \eqref{def-D}.
Including the reduction of $\I (\Phi^{\rm ev}_c)$ the action \eqref{SA_B} is completed as
\beq\label{Sphi_B}
     S_{\Phi^{\rm ev}_c} = \int_{M_{3,1}} G^{\rm Q}(\mathbf{d} \Phi^{\rm ev}_c, *_4\,\mathbf{d} \bar  \Phi^{\rm ev}_c)\ .
\eeq

The metric $G^{\rm Q}$ is shown to be the second derivative of the K\"ahler potential \footnote{%
As in the type IIA case we discard the non-trivial modes proportional to $\I(\Phi^{\rm ev}_c)$. These can 
be included if the K\"ahler potential is the logarithm of a volume form varying along $\cM_6$.}
\beq
    K^{\rm Q} (\Phi^{\rm ev}_c)= -2 \ln \big[i \int_{\cM_6} \big< \Phi^{\rm ev}, \bar\Phi^{\rm ev} \big> \big]
                                                   = -\ln \big[ e^{-4D} \big] \ ,
\eeq
where in the second equality we have used the definition of $\Phi^{\rm ev}$ as
given in \eqref{def-Phi}. Note that $K^{\rm Q}$ is a function of $\I(\Phi^{\rm ev}_c)$ only,
such that it depends on $\R(\Phi^{\rm ev})$ in $O3/O7$ orientifolds while it depends 
on $\I(\Phi^{\rm ev})$ in $O5/O9$ orientifolds. The functionals appearing 
in the logarithm are the Hitchin functionals (see also appendix \ref{GCG-HF}) \cite{Hitchin:2004ut} 
\beq \label{Hitchin_B}
   H[\R(\Phi^{\rm ev})] = i \int_{\cM_6}  \big< \Phi^{\rm ev}, \bar\Phi^{\rm ev} \big>\ , \qquad 
   H[\I(\Phi^{\rm ev})] = i \int_{\cM_6}  \big< \Phi^{\rm ev}, \bar\Phi^{\rm ev} \big>\ ,
\eeq
depending on whether we are dealing with $O3/O7$ and $O5/O9$ orientifolds.\footnote{%
In the first case one obtains the functional dependence of $H$ by evaluating $\I(\Phi^{\rm ev})$ 
as a function of the real part $\R(\Phi^{\rm ev})$, while in the second case 
one needs do find $\R(\Phi^{\rm ev})\big[\I(\Phi^{\rm ev}) \big]$.}
The metric $G^{\rm Q}$ given in eqn.~\eqref{def-GQ_B} is obtained by taking the 
second derivative of $K^{\rm Q}$ as
\beq
     G^{\rm Q}(\nu, \nu')=\big[ \partial_{\Phi^{\rm ev}_c} \partial_{\bar \Phi^{\rm ev}_c} K^{\rm Q} \big] (\nu, \nu')\ .
\eeq 
Due to the independence of $K^{\rm Q}$ of the R-R scalars in $A^{\rm ev}_{(0)}$ the 
metric $G^{\rm Q}$ possesses shift symmetries. 

It is straight forward to evaluate the K\"ahler metric $G^{\rm Q}$ for the 
finite basis of $\Delta^{\rm ev}_\pm$ introduced in \eqref{decomp_omega_B}. 
The coefficients are complex fields $ M^{\hat A}=(\tau,G^a,T_\alpha)$ for $O3/O7$ orientifolds
and $M^{\hat A}=(S,t^\alpha,A_a)$ for $O5/O9$ orientifolds as seen in eqns.~\eqref{exp_Phi3} and \eqref{exp_Phi5}.
Explicitly the metric $G^{\rm Q}$ is given by 
\beq
  \partial_{M^{\hat A}} \partial_{\bar M^{\hat B}} K = 
  e^{2D} \int \big<\nu_{\hat A} , *_B\, \nu_{\hat B}\big> \ ,
\eeq
where $\nu_{\hat A} = (1,\omega_a,\tilde \omega^\alpha)$ for
$O3/O7$ orientifolds while $\nu_{\hat A} = (\epsilon,\omega_\alpha,\tilde \omega^a)$
for $O5/O9$ orientifolds. These metrics are identical to the ones derived 
for type IIB Calabi-Yau orientifolds \cite{TGL1} if the finite basis $\Delta_{\rm finite}$ is 
consisting of harmonic forms only. Compared to the expression given in ref.~\cite{TGL1} 
we simplified the result considerably by introducing the B-twisted Hodge-star $*_B$.

To summarize we found that also in the type IIB setups the field 
space $\cM^{\rm K} \times \cM^{\rm Q}$ is a K\"ahler manifold 
with K\"ahler potentials given by the logarithm of the Hitchin functionals.
This fixes the kinetic terms of the chiral or dual linear multiplets. 
Surprisingly, our analysis can be performed in a 
rather general setting without specifying a finite 
reduction. To illustrate the results we nevertheless 
gave the reduction to the finite basis $\Delta_{\rm finite}$.
We will now turn to the analysis of the superpotential 
terms induced on $SU(3)$ structure orientifolds.

\subsection{The Superpotential of type II $SU(3)$ structure orientifolds
\label{Superpot}}
In this section we derive the superpotentials for type IIA and type 
IIB $SU(3)$ structure orientifolds in presence of fluxes and torsion. The 
calculation is most easily done on the level of the fermionic effective 
action. This is due to the fact that the superpotential $W$ appears linearly 
in a four-dimensional $N=1$  
supergravity theory as the mass of the gravitino $\psi_\mu$. The corresponding 
mass term reads 
\beq \label{standard-W}
   S_{\rm mass}= - \int_{M_{3,1}} e^{K/2} \big( W  \bar \psi_\mu \bar \sigma^{\mu\nu} \bar \psi_\nu + 
             \bar W  \psi_\mu \sigma^{\mu\nu} \psi_\nu \big) *_4 \mathbf{1}\ , 
\eeq 
where $*_4 \mathbf{1}=\sqrt{-g_4}\,d^4x$ is the four-dimensional volume element
and  $K$ is the K\"ahler potential on the chiral field space.
To determine \eqref{standard-W} for the orientifold setups one dimensionally 
reduces the fermionic part of the type IIA and type IIB actions. As in the 
bosonic part, the orientifold
projections ensure that the resulting four-dimensional theories possess $N=1$ 
supersymmetry. 

Let us start by recalling the relevant fermionic terms for our discussion in
the ten-dimensional 
type IIA and type IIB supergravity theories. We 
conveniently combine the two gravitinos into a 
two-vector $\hat\psi_N = (\hat\psi^1_N,\hat\psi^2_N)$. The 
effective action for the gravitinos in string frame takes the form \footnote{We only display
terms which are quadratic in the gravitinos $\hat \psi_N$ since we aim to calculate 
terms of the form \eqref{standard-W}. Moreover, note that the ten-dimensional 
fermions are Majorana-Weyl spinors and the conjugate spinor $\hat{\bar\psi}_M=\psi_M^\dagger \Gamma^0$ is obtained 
by hermitian conjugation and multiplication with the ten-dimensional 
gamma-matrix $\Gamma^0$.} 
\beq\label{Spsi}
    S_{\psi}=-\int_{\cM_{10}}  \big[
     e^{-2\hat \phi} \hat{\bar\psi}_M\Gx^{MNP}D_N \hat\psi_P\, * \mathbf{1}
      + \tfrac{1}{4} e^{-2\hat \phi}  \hat H_3 \wedge * \Psi
     + \tfrac{1}{8}\sum_n \hat F_n \wedge * \Psi_n\big]\ ,
\eeq
where we are using the democratic formulation of ref.~\cite{BKORvanP}.  The R-R
field strengths  $\hat F_n$ are defined as 
\beq\label{defFH}
    \hat F_n=d \hat C_{n-1}- \hat H_3\wedge \hat C_{n-3}\ ,\qquad \qquad  *\hat F_n  = \lambda  (\hat F_{10-n})\ , 
\eeq
where $n$ runs from $0$ to $8$ for type IIA and from $1$ to $9$ for type IIB and 
we set $\hat H_3 = d\hat B_2$. The self-duality condition in eqn.~\eqref{defFH} 
implies that half of the R-R fields in $\hat C^{\rm ev/odd}$ carry no extra degrees of freedom.
Furthermore, $\Psi$ and $\Psi_n$  are ten-dimensional  
three- and $n$-forms which are bilinear in $\hat \psi_M$ and have components 
\bea\label{defP}
(\Psi)_{M_1 M_2 M_3} &=& \hat{\bar\psi}_M \Gx^{[M}
    \Gx_{M_1 M_2 M_3} \Gx^{N]} \mathcal{P} \hat \psi_N \ , \nn\\
       (\Psi_n)_{M_1\ldots M_n}&=& e^{-\hat\phi} \hat{\bar\psi}_M \Gx^{[M}
    \Gx_{M_1 \ldots M_n} \Gx^{N]} \mathcal{P}_n \hat \psi_N\ ,
\eea
where $\mathcal{P}=\Gx_{11}$, $\mathcal{P}_n=(\Gx_{11})^n$ for type IIA while for 
type IIB  one has $\mathcal{P}=-\sigma^3$, $\mathcal{P}_n
=\sigma^1$ for $\tfrac{n+1}2$ even and $\mathcal{P}_n=\ii\sigma^2$ for 
$\tfrac{n+1}{2}$ odd.

In a next step we dimensionally reduce the action \eqref{Spsi} on the manifold 
$M_{3,1} \times \cM_6$ focusing on the derivation of 
four-dimensional mass terms of the form \eqref{standard-W}. In order to do that
we decompose the ten-dimensional gravitinos $\hat\psi_M$ into four-dimensional spinors
on $M_{3,1}$ times six-dimensional spinors on the $SU(3)$ structure manifold $\cM_6$.
Of particular interest is the reduction of $\hat \psi_\mu$ where $\mu$ labels the 
four space-time directions on $M_{3,1}$. In type IIB both ten-dimensional gravitinos 
have the same chirality and split as 
\beq\label{graviIIB}
   \hat \psi^{A}_\mu= \psi^{A}_\mu \otimes \eta_- +  \bar \psi^{A}_\mu \otimes \eta_+
\qquad\, A=1,2\ ,
\eeq
where $\eta$ denotes the globally defined spinor introduced in eqn.~\eqref{epsilon}
with six-dimensional chirality $\pm$. The four-dimensional spinors $\psi_\mu^{1,2}$ 
and $\bar \psi^{1,2}_\mu$ are Weyl spinors with positive and negative 
chiralities respectively.
In type IIA supergravity the gravitinos have different chiralities and hence  
decompose as
\beq\label{graviIIA}
    \hat \psi^1_\mu = \psi^{1}_\mu \otimes \eta_+ + \bar \psi^{1}_\mu\otimes \eta_-\ ,\qquad \qquad
    \hat \psi^2_\mu = \psi^{2}_\mu \otimes \eta_- + \bar \psi^{2}_\mu \otimes \eta_+\ .
\eeq
The spinor $\psi^{1,2}_\mu$ appearing in \eqref{graviIIB} and \eqref{graviIIA} yield
the four-dimensional gravitinos when appropriately combined with four-dimensional
spinors arising in the expansion of  $\hat \psi_m,\  m=1,\ldots ,6$. However, since 
they are combined linearly the mass terms of $\psi^{1,2}_\mu$ take the same form 
as the one for the four-dimensional gravitinos which label the $N=2$ supersymmetry.

The orientifold projections reduce the four-dimensional theory to an $N=1$
supergravity. Hence, the two four-dimensional gravitinos as well as the 
spinors $\psi_\mu^{1,2}$ are not independent, but rather combine into one 
four-dimensional spinor $\psi_\mu$ which parameterizes the $N=1$ supersymmetry.
This spinor is chosen in such a way that its ten-dimensional extension $\hat \psi_M$ 
is invariant under the projections $\cO$ and $\cO_{(1,2)}$ given in eqns.~\eqref{oproj} and \eqref{o3-projection} 
respectively. To investigate the transformation behavior of ten-dimensional spinors, 
recall that the world-sheet parity $\Omega_p$ exchanges  $\hat \psi^{1}_M$ and 
$\hat \psi^{2}_M$. If the orientifold projection contains the operator $(-1)^{F_L}$ 
one finds an additional minus sign when applied to $\hat \psi^2_M$. In this we asserted that 
 $\hat \psi^2_M$ is in the NS-R sector while $\hat \psi^1_M$ is in the R-NS sector.
 The geometric symmetry $\simga$ acts only on the internal space $\cM_6$ 
 which translates to a non-trivial transformation of the globally defined spinor
 $\eta$. The precise action of $\sigma^*$ is different for type IIA and 
 type IIB orientifolds. In the following we will discuss the reduction 
 of both ten-dimensional type II theories in turn and determine the induced superpotentials.
\\[0.4cm]
\underline{\textit{The type IIA superpotential}} \\[0.2cm]
Let us first determine the superpotential for type IIA orientifolds induced by 
non-trivial background fluxes and torsion. Background fluxes are vacuum 
expectation values for the R-R and the NS-NS field strengths. We denote
the background flux of $d \hat B_2$ by $ H_3$ while the fluxes of the R-R
forms $d \hat C_n$ are denoted by $F_{n+1}$.  In order that the four-dimensional 
background $M_{3,1}$ is maximally symmetric the fluxes have to be extended
in the internal manifold $\cM_6$  or correspond to a four-form on $M_{3,1}$.
In type IIA supergravity we additionally allow for a 
scalar parameter $F_0$, which corresponds to the mass in the 
massive type IIA theory introduced by Romans \cite{Romans}.
In order that the background fluxes respect the orientifold condition \eqref{trans_BC_A} 
they have to obey
\beq
  \sigma^* H_3 = - H_3\ ,\qquad \qquad  \sigma^* F_n = \lambda(F_n)\ .
\eeq
It is convenient to  combine the R-R background fluxes into an even form $F^{\rm ev}$ on $\cM_6$ as 
\beq
   F^{\rm ev} = F_0 + F_2 + F_4 + F_6\ .
\eeq
In addition to the background fluxes also a non-vanishing intrinsic torsion of the 
$SU(3)$ structure manifold will induce terms contributing to the $N=1$ superpotential. 
These arise due to the non-closedness of the globally defined two-form $J$ and three-form
$\Omega_\eta$ and can be parameterized as given in eqn.~\eqref{dJ}. 

In order to actually perform the reduction we need to specify the action of the orientifold 
projection $\cO= (-1)^{F_L} \Omega_p \sigma^*$ 
on the spinors $\hat \psi_\mu^1$ and $\hat \psi_\mu^2$. 
The transformation behavior of the ten-dimensional gravitinos under $(-1)^{F_L} \Omega_p$
was already discussed above. We supplement this by the action of $\sigma^*$ on the 
globally defined spinor $\eta$. In accord with condition \eqref{constrA} one has 
\beq
    \simga^* \eta_+ = e^{i\theta} \eta_-\ , \qquad \qquad  \simga^* \eta_- =  e^{-i\theta} \eta_+\ ,
\eeq
where $\theta$ is the phase introduced in eqn.~\eqref{constrA}.
Therefore, the invariant combination of the four-dimensional spinors 
is given by  $\psi_\mu =\tfrac{1}{2} (e^{i  \theta/2}\, \psi^{1}_\mu -  e^{-i \theta/2}\, \psi^{2}_\mu)$
with a similar expression for the Weyl spinors $\bar \psi^{1,2}_\mu$.
In order to ensure the correct form of the four-dimensional kinetic terms 
for $\psi_\mu$ we restrict to the specific choice 
\beq \label{def-psi_4}
  \psi_\mu = e^{i  \theta/2}\, \psi^1_\mu = -e^{-i  \theta/2}\,  \psi^2_\mu\ , \qquad
   \qquad \bar \psi_\mu = e^{-i  \theta/2}\, \bar \psi^1_\mu = -e^{i  \theta/2}\, \bar \psi^2_\mu\ .
\eeq
These conditions define a reduction of a four-dimensional $N=2$ to an
$N=1$ supergravity theory \cite{ADAF,D'Auria:2005yg,GLW}. Hence, the mass terms of the spinors 
$\psi_\mu$ take the standard $N=1$ form given in eqn.~\eqref{standard-W}.

Now we turn to the explicit reduction of the ten-dimensional effective action \eqref{Spsi}
focusing on the mass terms of $\psi_\mu$ induced by the background fluxes 
$H_3$ and $F_n$ and the torsion of $\cM_6$.
We use the decomposition \eqref{graviIIA} together with \eqref{def-psi_4} and 
the gamma-matrix conventions summarized in appendix \ref{Spinors} to derive 
\bea  \label{sup1}
 S_{\psi}
         &=& - \int_{M_{3,1}}  e^{\frac{K}2} \bar\psi_\mu
      \bar \sigma^{\mu \nu} \bar \psi_\nu  *_4 \mathbf{1} \int_{\cM_6} \Big[ 
       4 e^{-\hat \phi+i\theta} \eta^\dagger_+\gamma^m D_m\eta_-  +
       4 e^{-\hat \phi-i\theta} \eta^\dagger_-\gamma^m D_m\eta_+\nn
      \\ 
      &&  \quad +\  \tfrac{1}{3!}\ e^{-\hat\phi+i\theta}\, (\hat H_3)_{mnp}\,
      \,\eta^\dagger_+\gamma^{mnp}\eta_-\   -\ \tfrac{1}{3!}\  e^{-\hat \phi -i \theta}\,  (\hat H_3)_{mnp}\,
      \eta^\dagger_-\gamma^{mnp}\eta_+\\
     && \quad +\tfrac{1}{2} \sum_{k\ \rm even} \tfrac{1}{k!} \big((\lambda
    \hat F_{k})_{m_1 \ldots m_k} \eta^\dagger_+\gamma^{m_1 \ldots m_k}\eta_+ 
         +(\hat F_{k})_{m_1 \ldots m_k} \eta^\dagger_-\gamma^{m_1 \ldots m_k}\eta_- \big)\Big] *_6 \mathbf{1}
        + \ldots\ , \nn
\eea
where $e^{K/2} = e^{2D} e^{K^{\rm K}/2}$ with $K^{\rm K}$ as defined in eqn.~\eqref{def-KK_A}.
The  four-dimensional dilaton $e^{D}$ is introduced in \eqref{def-D}.
Note that after the reduction of the $D=10$
string frame action to four space-time dimensions we performed a Weyl-rescaling to 
obtain a standard Einstein-Hilbert term. More precisely, in the derivation of 
\eqref{sup1} we made the rescaling 
\beq\label{weyl4}
    g_{\mu\nu}\rightarrow e^{2D} g_{\mu\nu}\ ,\qquad
    \sigma^\mu\rightarrow e^{-D}\simga^\mu\ ,\qquad
    \psi_\mu\rightarrow e^{D/2}\psi_\mu\ .
\eeq
The rescaling of $\psi_\mu$ ensures that the four-dimensional theory 
has a standard kinetic term for the gravitino. The superpotential 
can be obtained by comparing the action \eqref{sup1} with 
the standard $N=1$ mass term \eqref{standard-W}. We will
discuss the arising terms in turn and rewrite them into the 
form language used in the previous sections.

Let us next express the result \eqref{sup1} in terms of the globally defined 
two-form $J$ and $\Omega_\eta$ defined in \eqref{JOdef}. 
First note that $\Omega_\eta$ is related to the $\Omega$ used in analysis 
of the bosonic terms (sections \ref{OSpectrum} and \ref{Kaehlerpot}) by a rescaling
\beq \label{def-OO}
    \Omega_\eta =  e^{(\Kcs-K^{\rm K})/2} \Omega\ , 
\eeq 
where $e^{-\Kcs} = i \int \Omega \wedge \bar \Omega$ and $K^{\rm K}$ is defined in \eqref{def-KK_A}. 
The three-form $ \Omega_\eta$ is defined in such a way, that it satisfies automatically the first condition
 in \eqref{SU(3)-cond} when integrated over $\cM_6$.
The quantities in the first line of \eqref{sup1} are expressed in terms of
the forms $\Omega_\eta$ and $J$ by using the 
identities \footnote{%
The expression \eqref{OdJ} can be shown by 
using the Fierz identity \eqref{Fierz-ex} and expression \eqref{D-d} for $\eta^1 = \eta^2 = \eta$.}
\beq \label{OdJ}
   \int_{\cM_6}\eta^\dagger_-\gamma^mD_m\eta_+ *_6 \mathbf{1} =-\tfrac{1}{8}\int_{\cM_6}\Ox_\eta\wedge dJ\ ,
   \qquad  \int_{\cM_6}\eta^\dagger_+\gamma^mD_m\eta_- *_6 \mathbf{1}=-\tfrac{1}{8}\int_{\cM_6}\bar \Ox_\eta\wedge dJ\ ,
\eeq
where $d$ is the six-dimensional exterior derivative.
Using these integrals as well as \eqref{def-OO} and the definition of  $\Pi^{\rm odd} = C\Omega$  displayed in 
\eqref{def-Pi}, \eqref{def-C} 
one finds 
\beq \label{aux0}
    4 \int_{\cM_6} e^{-\hat \phi} \Big[ 
       e^{i\theta} \eta^\dagger_+\gamma^m D_m\eta_-  +
       e^{-i\theta} \eta^\dagger_-\gamma^m D_m\eta_+ \Big]*_6 \mathbf{1} =  
       -\int_{\cM_6}  
       \big<d\R(\Pi^{\rm odd}), J \big>\ .
\eeq
Similarly, one expresses the remaining terms in the action \eqref{sup1} 
using the three-from $\Pi^{\rm odd}$ and the two-form $J$. 
More precisely, the terms in the second line of eqn.~\eqref{sup1} are
rewritten by applying eqns.~\eqref{JOdef}, \eqref{def-OO}, \eqref{def-Pi} and $* \Omega =-i \Omega$ as
\bea \label{aux1}
   && \tfrac{1}{3!}\int_{\cM_6} \big[e^{-\hat \phi +i\theta} (\hat H_3)_{mnp}
      \eta^\dagger_+\gamma^{mnp}\eta_-  - e^{-\hat \phi -i \theta} (\hat H_3)_{mnp}  \eta^\dagger_-\gamma^{mnp}\eta_+ \big] *_6 \mathbf{1}
       \\
        && \qquad \qquad \qquad \qquad \qquad =  -i\int_{\cM_6} \Big[ 
    \big< H_3 \wedge \R(\Pi^{\rm odd}),1\big> +
         \big< d  \R(\Pi^{\rm odd}), \hat B_2\big> \Big]\ , \nn
\eea
where we have used that $\hat H_3 = d \hat B_2 + H_3$ with $H_3$ being the background flux. 
Finally, we apply gamma-matrix identities 
and the definition  \eqref{JOdef} of $J$ to rewrite the  
terms appearing in the last line of \eqref{sup1} as
\bea \label{aux2}
  &&\tfrac{1}{2} \sum_{k\ \rm even} \tfrac{1}{k!} \int_{\cM_6} \Big[ 
   (\lambda \hat F_k)_{m_1 \ldots m_k} \eta^\dagger_+\gamma^{m_1 \ldots m_k} 
   \eta_+ +
    (\hat F_k)_{m_1 \ldots m_k} \eta^\dagger_-\gamma^{m_1 \ldots m_k} 
   \eta_- \Big]*_6 \mathbf{1}\\
   &&  \qquad  \qquad= \int_{\cM_6} \Big[ \big< F^{\rm ev}, e^{-\hat B_2 +iJ} \big> 
         - \big< H_3 \wedge C^{(0)}_3,1\big> - \big<dC^{(0)}_3,\hat B_2\big>  + i \big< dC^{(0)}_3, J \big> \Big]\ , \nn
\eea
where $C_3^{(0)}$ is defined in \eqref{decomp_C} as the part of $\hat C_3$ being 
a three-form on $\cM_6$ yielding scalar fields in $M_{3,1}$. In deriving this identity 
one uses the definition of $\hat F_k$ given in eqn.~\eqref{defFH} while eliminating half 
of the R-R fields by the duality condition  \eqref{defFH} . 

In summary one can now read off the complete type IIA superpotential induced by 
background fluxes and torsion. Introducing the differential operator $d_H = d - H_3\wedge$ one finds (see also refs.~\cite{GLW, Granarev})
\beq
     W^{O6} = \int_{\cM_6} \big< F^{\rm ev} + d_H \Pi^{\rm odd}_c, e^{J_c}\big> \ , 
\eeq
where we used the definitions of $J_c = -\hat B_2 + iJ$ and $\Pi^{\rm odd}_c = C^{(0)}_3+i\R(\Pi^{\rm odd})$
given in eqns.~\eqref{def-J_c} and \eqref{def-Pi_c}. The superpotential extends the results 
of refs.~\cite{Derendinger:2004jn,VZ,Berglund, GLW, Camara:2005dc,Granarev} and together with the 
discussions above it is readily checked to be holomorphic in the $N=1$ coordinates.
As discussed in section \ref{OSpectrum} the complex forms $J_c$ and $\Pi^{\rm ev}_c$ are
linear in the complex $N=1$ coordinates. 
This is also the case for their derivatives $dJ_c$ and $d\Pi^{\rm ev}_c$,
where $d$ is the exterior derivative along $\cM_6$.
Therefore we deduce that $W$ is a polynomial of cubic order in $J_c$ times a linear polynomial in $\Pi^{\rm odd}_c$. 
 Let us now determine $W$ for the type IIB orientifold compactifications. 
\\[.4cm]
%
\underline{\textit{The type IIB superpotential}} \\[0.2cm]
%
In the following we will determine the superpotential of the 
type IIB orientifolds induced by the background fluxes and 
torsion. In the type IIB theory we allow for a non-trivial 
NS-NS flux $H_3$ as well as odd R-R fluxes. Due to the 
fact that we do not expand in one- or five-forms on $\cM_6$
the only non-vanishing R-R is the three-form $F_3$. The 
equation \eqref{trans_BC_B} implies that $H_3,F_3$ transform under 
the orientifold projection as
\beq
   \simga^* H_3 = - H_3 \ , \qquad \qquad \qquad \simga^* F_3 = \mp F_3\ ,
\eeq
where the minus sign in the second condition applies to type IIB orientifolds with $O3/O7$ planes
while the plus sign is chosen for $O5/O9$ orientifolds. 
Since, there are some qualitative differences between both cases we will 
discuss them in the following separately. 

\underline{$O3/O7:$} Our analysis starts with the $O3/O7$ orientifolds.  
As in the type IIA case we need to specify the spinor invariant under the orientifold
projections $\cO_{(1)}$ defined in eqn.~\eqref{o3-projection}.
We already gave the transformation of the ten-dimensional spinor
under the world-sheet parity $\Omega_p$ and $(-1)^{F_L}$. It 
remains to specify how $\simga^*$ acts on the internal spinor $\eta_\pm$.
Using eqns.~\eqref{transJB} and \eqref{Omegatransf} one infers \cite{Jockers:2005zy}
\beq
   \sigma^* \eta_+ =  i \eta_+\ , \qquad \qquad \sigma^* \eta_- = - i \eta_-\ ,
\eeq
such that $(\sigma^*)^2 \eta_\pm = - \eta_\pm$ consistent with the 
fact the $(-1)^{F_L}\Omega_p$ squares to $-1$ on the ten-dimensional gravitinos.
With these identities at hand one defines the four-dimensional linear combinations
$\psi_\mu = \frac{1}{2}(\psi^1_\mu + i \psi^2_\mu)$ together with the conjugate 
expression for $\bar \psi_\mu$. Combining $\psi_\mu,\bar \psi_\mu$ into a 
ten-dimensional spinor $\hat \psi_\mu$  by multiplication with $\eta_-$ and $\eta_+$ 
respectively it is readily shown that $\hat \psi_\mu$ is invariant under $\cO_{(1)}$. 
It turns out to be sufficient to determine $W$ for a more simple choice of the four-dimensional 
spinor $\psi_\mu$ given by 
\beq \label{four-grav_B}
   \psi_\mu = \psi^1_\mu = -i \psi^2_\mu\ , \qquad  \bar \psi_\mu = \bar  \psi^1_\mu = i \bar \psi^2_\mu\ .
\eeq 
These conditions define the reduction of the $N=2$ theory to $N=1$ induced by the 
orientifold projection. Inserting the decompositions \eqref{graviIIB} together with \eqref{four-grav_B}
into the ten-dimensional action \eqref{Spsi} one determines the 
$\psi_\mu$ mass terms 
\beq \label{sup2}
 S_{\psi} = - \int_{M_{3,1}}  e^{\frac{K}2} \bar\psi_\mu
      \bar \sigma^{\mu \nu} \bar \psi_\nu  *_4 \mathbf{1} \int_{\cM_6} 
     \tfrac{1}{3!}\Big[( e^{-\hat \phi} (\hat H_3)_{mnp}
      +i (\hat F_{3})_{m n p} )\Omega^{mnp}\Big] *_6 \mathbf{1}
        + \ldots , 
\eeq
where $e^{K/2} = e^{2D} e^{K^{\rm cs}/2}$ with $K^{\rm cs}$ as defined in eqn.~\eqref{def-KK_B}. 
In order to derive this four-dimensional action we performed the Weyl-rescaling \eqref{weyl4} to 
obtain a standard Einstein-Hilbert term. Moreover,  we used the identities \eqref{JOdef} and \eqref{def-OO} to 
replace the gamma-matrix expressions $\eta^\dagger_- \gamma^{mnp}\eta_+$  with the complex 
three-form $\Omega^{mnp}$ and absorbed a factor arising due to the Weyl-rescaling \eqref{weyl4}
into $e^{K/2}$.
It is interesting to note that there is no contribution from the reduction of the ten-dimensional 
kinetic term in the action \eqref{Spsi}. This can be traced back to the fact that in type IIB orientifolds
with $O3/O7$ planes the globally defined three- and two-forms $\Omega$ and $J$ 
transform with opposite signs under the map $\simga^*$. However, since the volume 
form is positive under the orientation preserving map $\simga$ the integral 
over terms like $d\Omega \wedge J$ vanishes. The non-closed forms $dJ$ and
$d\Omega$ nevertheless yield a potential for the four-dimensional scalars which 
is encoded by non-trivial D-terms.

Let us now express  the action \eqref{sup2} in terms of the globally defined three-form $\Omega$
and the form $\Phi^{\rm ev}$. Using the definition \eqref{def-Phi} of $\Phi^{\rm ev}$ one infers
\bea \label{aux3}
  \tfrac{1}{3!} \int_{\cM_6}\big[\, e^{-\hat \phi -i \theta} (\hat H_3)_{mnp} \Omega^{mnp} \big] *_6 \mathbf{1} 
     &=&-i \int_{\cM_6} e^{-\hat \phi} \big[ \big< H_3, \Omega \big> 
     + \big< d\hat B_2, \Omega \big>\big] \\
       &=& -i \int_{\cM_6} \big[ \big< H_3 \wedge \R(\Phi^{\rm ev}), \Omega \big> - \big< d\R(\Phi^{\rm ev}), \Omega \big>\big] \ , \nn
\eea
where we have used $\R(\Phi^{\rm ev})_0 = e^{-\hat \phi}$ and $\R(\Phi^{\rm ev})_2 = - e^{-\hat \phi} \hat B_2$ as
simply deduced from the definition \eqref{def-Phi}. For the R-R term in \eqref{sup2} one derives 
\beq \label{aux4}
  \tfrac{i}{3!} \int_{\cM_6}  (\hat F_{3})_{m n p} \Omega^{mnp} *_6 \mathbf{1} = \int_{\cM_6} \big[ \big< F_3 , \Omega \big>
  + \big< dA^{(0)}_2, \Omega\big> - \big< H_3 \wedge A^{(0)}_{0}, \Omega \big> \big]\ ,
\eeq
where $A^{(0)}_{2}$ and $A^{(0)}_0$ 
denote the two- and zero- forms in $A^{\rm ev}_{(0)}$ defined in \eqref{decomp_A}.%
\footnote{Expanding $A^{\rm ev}_{(0)}$ in \eqref{decomp_A} one 
finds $A^{(0)}_{2} = \hat C_2 - \hat C_0 \hat B_2$ and $A^{(0)}_0 = \hat C_0$.}
Together the two terms \eqref{aux3} and \eqref{aux4} combine into the superpotential 
\beq \label{def-W37}
    W^{O3/O7} = \int_{\cM_6}  \big<F_3 + d_H \Phi^{\rm ev}_c, \Omega \big> 
\eeq  
where $d_H = d - H_3 \wedge$ and  $\Phi^{\rm ev}_c$ is defined in eqn.~\eqref{def-Phi_c}. 
This superpotential contains the well-known Gukov-Vafa-Witten superpotential \cite{GVW,Taylor:1999ii} 
as well as contributions due to non-closed two-forms $\hat B_2$ and $\hat C_2$.
Also in this type IIB case it is 
straight forward to check the holomorphicity of $W$. As mentioned in section \ref{Kaehlerpot} 
the form $\Omega(z)$ is in general a complicated holomorphic function of the chiral coordinates $z$.
On the other hand $\Phi^{\rm ev}_c$ as well as $d \Phi^{\rm ev}_c$ 
depends linearly on the $N=1$ chiral coordinates. Hence,  
the superpotential $W^{O3/O7}$ is a linear function in $\Phi^{\rm ev}_c$ 
times a holomorphic function in the fields $z$ and contains no conjugate fields. 
Let us complete the discussion of the type IIB orientifolds by determining the 
$O5/O9$ superpotential.

\underline{$O5/O9 : $} To derive the superpotential for the $O5/O9$ orientifolds we first 
specify the combination of the two ten-dimensional gravitinos invariant under $\cO_{(2)}$ defined in
eqn.~\eqref{o3-projection}. We deduce the action of $\sigma^*$ on the globally defined 
spinor $\eta$ by examining the expressions \eqref{transJB} and \eqref{Omegatransf}, which yield
\beq
   \sigma^* \eta_+ = \eta_+\ , \qquad \qquad \sigma^* \eta_- =  \eta_-\ .
\eeq
The invariant combination of the four-dimensional spinors is given by $\psi_\mu=\tfrac12(\psi^1_\mu+\psi^2_\mu)$
with a similar relation for $\bar \psi_\mu$. As a specific choice for this combination we 
simplify to
\beq
   \psi_\mu=\psi^1_\mu=\psi^2_\mu\ , \qquad \quad \bar \psi_\mu=\bar \psi^1_\mu = \bar \psi^2_\mu\ .
\eeq
Together with the decomposition \eqref{graviIIB} we reduce the action \eqref{Spsi} to determine the mass 
term of $\psi_\mu$ as 
\beq\label{sup3}
 S_{\psi} = - \int_{M_{3,1}}  e^{\frac{K}2} \bar\psi_\mu
      \bar \sigma^{\mu \nu} \bar \psi_\nu  *_4 \mathbf{1} \int_{\cM_6} 
      \big< -i \hat F_{3} +  d(e^{-\hat \phi} J), \Omega \big> *_6 \mathbf{1}
        + \ldots ,     
\eeq  
where we have applied \eqref{OdJ} and performed the Weyl rescaling \eqref{weyl4}.
Note that the term involving the NS-NS fluxes vanishes 
in the case of $O5/O9$ orientifolds since $\Omega$ and $\hat H_3$ transform 
with an opposite sign under the symmetry $\simga^*$ as can be deduced from 
eqns.~\eqref{Omegatransf} and \eqref{trans_BC_B}.
Inserting the definition \eqref{defFH} of $\hat F_3$ into \eqref{sup3} one obtains the superpotential \cite{Granarev}
\beq
    W^{O5/O9} =-i \int_{\cM_6}  \big<F_3 +  d \Phi^{\rm ev}_c, \Omega \big> \ ,
\eeq
where we have used $\I(\Phi^{\rm ev})_2 = e^{-\hat \phi} J$ and the definition \eqref{def-Phi_c} of $\Phi^{\rm ev}_c$. 
The superpotential $W^{O5/O9}$ is a linear function in the $N=1$ fields encoded by $\Phi^{\rm ev}_c$
times a holomorphic function of the fields $z$. $W^{O5/O9}$ is independent of the NS-NS flux 
$H_3$ which was shown in ref.~\cite{TGL1} to contribute a D-term potential to the four-dimensional theory.

\section{Generalized orientifolds and mirror symmetry \label{Mirror}}

In this section we discuss $SU(3)\times SU(3)$ structure 
orientifolds and investigate mirror symmetry of the 
type IIA and type IIB setups. More precisely, we aim to 
specify setups dual to an orientifold compactification on a 
Calabi-Yau manifold $Y$ with background fluxes.
In doing that our main 
focus will be the identification of the $N=1$ superpotentials.
The superpotentials are holomorphic functions of the chiral fields 
of the four-dimensional theory and do not receive perturbative corrections. Hence, they 
yield a good testing ground for the mirror relations we will propose 
below. Note however, that the potential for Calabi-Yau orientifolds with $O5$ planes 
contains in addition to a superpotential contribution also a D-term potential which 
arises due to the presents of a gauged linear multiplet \cite{TGL1,mass_tensors}.
We will therefore focus on the mirror identifications between the 
type IIA orientifolds and the type IIB orientifolds with $O3/O7$ planes.
In general Calabi-Yau orientifolds with $O3/O7$ or $O6$ planes the potential induced 
by non-trivial NS-NS and R-R background fluxes 
is entirely encoded by a superpotential and the K\"ahler potential \cite{TGL1,TGL2,Granarev}.
We propose a possible mirror space $\cM_{\tilde Y}$ 
which possesses  a geometry 
dual to part of the electric and  magnetic NS-NS fluxes. In other words, 
we identify a space $\cM_{\tilde Y}$ such that
\beq \label{mirror1}
   \text{Typ IIB$_{O3/7}$}/ Y \text{ with } H^Q_3 \qquad \leftarrow\! \xrightarrow{ \text{mirror}\ } \qquad 
   \text{Typ IIA$_{O(\text{even})}$}/ \cM_{\tilde Y} \ . \eeq 
where the precise definition of the NS-NS flux $H^Q_3$ will be given shortly and 
the manifold $\cM_{\tilde Y}$ is specified in section \ref{gen-half}.
The evidence for the identification \eqref{mirror1} is discussed in section \ref{Ocheck}, where 
we also check consistency by analyzing the mirror relation 
\beq \label{mirror2}
   \text{Typ IIA$_{O6}$}/ Y \text{ with } H^Q_3 \qquad \leftarrow\! \xrightarrow{ \text{mirror}\ } \qquad 
   \text{Typ IIB$_{O(\text{odd})}$}/ \cM_{\tilde Y} \ ,
\eeq 
In both cases we concentrate on the superpotentials induced by the NS-NS flux $H^Q_3$ only.

In order to make the mirror conjectures \eqref{mirror1} and \eqref{mirror2} more 
precise we have to define the background flux $H^{\rm Q}_3$ as well as the properties 
of $\cM_{\tilde Y}$. Let us start with $H^{\rm Q}_3$. Recall 
that the background fluxes in Calabi-Yau compactifications are demanded 
to be harmonic forms in order to obey the equations of motion and Bianchi identities.  
This implies that before imposing the orientifold projections 
the general expansion of the NS-NS flux $H_3$ reads
\beq 
   H_3 = m^\Kh \alpha_\Kh - e_\Kh \beta^\Kh\ , \qquad \quad \Kh = 0, \ldots , \dim H^{(2,1)}\ ,
\eeq
where $(\alpha_\Kh , \beta^\Kh)$ is a real symplectic basis of $H^{3}(Y)$ satisfying \eqref{int_alpha_beta}.
We denoted the magnetic and electric flux quanta of $H_3$ by $(m^\Kh,e_\Kh)$. 
Different choices of the symplectic basis $(\alpha_\Kh, \beta^\Kh)$ are related by 
a symplectic rotation which also acts on the vector of flux quanta guaranteeing 
invariance of $H_3$. Note however, that due to the fact the supergravity 
reduction is only valid in the large volume limit the mirror symmetric theory 
has to be evaluated in the `large complex structure limit'.
Around this point of the moduli space the holomorphic three-form $\Omega$ on $Y$
admits a simple dependence on the complex structure moduli $z^K,\ K = 1,\ldots, \dim H^{(2,1)}$ 
explicitly given by \cite{Klemm:2005tw}
\beq \label{Omega_large}
  \Omega(z) = \alpha_0 + z^K \alpha_K + \tfrac{1}{2!} z^K z^L \kappa_{KLM} \beta^M - \tfrac{1}{3!} z^K z^L z^M \kappa_{KLM} \beta^0\ ,
\eeq
where $\kappa_{KLM}$ are intersection numbers on $Y$ defined, for example, in ref.~\cite{Candelas:1990pi}.
The expression \eqref{Omega_large} specifies a certain basis $(\alpha_\Kh, \beta^\Kh)$ of $H^{(3)}(Y)$. In particular it
singles out the elements $\alpha_0$ and $\beta^0$ with coefficients constant and cubic 
in the complex fields $z^K$.
Using this specification we are now in the position to define the NS-NS flux $H^Q_3$ 
by demanding that the flux quanta $e_0,\, m^0$ along $\beta^0,\, \alpha_0$ vanish. In other words 
we set 
\beq \label{def-HQ}
    H^{Q}_3 = m^K \alpha_K - e_K \beta^K\ , \qquad \qquad e_0=0\ ,\quad m^0 =0 \ ,
\eeq
where the index $K$ runs from $K = 1,\ldots, \dim H^{(2,1)}$. An equivalent  
definition of $H^{Q}_3$ can be given by interpreting mirror symmetry 
as T-duality along three directions of $Y$ \cite{Strominger:1996it}. One demands that the
components of the NS-NS flux $(H^Q_3)_{mnp}$ have never zero or three indices in the T-dualized directions
which corresponds to $e_0=m^0 = 0$.

\subsection{Generalized half-flat manifolds \label{gen-half}}

Let us now turn to the definition of the manifold $\cM_{\tilde Y}$. 
In reference \cite{GLMW,Fidanza:2003zi} it was argued that type II compactifications on Calabi-Yau manifolds 
with electric NS-NS fluxes are the mirror symmetric duals of compactifications 
on half-flat manifolds \eqref{def-half}. In order to also include magnetic fluxes into this 
mirror identification it is inevitable to generalize away from the $SU(3)$ structure 
compactifications \cite{Mathai:2005fd, Dabholkar:2005ve,GLW,Shelton:2005cf}. This might also 
lead to the application of the generalized manifolds $\cM_{\tilde Y}$ with $SU(3) \times SU(3)$ structure 
\cite{GLW}. In the remainder of this section we discuss 
some of the properties of the spaces $\cM_{\tilde Y}$, which 
might be mirror dual to Calabi-Yau manifolds with NS-NS fluxes $H^Q_3$.
We term these spaces `generalized half-flat manifolds'.
Some evidence for the mirror identifications \eqref{mirror1} and \eqref{mirror2}
will be provided in section \ref{Ocheck}. 

To start with let us recall the definition of a manifold with $SU(3) \times SU(3)$ structure \cite{Hitchin:2004ut,Gualtieri}.
Clearly, the group $SU(3) \times SU(3)$ cannot act on the tangent bundle alone
and one has to introduce a generalized tangent bundle $E$.
Following the work of Hitchin \cite{Hitchin:2004ut,Hitchin:2005in}, the  
generalized tangent bundle $E$ is given by
\beq \label{def-E}
      E_p \cong T_p \cM_6 \oplus T^*_p \cM_6\ , \qquad \quad p \in \cM_{\tilde Y}\ ,
\eeq
where $E$ is locally identified with the sum of the tangent and cotangent space.
Its global definition is more involved, since the spaces $E_p$ might be glued together 
non-trivially along $\cM_{\tilde Y}$ \cite{Hitchin:2005in}. To make this more precise one introduces 
a natural $O(6,6)$ metric on $E_p$ defined by 
  $(v+ \xi, u + \zeta) = \tfrac{1}{2}(\xi(u) + \zeta(v))$,
for $v,u \in T_p $ and $\xi,\zeta \in T_p^*$. Restricting further to 
transformations preserving the (natural) orientation of $E$
reduces the group down to $SO(6,6)$. A global definition 
can then be given by specifying elements of this group 
serving as transition function on overlapping patches on $\cM_{\tilde Y}$.
We are now in the position to define an $SU(3) \times SU(3)$ structure manifold by demanding 
the structure group of the bundle $E$ to reduce to 
$SU(3) \times SU(3) \subset SO(6,6)$. As in the case of $SU(3)$ structure 
manifolds discussed in section \ref{SU3}, this reduction can be specified in 
terms of two globally defined forms or two globally defined spinors on $\cM_{\tilde Y}$. 
We comment on the spinor picture in section \ref{Ocheck}, where it will also 
become clear that the structure group $SU(3) \times SU(3)$ 
is dictated by demanding that type II compactifications 
on $\cM_{\tilde Y}$ yield four-dimensional $N=2$ supergravity theories. 
Let us analyze here the characterization in terms of globally 
defined forms \cite{Hitchin:2004ut,Gualtieri,GLW}. 

Note that the group $SO(6,6)$ naturally admits spin representations on 
even and odd forms of $\cM_{\tilde Y}$. More precisely, one finds
two irreducible  Majorana-Weyl 
representations $S^{\rm ev}$ and $S^{\rm odd}$ given by
\beq \label{def-S}
   S^{\rm ev} \cong \Lambda^{\rm ev}T^* \otimes |\det T|^{1/2} \ ,\qquad 
   S^{\rm odd} \cong \Lambda^{\rm odd}T^* \otimes |\det T|^{1/2}\ ,
\eeq
where $\det T \cong \Lambda^6 T$ is fixed once a particular volume 
form is chosen. On elements $\Phi \in S^{\rm ev/odd}$ 
the group $SO(6,6)$ acts  with the Clifford 
multiplication 
\beq \label{Cliff}
  (v+\xi)\cdot \Pi = {v}   \lrcorner \Pi + \xi \wedge \Pi\ ,
\eeq
where $v \lrcorner$ indicates insertion of the vector $v \in T$
and $\xi \in T^*$ is a one-form.  
Using these definitions an 
$SU(3) \times SU(3)$ structure on $\cM_{\tilde Y}$ is specified by 
two complex globally defined even and odd forms $\Pi'^{\, \rm ev}$ and $\Pi'^{\rm odd}$
which are annihilated by half of the elements in $E$.\footnote{%
More precisely, each form $\Pi'^{\, \rm ev}$ and $\Pi'^{\rm odd}$ is demanded to be annihilated 
by a maximally isotropic subspace $E^{\rm ev}$ and $E^{\rm odd}$ of $E$. Isotropy implies that 
elements $v+\xi,u + \zeta \in E^{\rm ev/odd}$ obey $(v+ \xi, u + \zeta)=0$, while maximality 
corresponds to $\dim E^{\rm ev/odd}=6$.} 
Furthermore, in order to ensure the reduction of
$SO(6,6)$ to the direct product $SU(3)\times SU(3)$ the globally defined forms 
also have to obey \cite{Gualtieri,GLW}
\beq
  \big<\Pi'^{\, \rm ev}, \bar \Pi'^{\, \rm ev} \big> = \tfrac{3}{4} \big<\Pi'^{\rm odd},\bar \Pi'^{\rm odd} \big>\ , \qquad \quad 
  \big<\Pi'^{\, \rm ev}, (v + \xi) \cdot \Pi'^{\rm odd}\big> = 0\ ,
\eeq
for all elements $v + \xi \in E$. The pairing $\big<\cdot,\cdot \big>$ appearing in 
this expression is defined in \eqref{def-Mukai}. These conditions 
reduce to the standard $SU(3)$ structure conditions \eqref{SU(3)-cond} 
in case we identify 
\beq \label{sp_case}
   \Pi'^{\, \rm ev} = e^{-\hat \phi} e^{iJ}\ , \qquad \Pi'^{\rm odd} = C \Omega\ ,
\eeq
where $J$ and $\Omega$ are the globally defined two- and three-form.
In this expressions the additional degree of freedom in $|\det T|^{1/2}$ 
is labeled by the ten-dimensional dilaton $e^{-\hat \phi}$
also linearly appearing in the definition \eqref{def-C}
of $C$. Note however, that in the general $SU(3) \times SU(3)$ structure case
the odd form $\Pi^{\rm odd}$ also contains a 
one- and five-form contribution such that $\Pi'^{\rm odd} = \Pi'_1 + \Pi'_3 + \Pi'_5$.
It was shown in ref.~\cite{Gualtieri} that each of these forms 
locally admits the expression 
\beq \label{gen-exp}
    \Pi^{\rm odd} = e^{-\hat B_2} \wedge \Pi'^{\rm odd} = e^{-\hat B_2+iJ} \wedge C\Omega_k \ ,
\eeq
where $J$ is a real two-form and we also included a possible B-field on the internal manifold $\cM_{\tilde Y}$. 
The index $k$ is the degree of the complex form $\Omega_k$. In the special case that 
$k=3$ on all of $\cM_{\tilde Y}$ the form $\Pi'^{\rm odd}$ descends to the 
form \eqref{sp_case}. 
However, the degree of $\Omega_k $ can change when moving along $\cM_{\tilde Y}$ \cite{Gualtieri}.\footnote{%
Interesting examples of manifolds allowing such transitions were recently constructed in ref.~\cite{Gualtieri2}.}
In other words, the form $\Pi^{\rm odd}$ can locally contain a one-form $C\Omega_1$.
The presents of this one-form in the expansion \eqref{gen-exp} 
will be the key to encode the mirror of the magnetic fluxes in $H^Q_3$
given in \eqref{def-HQ}.

To make this more precise, one notes that the 
globally defined forms $ \Pi'^{\, \rm ev}$ and $\Pi'^{\rm odd}$
are not necessarily closed. This is already the 
case for $SU(3)$ structure manifolds which are 
half-flat and hence obey \eqref{def-half}.
For these manifolds the special forms \eqref{sp_case} 
are no longer closed, since $d\R(e^{i \theta}\Pi'^{\rm odd})$ and 
$d\I(\Pi'^{\, \rm ev})$ are non-vanishing. 
This obstruction of the internal manifold $\cM_{\tilde Y}$ to be Calabi-Yau is interpreted 
as mirror dual of the electric NS-NS fluxes $e_K$ appearing in the expansion \eqref{def-HQ}
of $H_3$ \cite{GLMW}.\footnote{The remaining flux parameter $e_0$ in eqn.~\eqref{def-HQ} 
induces a non-trivial $H_3$ flux on the mirror $\cM_{\tilde Y}$.} 
In order to also encode dual magnetic fluxes 
we generalize the half-flat conditions 
to the general odd form $\Pi^{\rm odd}$ given in eqn.~\eqref{gen-exp}. 
These generalized half-flat manifolds
are $SU(3) \times SU(3)$ structure manifolds 
for which 
\beq
    d \I(e^{i \theta}\Pi'^{\rm odd})=0
\eeq
where as above $e^{-i\theta}$ is the phase of $C$.
The real part of $e^{i\theta} \Pi'^{\rm odd}$ and the form $\Pi'^{\, \rm ev}$ are non-closed.
We conjecture that in a finite reduction the differentials 
$d\R(e^{i\theta}\Pi'^{\rm odd})$ and $d\Pi'^{\, \rm ev}$ 
are identified under mirror symmetry with the NS-NS fluxes $H^Q_3$.

Let us now make the mirror map between the type II theories on a manifold $\cM^{Y}_6$
and the Calabi-Yau compactifications with NS-NS fluxes explicit. 
In order to do that, we perform a finite reduction by specifying 
a set of forms $\Delta_{\rm finite}$. In contrast to the $SU(3)$
case discussed in section \ref{OSpectrum}, the forms in
the set $\Delta_{\rm finite}$ cannot anymore be distinguished 
by their degree. In the generalized manifolds only a distinction 
of even and odd forms is possible, such that 
\beq
   \Delta_{\rm finite} = \Delta^{\rm odd} \oplus \Delta^{\rm ev}\ ,
\eeq
where $\Delta^{\rm odd}$ now contains forms of all odd degrees.
In particular, the one-, three and five-form components of the 
form $\Pi^{\rm odd}$ given in eqn.~\eqref{gen-exp} 
can mix once one moves along $\cM_{\tilde Y}$. Nevertheless,
we are able to specify a basis of $\Delta_{\rm finite}$ 
such that a Kaluza-Klein reduction on these forms precisely 
yields the mirror theory obtained by a Calabi-Yau reduction 
with NS-NS fluxes. 

To make this more precise, we first specify a finite real symplectic basis of $\Delta^{\rm odd}$. 
We demand that it contains the non-trivial odd forms $(\gamma_\Kh,\tilde \gamma^\Kh)$ defined as 
\beq \label{gamma-basis}
   \gamma_\Kh = \big((\alpha_0 + \alpha_{(1)}), \alpha_K \big)\ ,\qquad
    \tilde \gamma^\Kh = \big((\beta^0+\beta^{(5)}), \beta^K\big)\ , \qquad  \int_{\cM_{\tilde Y}} \big<\gamma_\Kh,\tilde \gamma^\Lh \big>= \delta^\Lh_\Kh\ ,
\eeq
where $\alpha_{(1)}$ and $\beta^{(5)}$ are a one-form and five-form respectively. 
Note that as remarked above, the basis elements $(\gamma_\Kh,\tilde \gamma^\Kh)$ 
carry no definite form degree since $\gamma_0$ and $\tilde \gamma^0$ consist of 
a sum of one-, three- and five-forms. Using this basis the odd form $\Pi^{\rm odd}$
admits the expansion \footnote{Note that the precise moduli dependence 
of the expansion \eqref{exp_Piodd} will be not relevant in the following. 
The essential part is that $\Pi^{\rm odd}$ contains a part $C\gamma^0$ which is linear in C.
This can be always achieved by an appropriate rescaling of $C$.}
\beq \label{exp_Piodd}
    \Pi^{\rm odd} = C\big(\gamma_0 + z^K \gamma_K + \tfrac{1}{2!} z^K z^L \kappa_{KLM} \tilde \gamma^M - \tfrac{1}{3!} z^K z^L z^M \kappa_{KLM} \tilde \gamma^0 \big)\ , 
\eeq
which generalizes the expansion \eqref{Omega_large} for the three-form $\Omega$.
In order to identify the fields $z^K$ under the mirror map with the complexified K\"ahler structure 
deformations of $Y$ one has $K = 1,\ldots, \dim H^{(1,1)}(Y)$, while $\Kh$ takes an additional 
value $0$.
We also introduce a basis $\Delta^{\rm ev}$ of even forms on $\cM_{\tilde Y}$
denoted by $\omega_\Ah=(1,\omega_A)$ and $\tilde \omega^\Ah=(\tilde \omega^A,\epsilon)$,
with intersections as in equation \eqref{int_omega}. Mirror symmetry imposes that 
$A= 1,\ldots, \dim H^{(2,1)}(Y)$.
Note that due to the fact that $\Pi^{\rm odd},\ \Pi'^{\, \rm ev} \in \Delta_{\rm finite}$ are no longer 
closed not all basis elements $( \gamma_{\Kh},\tilde \gamma^{\Kh})$ and $(\omega_\Ah,\tilde \omega^\Ah)$
are annihilated by the exterior differential.
More precisely, we assign that 
\beq \label{devodd}
    d \gamma_0 = -m^A \omega_A - e_A \tilde \omega^A\ , \qquad d\omega_A = -e_A \beta^0\ ,\qquad d \tilde \omega^A = -m^A \beta^{(5)}\ ,
\eeq
which is in accord with the non-vanishing intersections \eqref{int_omega}
and \eqref{gamma-basis}. It is now clear from eqn.~\eqref{devodd} that  the existence 
of one- and five forms in $\gamma_0,\tilde \gamma^0$ is essential to encode
non-vanishing magnetic fluxes. 
In the case, one evaluates \eqref{devodd} for $\alpha_{(1)}=\beta^{(5)} = 0$
one encounters set-ups with dual electric fluxes only \cite{GLMW}.

In the finite reduction on $\cM_{\tilde Y}$ the equation \eqref{devodd} 
parameterizes the deviation of $\cM_{\tilde Y}$ to be Calabi-Yau. 
Using the expansion \eqref{exp_Piodd} of the globally defined forms $\Pi^{\rm odd}$ 
one easily applies \eqref{devodd} to derive 
\beq \label{non-closed}
  d \R(e^{i\theta}\Pi^{\rm odd}) = -|C|(m^A \omega_A + e_A \tilde \omega^A)\ ,\qquad  d \I(e^{i\theta}\Pi^{\rm odd}) = 0\ .
\eeq
In order that the low energy theories of compactifications 
on $\cM_{\tilde Y}$ coincide with the mirror reductions on the 
Calabi-Yau space $Y$ with fluxes the scale of 
torsion on $\cM_{\tilde Y}$ has to be below the Kaluza-Klein 
scale. In other words, the generalized half-flat manifold 
should be understood as a `small' deviation from the Calabi-Yau 
space $\tilde Y$ which is the mirror of $Y$ in the absence of fluxes.
Note however, that the topology of $\cM_{\tilde Y}$ differs 
from the one of the Calabi-Yau space $\tilde Y$ since $\Delta_{\rm finite}$
contains various non-harmonic forms. This suggests that an explicit 
construction of $\cM_{\tilde Y}$ might involve the shrinking 
of cycles in homology, which are later resolved with a non-trivial deformation.  
Unfortunately, an explicit construction of the manifolds  $\cM_{\tilde Y}$  is still missing. 
Moreover, it remains challenging to investigate the geometric structure of these manifolds 
in more detail. Despite of the fact that $\cM_{\tilde Y}$ possesses two 
globally defined forms $\Pi^{\rm odd}$ and $\Pi'^{\rm ev}$ it remains to be 
investigated if this allows to define the mirror of the Riemannian metric. 
Note however, that from a four-dimensional point of view the globally defined 
even and odd forms are sufficient to encode the $N=2$ or $N=1$ characteristic 
data. 

In the final section we provide some evidence for the conjecture that the generalized 
half-flat manifolds are the mirrors of Calabi-Yau compactifications with NS-NS fluxes. We
do this by deriving the superpotentials induced by the general odd forms $\Pi^{\rm odd}$.
The type IIA and type IIB cases will be analyzed in turn.

\subsection{The mirrors of type II Calabi-Yau orientifolds with fluxes \label{Ocheck}}

In this section we dimensionally reduce the fermionic action \eqref{Spsi} on 
a generalized $SU(3)\times SU(3)$ structure manifold $\cM_{\tilde Y}$. In addition we will impose the 
orientifold projections ensuring that the four-dimensional theory is an $N=1$
supergravity. This will allow us to derive the superpotentials arising due to 
the non-closed forms $\Pi^{\rm odd}$ and $\Pi^{\rm ev}$ and the background fluxes.
These can be evaluated for the generalized half-flat manifolds introduced in the 
previous section. We use the finite expansion \eqref{non-closed} to compare the 
superpotentials depending 
on $d\Pi^{\rm odd}$ to their mirror partners 
arising due to NS-NS  flux. 

In order to perform the dimensional reduction of the fermionic action 
\eqref{Spsi} the two ten-dimensional gravitinos $\psi^{1,2}_M$
are decomposed on the background $M_{3,1} \times \cM_{\tilde Y}$.
Hence, we are looking for a generalization of the decompositions 
\eqref{graviIIB} and \eqref{graviIIA}. Note however, that the internal manifold $\cM_{\tilde Y}$
possesses an $SU(3)\times SU(3)$ structure implying that one generically 
finds two globally defined spinors $\eta^1$ and $\eta^2$ on this space \cite{Gualtieri,GLW}. 
In terms of these two globally defined spinors $\eta^{1,2}$
the globally defined forms $\Pi'^{\rm ev}$ and $\Pi'^{\rm odd}$ are 
expressed as
\beq \label{component}
  \Pi'^{\rm ev}  =  2e^{-\hat \phi}\sum_{n \text{\ even}} \tfrac{1}{n!} \eta^{\dagger\, 2}_+
  \gamma_{p_1\ldots p_n} \eta^1_+ e^{p_n \ldots p_1} ,
  \qquad \Pi'^{\rm odd} = -2 C \sum_{n \text{\ odd}} \tfrac{1}{n!}\eta^{\dagger\, 2}_- \gamma_{p_1\ldots p_n} \eta^1_+ e^{p_n  \ldots p_1} ,
\eeq 
where $e^{p_1  \ldots p_n} = e^{p_1} \wedge \ldots \wedge e^{p_n}$ is a basis of $n$-forms $\Lambda^n T^*$ 
on the manifold $\cM_{\tilde Y}$. The presence of the two spinors $\eta^{1,2}$ ensures that 
the four-dimensional theory obtained by compactifying on the space $\cM_{\tilde Y}$ 
possesses $N=2$ supersymmetry. More precisely, the type IIB ten-dimensional 
gravitinos decompose on $M_{3,1} \times \cM_{\tilde Y}$ as
\beq\label{graviIIB_33}
   \hat \psi^{A}_\mu= \psi^{A}_\mu \otimes \eta^A_- +  \bar \psi^{A}_\mu \otimes \eta^A_+
\qquad\, A=1,2\ ,
\eeq
while the type IIA decomposition is given by
\beq\label{graviIIA_33}
    \hat \psi^1_\mu = \psi^{1}_\mu \otimes \eta^1_+ + \bar \psi^{1}_\mu\otimes \eta^1_-\ ,\qquad \qquad
    \hat \psi^2_\mu = \psi^{2}_\mu \otimes \eta^2_- + \bar \psi^{2}_\mu \otimes \eta^2_+\ .
\eeq
As in section \ref{Superpot}, the Weyl spinors $\psi_\mu^{1,2}$ 
and $\bar \psi^{1,2}_\mu$ yield the four-dimensional gravitinos parameterizing 
the $N=2$ supersymmetry of the theory. Clearly, the decompositions \eqref{graviIIB_33},  
\eqref{graviIIA_33} reduce on an $SU(3)$ structure manifold to the expressions \eqref{graviIIB}, \eqref{graviIIA},
if $\eta = \eta^1 = \eta^2$ is the only globally defined spinor. In general $\eta^1$
and $\eta^2$ are not necessarily parallel along all of $\cM_{\tilde Y}$. It is precisely this 
deviation which allows the general odd
forms \eqref{gen-exp} to locally contain a one-form component $\eta^{1\, \dagger} \gamma_m \eta^2$.

A dimensional reduction on backgrounds $\cM_{\tilde Y}$ with $SU(3)\times SU(3)$ structure 
yields a four-dimensional $N=2$ supergravity \cite{GLW}. The number of supersymmetries 
is further reduced to $N=1$ by imposing appropriate orientifold projections. 
To perform the four-dimensional $N=1$ reductions we discuss the type IIA 
and type IIB cases in turn.\\[.4cm]
%
\underline{\textit{The type IIA mirror of type IIB Calabi-Yau orientifolds with NS-NS flux}}\\[.2cm]
%
Let us first derive the four-dimensional superpotential for type IIA orientifolds 
on $\cM_{\tilde Y}$. It is most conveniently read off from the mass term \eqref{standard-W}
arising in the reduction of the fermionic action \eqref{Spsi}. In this derivation 
we have to impose the type IIA orientifold projection \eqref{oproj}. It is straight forward to 
extend the conditions \eqref{const_PA_A} to the more general odd and even forms 
\beq
   \Pi^{\rm odd} = e^{- \hat B_2} \wedge \Pi'^{\rm odd} \ ,\qquad \quad \Pi^{\rm ev}= e^{\hat \phi} e^{-\hat B_2} \wedge \Pi'^{\rm ev}\ ,
\eeq   
where $\Pi'^{\rm odd}$ and $\Pi'^{\rm ev}$ are given in expression \eqref{component}. One has
\beq \label{cond1}
   \sigma^* \Pi^{\rm odd} = \lambda(\bar \Pi^{\rm odd})\ ,\qquad \quad \sigma^* \Pi^{\rm ev} = \lambda(\Pi^{\rm ev})\ .
\eeq
In complete analogy to section \ref{Superpot} the transformations \eqref{cond1} impose 
constraints on the spinors $\eta^{1,2}$ appearing in the component expansion \eqref{component}.
Eventually, this implies that the two four-dimensional spinors $\psi^{1}_\mu$ and $\psi^2_\mu$
are related as in eqn.~\eqref{def-psi_4}. 

We are now in the position to perform the reduction of the action \eqref{Spsi}. 
The ten-dimensional terms only depending on NS-NS fields reduce using \eqref{graviIIA_33} 
and \eqref{def-psi_4} as 
\bea \label{Spsigen}
  S_{\psi- \text{NS}}  &=& - \int_{M_{3,1}}  e^{\frac{K}2} \bar\psi_\mu
      \bar \sigma^{\mu \nu} \bar \psi_\nu  *_4 \mathbf{1} \int_{\cM_6} \big<d \R (\Pi'^{\rm odd}) - \hat H_3\wedge \R (\Pi'^{\rm odd}), \Pi'^{\rm ev} \big>+ \ldots  \nn \\
      &=& - \int_{M_{3,1}}  e^{\frac{K}2} \bar\psi_\mu
      \bar \sigma^{\mu \nu} \bar \psi_\nu  *_4 \mathbf{1} \int_{\cM_6} \big<d_H \R (\Pi^{\rm odd}), \Pi^{\rm ev} \big>+ \ldots\ ,  
\eea
where $ \hat H_3=d\hat B_2 + H_3$ and the dots indicate terms depending on the R-R fields or
 not contributing to the mass term \eqref{standard-W}.
The expression \eqref{Spsigen} is a generalization of eqn.~\eqref{aux0} and \eqref{aux1} for the 
globally defined forms \eqref{component}. However, the derivation of \eqref{Spsigen} is 
slightly more involved, since terms proportional to $\eta^{A\, \dagger} \gamma^m D_m \eta^A$
or $H_{mnp}  \eta^{A\, \dagger} \gamma^{mnp} \eta^A$ for $A=1,2$ need to be converted to the sum of
forms \eqref{component}.
In order to do that one repeatedly uses the Fierz identities \eqref{Fierz} and \eqref{Fierz-ex} \cite{VanProeyen:1999ni}. 
Furthermore, the derivatives on the spinors $\eta^{1,2}$ translate to differentials on the forms 
$\Pi^{\rm ev/odd}$ defined in eqn.~\eqref{component} by using the identity \eqref{D-d}. 
The K\"ahler potential $K$ appearing in the action \eqref{Spsigen} takes the same form as the one 
for $SU(3)$ structure manifolds \eqref{Kaehler_A} if one substitutes the general odd and even forms $\Pi^{\rm ev}$ and 
$\Pi^{\rm odd}$. More precisely, the K\"ahler potential $K$ consists of the logarithms of the 
extended Hitchin functionals introduced in ref.~\cite{Hitchin:2004ut}.\footnote{The relevance of the 
extended Hitchin functional as a space-time action for the topological string was recently discussed in 
ref.~\cite{Pestun:2005rp}.}  A brief review of the relevant 
mathematical definitions can be found in appendix \ref{GCG-HF}. As in section \ref{Superpot}, the factor $e^{K/2}$
in the expression \eqref{Spsigen} arises after a four-dimensional Weyl rescaling \eqref{weyl4}.

It is straight forward to include the R-R fields into the fermionic reduction in full 
anology to section \ref{Superpot}. Together with 
the terms \eqref{Spsigen} the four-dimensional superpotential takes 
the form 
\beq \label{W-full_A}
    W =  \int_{\cM_6} \big< F^{\rm ev} + d_H \Pi^{\rm odd}_c, \Pi^{\rm ev} \big> \ ,
\eeq
where 
\beq
    \Pi^{\rm odd}_c = A^{\rm odd}_{(0)}+ i\R(\Pi^{\rm odd})\ , \qquad \quad \hat A^{\rm odd} = e^{-\hat B_2} \wedge \hat C^{\rm odd}\ .
\eeq
The complex form $\Pi^{\rm odd}_c$ can locally contain a one- and five-form contribution. 
The compex chiral multiplets parametrized by $\Pi^{\rm odd}_c$ arise as complex coeficients 
of an expansion into real forms $\Lambda^{\rm odd}_+$.
 
In order to compare this result to the mirror result for type IIB Calabi-Yau 
orientifolds with $O3/O7$ planes we perform the finite reduction discussed 
in the previous section. 
Note that we also have to impose the orientifold condition, such that the 
expansion of $\Pi^{\rm ev/odd}$ in performed into the appropriate subset 
of $\Delta_{\rm finite} = \Delta_+ \oplus \Delta_-$. As in eqn.~\eqref{split-Delta}
this splitting is with respect to the geometric symmetry $\cP_6=\sigma^* \lambda$.
Using the finite basis $(\gamma_\Kh,\gamma^\Kh)$ introduced in eqn.~\eqref{gamma-basis}
one expands 
\beq \label{exp_Pi_1}
  \Pi^{\rm odd}_c = N^k \gamma_k + T_\lambda \tilde \gamma^\lambda\ , \qquad (\gamma_k,\tilde \gamma^\lambda) \in \Delta^{\rm odd}_{+}\ .
\eeq
It is an important requirement that the form $\gamma_0$ is an 
element of $\Delta^{\rm odd}_{+}$ in order that the type IIA setups are
mirror dual to type IIB setups with $O3/O7$ planes \cite{TGL2}. The 
even form $\Pi^{\rm ev}$ is expanded in a basis $(1,\omega_a,\tilde \omega^b,\epsilon)$ 
of $\Delta^{\rm ev}_+$ as
\beq \label{exp_Pi_2}
  \Pi^{\rm ev} = 1 + t^a \omega_a + \tfrac{1}{2!} t^a t^b \cK_{abc} \tilde \omega^c + \tfrac{1}{3!} t^a t^b t^c \cK_{abc} \epsilon 
\eeq
where the $\cK_{abc}=\int \omega_a \wedge \omega_b \wedge \omega_c$ are the intersection numbers on $\Delta^2_-$.
On the type IIA side the NS-NS and R-R fluxes are set to be zero. 
Inserting the expressions \eqref{exp_Pi_1}, \eqref{exp_Pi_2} and \eqref{devodd}
into the superpotential \eqref{W-full_A} one finds 
\beq \label{m-GVW}
  W = - N^0 (e_b t^b + \tfrac{1}{2!} t^a t^b m^c \cK_{abc})\ .
\eeq 
The superpotential depends on the `electric' flux parameters $e_{0},e_a$ as
well as the `magnetic' fluxes $m^a$. Under the mirror map these parameters 
are identified with the NS-NS flux quanta in $H^Q_3$.

It is not hard to see, that the superpotential \eqref{m-GVW} is 
precisely the mirror superpotential to the well-known 
Gukov-Vafa-Witten superpotential for type IIB Calabi-Yau orientifolds 
with $O3/O7$ planes \cite{GVW,Taylor:1999ii}.
Denoting by $\tau$, the type IIB dilaton-axion the Gukov-Vafa-Witten superpotential 
for vanishing R-R fluxes reads \footnote{The superpotential \eqref{GVW} is a special 
case of the superpotential \eqref{sup_IIB} derived in section \ref{Superpot}.} 
\beq \label{GVW}
  W= - \tau \int_Y H^Q_3 \wedge \Omega = - \tau \big(e_k z^k + \tfrac{1}{2!} m^k z^l z^m \kappa_{klm}\big) \ ,
\eeq
where the NS-NS flux $H^Q_3$ is given in eqn.~\eqref{def-HQ} and $\Omega$ takes the form \eqref{Omega_large} 
in the large complex structure limit. Note that we additionally imposed the 
orientifold projection on the type IIB Calabi-Yau compactification, such that 
$H^Q_3 \in H^3_-(Y)$ contains flux quanta $(e_k,m^k)$, while $\Omega \in H^3_-(Y)$ is 
parameterized by fields $z^k$ only. It is now straight forward to identify the 
superpotentials \eqref{m-GVW} with \eqref{GVW} by applying the mirror map 
$t^a \cong z^k$, $N^0 \cong \tau$ and $\cK_{abc} \cong \kappa_{klm}$. The fluxes 
are identified as 
\beq
   d\Pi^{\rm odd}_c \ \cong \ -\tau H^3_Q\ .
\eeq
The fact that the two superpotentials can be identified gives some 
evidence for the chosen mirror geometry $\cM_{\tilde Y}$. Next, we will 
perform a similar analysis for the type IIB theories on $\cM_{\tilde Y}$
and check the consistency of our assertions.\\[.4cm]
\underline{\textit{The type IIB mirror of type IIA Calabi-Yau orientifolds with NS-NS flux}}\\[.2cm]
Let us now give a brief check of the second mirror identification displayed in eqn.~\eqref{mirror2}
by comparing the induced superpotentials.
In order to do that we perform a four-dimensional reduction of the type IIB effective 
action \eqref{Spsi} on the generalized manifolds $\cM_{\tilde Y}$.
In addition we impose the orientifold projection $\cO_{(1)}$ given in eqn.~\eqref{o3-projection}
such that the four-dimensional theory has $N=1$ supersymmetry.
We define the forms 
\beq
    \Phi^{\rm ev} = e^{-\hat B_2}\wedge \Pi'^{\rm ev}\ , \qquad \quad
    \Phi^{\rm odd} = C^{-1} e^{-\hat B_2} \wedge \Pi'^{\rm odd}\ ,
\eeq
where $\Pi'^{\rm ev}$ and $\Pi'^{\rm odd}$
are defined in eqn.~\eqref{component}. The orientifold symmetry 
$\sigma^*$ acts on these forms as 
\beq \label{gen_B}
   \sigma^* \Phi^{\rm odd} = - \lambda(\Phi^{\rm odd})\ , \qquad \qquad\sigma^* \Phi^{\rm ev} =  \lambda(\bar \Phi^{\rm ev})\ ,
\eeq
generalizing the conditions \eqref{const_PA_B}. These constraints translate 
into conditions on the spinors $\eta^{1,2}$. It is then consistent to identify 
the four-dimensional gravitinos parameterizing the original $N=2$
supersymmetry as in eqn.~\eqref{four-grav_B}. The single spinor $\psi_\mu$  
acquires a mass term due to background fluxes and 
the non-closedness of the forms $\Phi^{\rm ev}$ and $\Phi^{\rm odd}$. 

To derive the mass term \eqref{standard-W} for the spinor $\psi_\mu$
we dimensionally reduce the fermionic action \eqref{Spsi} for type 
IIB supergravity. Using the Fierz identities \eqref{Fierz}, \eqref{Fierz-ex} and the expression 
\eqref{D-d} one derives the superpotential 
\beq \label{gen_W_B}
  W = \int_{\cM_{\tilde Y}} \big<F^{\rm odd} + d_H \Phi^{\rm ev}_c,\Phi^{\rm odd} \big>\ ,
\eeq
where the even form $\Phi^{\rm ev}_c = A^{\rm ev}_{(0)} + i\R(\Phi^{\rm ev})$ is 
defined as in eqn.~\eqref{def-Phi_c}. The odd form $\Phi^{\rm odd}$ generically 
contains a one- and five-form. 

This superpotential can be compared with the 
NS-NS superpotential arising in type IIA Calabi-Yau orientifolds when performing 
a finite reduction outlined in section \ref{gen-half}. However, to also incorporate the 
orientifold constraints \eqref{gen_B} the expansion of $\Phi^{\rm ev}_c$ and $\Phi^{\rm odd}$ is in forms 
of the appropriate eigenspace of  $\Delta_{\rm finite}= \Delta_+ \oplus  \Delta_- $. 
More precisely, we have 
\beq \label{ev_ex_B}
   \Phi^{\rm ev}_c = \tau + G^a \omega_a + T_\alpha \tilde \omega^\alpha\ ,
\eeq 
where $(1,\omega_a,\tilde \omega^\alpha)$ is a basis of $\Delta^{\rm ev}_+$.
The expansion of $\Phi^{\rm odd}$ is given in eqn.~\eqref{exp_Piodd} and reduces under the 
orientifold projection to 
\beq \label{odd_ex_B}
    \Phi^{\rm odd} = \gamma_0 + z^k \gamma_k + \tfrac{1}{2!} z^k z^l \kappa_{klm} \tilde \gamma^m 
                    - \tfrac{1}{3!} z^k z^l z^m \kappa_{klm} \tilde \gamma^0\ , 
\eeq
where $(\gamma_0,\gamma_k, \gamma^k, \gamma^0 )$ is a basis of $\Delta^{\rm odd}_-$.
It is now straight forward to evaluate the general superpotential \eqref{gen_W_B}
for the even and odd forms \eqref{ev_ex_B} and \eqref{odd_ex_B}. Setting
$F^{\rm odd}=0$ and $H_3 = 0$ and using the expression \eqref{devodd} we find  
\beq \label{W-IIB}
  W^{\rm B}_{\cM_{\tilde Y}} = -G^a e_a - T_\alpha m^\alpha\ .
\eeq
Let us now recall the superpotential for type IIA Calabi-Yau orientifolds 
with NS-NS background flux $H^{\rm Q}_3$. It was shown in refs.~\cite{Derendinger:2004jn,TGL2}
that $W^{\rm A}_{H_3}$ takes the form
\beq \label{W-IIA}
  W^{\rm A}_{H_3} = - \int H_3^Q \wedge \Pi^{\rm odd}_c = - N^k e_k - T_\lambda m^\lambda \ , 
\eeq
where the expansion of $\Pi^{\rm odd}_c = N^k \alpha_k + T_\lambda \beta^\lambda$
is in harmonic three-forms $(\alpha_k,\beta^\lambda) \in H^3_+(Y)$.
The decomposition of $H_3^Q$ is given in eqn.~\eqref{def-HQ} and we appropriately 
imposed the orientifold constraint $H_3^Q \in H^{3}_-(Y)$. The two superpotentials \eqref{W-IIB} and \eqref{W-IIA}
coincide if applies the mirror map $G^a \cong N^k$ and $T_\alpha \cong T_\lambda$.

In summary, we conclude that the mirror identifications \eqref{mirror1} and \eqref{mirror2} might be correct 
for the special generalized half flat manifolds with finite reduction \eqref{devodd}. Clearly, this is only a first 
step and more involved checks are necessary to make the identifications  \eqref{mirror1} and \eqref{mirror2}
more precise. Moreover, it is a challenging task to explore more general orientifold compactifications 
on non-trivial $SU(3) \times SU(3)$ manifolds. Work along these lines is in progress.
 
\section{Conclusions and Discussion}

In this paper we discussed the four-dimensional $N=1$ supergravity 
theories arising in generalized orientifold compactifications of type IIA and 
type IIB supergravities. After defining the orientifold projection the $N=1$
spectrum of the four-dimensional theory was determined. As we have argued,
this can be done before specifying a particular finite reduction.
The degrees of freedom of the bosonic 
NS-NS fields encoded by the ten-dimensional metric, the B-field and the dilaton, decompose 
on $M_{3,1} \times \cM_6$ into a four-dimensional metric $g_4$ and two complex forms on $\cM_6$,
\beq \label{form-list}
    \underline{\text{Type IIA:}} \quad \Pi^{\rm ev}\ , \ \Pi^{\rm odd}\ ,\qquad \qquad \qquad
    \underline{\text{Type IIB:}}\quad \Phi^{\rm ev}\ ,  \ \Phi^{\rm odd}\ .
\eeq
The normalization of $ \Pi^{\rm odd}$ and $\Phi^{\rm ev}$ is set by the ten-dimensional 
dilaton, while the normalization of $\Pi^{\rm ev},\ \Phi^{\rm odd}$ is a unphysical scaling freedom. 
The forms $ \Pi^{\rm ev/odd}$ as well as $\Phi^{\rm ev/odd}$ obey various compatibility conditions
ensuring that the four-dimensional theory is supersymmetric. 
Moreover, the real and imaginary parts of these forms are not independent such that, at least formally, 
the real part can be expressed as a function of the imaginary part and vice versa. 
In case $\cM_6$ is an $SU(3)$ structure manifold the odd forms $\Pi^{\rm odd},\ \Phi^{\rm odd}$
only contain a three-form contribution, while the forms $\Pi^{\rm ev},\ \Phi^{\rm ev}$ 
are of general even degree.

From a four-dimensional point of view, 
the introduction of the odd and even forms \eqref{form-list} is appropriate to encode 
the bosonic degrees of freedom in the NS-NS sector. The bosonic fields 
in the R-R sector are captured by the ten-dimensional forms $\hat A^{\rm odd/ev}= e^{-\hat B_2} \wedge \hat C^{\rm odd/ev}$ for 
type IIA and type IIB respectively. Once again, not all degrees of freedom in these forms are independent 
since the duality condition \eqref{defFH} on the field strengths of  $\hat A^{\rm odd/ev}$ needs to be imposed.
The four-dimensional spectrum arises by expanding these ten-dimensional fields
into forms on the internal manifold $\cM_6$.
Despite the fact that forms on $\cM_6$ might only possess a grading into 
odd and even forms the orientifold projection allows to distinguish 
four-dimensional scalars and two-forms as well as vectors and three-forms. 
Altogether, the fields arrange into $N=1$ supermultiplets.

In determining the kinetic terms of the four-dimensional 
supergravity theory we focused on the metric on the chiral 
field space. Supersymmetry implies that this metric is K\"ahler and
we argued that the K\"ahler potential consists of the two Hitchin functionals
on $\cM_6$. These are functions of the odd and even forms listed in eqn.~\eqref{form-list}
respectively. The K\"ahler potentials are independent of the R-R fields which are protected by 
continuous shift symmetries. This will no longer be the case when D-instanton 
corrections are included. Given the K\"ahler potentials in the chiral description, the
kinetic potentials for the dual linear multiplets are determined by a Legendre transform \cite{BGG}.
In general, the theory consists of a set of (possibly massive) chiral and linear multiplets.
In this work we did not analyze the vector sector and three-form sector of the 
four-dimensional theory. In order to gain a full picture of possible supergravity
theories in four-dimensions it will be necessary to carefully include these fields.

Due to background fluxes and torsion the chiral multiplets can acquire a scalar 
potential. This scalar potential consists of an F-term contribution encoded by 
a holomorphic superpotential and possible D-term contributions due to non-trivial 
gaugings. Using a fermionic reduction we derived the general form of the superpotential
on $SU(3)$ structure manifolds. Together with the K\"ahler potential this allows to 
determine the chiral supersymmetry conditions on four-dimensional vacua and their cosmological 
constant $\Lambda = -3 e^K |W|^2$. In order to derive these data and to study 
moduli stabilization the explicitly construction of non-Calabi-Yau backgrounds is essential. 
Moreover, the inclusion of matter and moduli fields due to space-time filling 
D-branes will be needed in attempts to construct specific models for particle 
physics and cosmology.

We also presented some first results on type II compactifications 
on $SU(3)\times SU(3)$ structure orientifolds. Even though many of the 
$SU(3)$ structure results naturally generalize to the $SU(3)\times SU(3)$
structure case the consequences of this extension are enormous. 
The even and odd forms listed in eqn.~\eqref{form-list} are in these
generalized settings of generic even and odd degree. Moreover, the notion
of a specific form degree is not anymore well defined and can change on 
different patches of $\cM_6$. This can be traced back to the fact that the tangent 
and cotangent bundles $T$, $T^*$ are no longer the central geometric objects, but rather 
get replaced with the generalized  tangent bundle $E$ locally given by $T \oplus T^*$.
A non-closed NS-NS B-field has a natural interpretation in this formalism as 
a twisting of the forms $\Pi^{\rm ev/odd}$ and $\Phi^{\rm ev/odd}$ with a gerbe \cite{Hitchin:2004ut,Gualtieri,Hitchin:2005in}.
At least from a four-dimensional point of view one may attempt to formulate 
the supergravity in this generalized language providing a natural 
unification of all NS-NS fields. However, it should be clear that the generalized 
set-ups are not anymore `geometric' in the standard Riemannian sense. The 
metric $g_6$ on the tangent bundle is replaced by a metric on the extended  
tangent bundle $E$, which supports higher symmetry group then diffeomorphisms
of $\cM_6$. In general, this might also imply that the metric $g_6$ and the B-field 
mix as one moves along the internal manifold.

In this work we explored an interesting application of the generalized geometries 
as mirrors of Calabi-Yau compactifications with NS-NS fluxes. We characterized 
properties of generalized half-flat manifolds which might serve as a mirror of 
NS-NS fluxes $H^Q$. The superpotentials of $SU(3)\times SU(3)$ structure 
compactifications were derived from the reduction of the fermionic type IIA 
and type IIB actions. In a specific finite truncation the mirror fluxes can be 
identified as contributions from non-closed one- and three-forms in $\Pi^{\rm odd}$
and $\Phi^{\rm odd}$. Clearly, this is only a first step in the study of compactifications 
on generalized manifolds with $SU(3)\times SU(3)$ structure. 
It remains a challenging task to explore the pattern of fluxes supported in generalized 
compactifications and to determine the complete classical 
four-dimensional gauged supergravity.

\section*{Acknowledgments}

We would like to thank Dmitriy Belov, Frederik Denef, Ian Ellwood, Olaf Hohm, Hans Jockers, 
Albrecht Klemm, Yi Li, Jan Louis and Washington Taylor for valuable discussions and useful
comments. TWG 
also likes to thank Michael Douglas and the Rutgers University for hospitality 
and financial support.
TWG was supported in part by NSF CAREER Award No. PHY-0348093, 
DOE grant DE-FG-02-95ER40896, a Research Innovation Award and a Cottrell 
Scholar Award from Research Corporation. 
IB likes to thank  the DFG -- The German Science Foundation,
European RTN Program MRTN-CT-2004-503369  and the
DAAD -- the German Academic Exchange Service.\\[.5cm]
%
%
%
\section*{Appendix}
\renewcommand{\theequation}{\Alph{section}.\arabic{equation}}
\appendix

\section{The Clifford Algebra in 4 and 6 dimensions \label{Spinors}}
In this appendix we assemble the spinor conventions used throughout
the paper.

In $D=10$ the $\Gamma$-matrices are hermitian and satisfy the Clifford algebra
\beq
    \{\Gx^M,\Gx^N\}=2g^{MN}\ , \qquad M,N =0,\ldots, 9\ .
\eeq
One defines \cite{JPbook}
\beq
    \Gx^{11}=\Gx^0\ldots\Gx^9\ ,
\eeq
which has the properties
\beq
    (\Gx^{11})^2=1\ ,\quad \{\Gx^{11},\Gx^M\}=0\ .
\eeq
This implies that 
the Dirac representation can be split into  two Weyl 
representations 
\beq
    \mathbf{32}_{Dirac}= \mathbf{16}+\mathbf{16'}
\eeq
with 
eigenvalue $ +1$ and $-1 $ under $ \Gx^{11}$.

In backgrounds of the form \eqref{metricsplit} the 10-dimensional  Lorentz
group decomposes as 
\beq
    SO(9,1)\rightarrow SO(3,1) \times SO(6)\ ,
\eeq
implying a decomposition of the spinor representations as
\beq
    \mathbf{16}=(\mathbf{2},\mathbf{4})+(\mathbf{\bar 2},\mathbf{\bar
4})\ .
\eeq 
Here $\mathbf{2},\mathbf{4}$ are the Weyl representations of $SO(3,1)$
and $SO(6)$ respectively.

In the background \eqref{metricsplit} the ten-dimensional $\Gx$-matrices
 can be chosen block-diagonal as
\beq\label{gammadec}
\Gx^M = (\gamma^\mu\otimes {\bf 1}, \gamma^5\otimes \gamma^m) , \quad
\mu=0,\ldots, 3,\ m=1,\ldots, 6\ ,
\eeq
where $\gamma^5$ defines the Weyl representation in $D=4$.
In this basis $\Gx^{11}$ splits as
\cite{JPbook} 
\beq
    \Gx^{11}=-\gamma^{5}\otimes\gamma^{7}\ ,
\eeq  
where $\gamma^{7}$ 
defines the Weyl representations in $D=6$.

Let us now turn to our spinor convention in $D=6$ and  $D=4$ respectively.
%
\subsection{Clifford algebra in 6 dimensions }
%
%
In $D=6$ the gamma matrices are chosen hermitian 
$ \gamma^{m\dagger}=\gamma^m$ and they obey the Clifford algebra
\beq
    \{\gamma^m,\gamma^n\}=2g^{mn}\ ,\quad m,n=1,\ldots, 6\ .
\eeq
The Majorana condition on a spinor $\eta$ reads
\beq
    \eta^\dagger=\eta^T C\ ,
\eeq
where $C$ is the charge conjugation matrix 
\beq
    C^T=C\ ,\quad \gamma_m^T=-C\gamma_mC^{-1}\ .
\eeq
The following Fierz identity holds for spinors on $\cM_6$ \cite{VanProeyen:1999ni}
\beq \label{Fierz}
    M = \tfrac{1}{8} \sum_{k=0}^6 \tfrac{1}{k!} \gamma_{p_1 \ldots p_k} \text{Tr}( \gamma^{p_k \ldots p_1}M)\ ,  
\eeq
where $M$ is an arbitrary matrix in spinor space. Relevant examples used in the calculation of \eqref{Spsigen} are
$M = \eta_1 \otimes \eta^\dagger_2$, $M = (\gamma^m D_m \eta_1) \otimes \eta^\dagger_2$, etc. 
Using eqn.~\eqref{Fierz} it is not hard to show that 
\bea \label{Fierz-ex}
   \eta^{1\, \dagger} \gamma^m D_m \eta^1 &=&
    \tfrac18 \sum^6_{n=0} \tfrac{1}{n!}  \eta^{2\, \dagger} \gamma_{p_1 \ldots p_n}  \gamma^m D_m \eta^1 
   \ \eta^{1\, \dagger}\gamma^{p_n \ldots p_1} \eta^{2} \ , \\
   \eta^{1\, \dagger} \gamma^{mnp} \eta^1 &=&
    \tfrac18 \sum^6_{n=0} \tfrac{1}{n!}  \eta^{2\, \dagger} \gamma_{p_1 \ldots p_n}  \gamma^{mnp} \eta^1 
   \ \eta^{1\, \dagger}\gamma^{p_n \ldots p_1} \eta^{2} \ , \nn
\eea
with similar expressions for $\eta^2$. 
A second important identity encodes how derivatives on spinors translate into 
exterior derivatives on forms. Explicitly one has (see for example \cite{Jeschek:2005ek})
\beq \label{D-d}
   \sum_n \tfrac{1}{n!} \eta^{2\, \dagger} \{\gamma_{p_1 \ldots p_n}, \gamma^m \} D_m\eta^1 e^{p_n \ldots p_1} 
   = (d + d^*) \sum_n \tfrac{1}{n!}  \eta^{2\, \dagger} \gamma_{p_1 \ldots p_n} \eta^1 e^{p_n \ldots p_1} 
\eeq
where $d^*= -*_6\, d\, *_6$ is the formal adjoint of $d$, with $*_6$ being the six-dimensional Hodge-star.

\subsection{Clifford algebra in 4 dimensions}
%
In $D=4$ we adopt the conventions of \cite{WB} 
and choose
\bea
    \gamma^\mu=-i\left( \begin{array}{cc}
     0 &\sigma^\mu\\
    \bar\sigma^\mu &0 
    \end{array}\right),
    \qquad
    \gamma^5=\left(\begin{array}{cc}
    \mathbf{1} & 0\\
    0 & -\mathbf{1}
    \end{array}\right)
\eea
where the $\sx^\mu$ are the $2\times 2$ Pauli matrices 
\beq
    \sx^0=\left( \begin{array}{cc}
     -i &0\\
    0 &-i 
    \end{array}\right),
    \quad
    \sx^1=\left(\begin{array}{cc}
    0 & 1\\
    1 & 0
    \end{array}\right)\ ,
    \quad
    \sx^2=\left( \begin{array}{cc}
    0 & -i\\
    i & 0
    \end{array}\right),
    \quad
    \sx^3=\left(\begin{array}{cc}
      1 &0\\
    0 &-1 
    \end{array}\right)\ ,
\eeq
and $\bar\sx^0=\sx^0,\ \bar\sx^{1,2,3}=-\sx^{1,2,3}$. We define 
\beq
    \sigma^{\mu\nu}= \tfrac14 (\sigma^\mu \bar\sigma^\nu - \sigma^\nu \bar
    \sigma^\mu)\ ,\qquad
    \bar\sigma^{\mu\nu}= \tfrac14 (\bar\sigma^\mu \sigma^\nu - \bar\sigma^\nu 
    \sigma^\mu)\ .
\eeq

\section{Stable forms and the Hitchin functional 
\label{GCG-HF}}

In this appendix we collect some basic facts about the geometry 
of stable even and odd forms on a six-dimensional manifold $\cM_6$.
The definition of the Hitchin functionals will be recalled. The
case of stable three-forms and the general definition of 
stable odd and even forms will be reviewed in turn.
A more exhaustive discussion of these issues can be found in refs.~\cite{Hitchin:2000jd,Hitchin:2004ut,Gualtieri,Hitchin:2005in}.
We also comment on the derivation of the expression \eqref{second_der}.
\subsection{Stable three-forms and the standard Hitchin functional}
Let us first consider a six-dimensional manifold $\cM_6$ with 
a real globally defined  three-form $\rho \in \Lambda^3 T^*$.
A natural notion of non-degeneracy is that the form $\rho$
is stable. From an abstract point of view 
a stable form $\rho$ is defined by demanding that the natural 
action of $GL(6)$ on $\rho$ spans an open orbit in $\Lambda^3_p T^*$ 
at each point $p$ of $\cM_6$. 
This condition can also be formulated in terms 
of the map $q:\ \Lambda^3T^*\ \rightarrow\ \Lambda^6T^*\otimes\Lambda^6T^*$ defined as \cite{Hitchin:2000jd}
\beq \label{def-q}
     q(\rho)=\big<e^m\wedge f_n \lrcorner \rho,\rho\big> \big<e^n\wedge f_m \lrcorner \rho\wedge\rho\big>\ ,
\eeq 
where $e^m$ is a basis of $T^*\cM_6$ and $f_m$ is a basis of $T\cM_6$. 
The set of stable three-forms on $\cM_6$ is then shown to be
\beq
    U^3 = \big\{\rho \in \Lambda^{3}T^*: q(\rho) < 0 \big\}\ ,
\eeq 
where $q(\rho) < 0$ holds if $q(\rho) = - s \otimes s$ for some $s \in \Lambda^6T^*$.
Clearly, since $\Lambda^6T^* \cong \bbR$ this means that the product 
of the coefficients of the volume forms in \eqref{def-q} is negative.

It was shown in ref.~\cite{Hitchin:2000jd} that each real stable form $\rho \in U^3$
is written as 
\beq
  \rho = \tfrac{1}{2}(\Omega + \bar \Omega)\ ,
\eeq
where $\Omega$ is a complex three-form satisfying $\big<\Omega,\bar \Omega \big> \neq 0$.
The imaginary part of $\Omega$ is unique up to ordering and we denote it 
by $\hat \rho = \I(\Omega)$. The real three-forms 
$\hat \rho(\rho)$ can also be defined by using the map $q$
introduced in eqn.~\eqref{def-q}.  On forms $\rho \in U^3$
one defines the Hitchin function 
\beq \label{Hitchin-function}   
   \cH(\rho) := \sqrt{-\tfrac{1}{3} q(\rho)} \quad \in \ \Lambda^6 TY\ ,
\eeq
which is well defined since $ q(\chi) < 0$. 
The form $\hat \rho$ is then defined to be the 
Hamiltonian vector field on $TU^3 \cong \Lambda^3 T^*$ \footnote{The factor $4$ is not 
present in the corresponding expression in ref.~\cite{Hitchin:2000jd}. It arises due to the fact that 
we have set $\rho = \R(\Omega)$ and not $\rho_{\rm Hitchin} = 2 \R(\Omega)$ as in ref.~\cite{Hitchin:2000jd}}
\beq
   4 \big< \hat \rho, \alpha \big>= - \cD \cH(\alpha)\ ,\qquad \forall \alpha\ \in\ \Lambda^{3}T^*\ ,
\eeq
where $\cD$ is the differential on $TU^3$. Note that $H(\rho)$ can be 
rewritten as $\cH(\rho)=i \big<\Omega, \bar \Omega \big>$.

In this paper we mostly use the integrated version of the Hitchin function $\cH(\rho)$.
Since $\cH(\rho)$ is a volume form it is natural to define the Hitchin functional
\beq
   H[\rho] = \int_{\cM_6}\cH(\rho) =  i \int_{\cM_6} \big<\Omega, \bar \Omega \big>\ .
\eeq
Its first (variational) derivative is precisely the form $\hat \rho$ such that
\beq \label{first-der}
   \partial_\rho H = -4 \hat \rho\ , \qquad \qquad \partial_\rho H(\alpha) =-4 \int_{\cM_6} \big< \hat \rho, \alpha \big>\ .
\eeq
Here we also displayed how $\partial_\rho H$ is evaluated on some real form $\alpha \in\Lambda^3 T^*$.
The second derivative of $H[\rho]$ is given by 
\beq \label{second-der}
  \partial_\rho \partial_\rho H =- 4 \cI\ , \qquad \partial_\rho \partial_\rho H(\alpha,\beta) = -4\int_{\cM_6} \big< \alpha, \cI \beta \big> \ .
\eeq 
The map $\cI: \Lambda^3 T^* \rightarrow \Lambda^3 T^*$ is shown to be an almost complex structure
on $U^3$. It is used to prove that $U^3$ is actually a rigid special K\"ahler manifold \cite{Hitchin:2000jd}. 
The real form $\rho$ can be also used to define an almost complex structure $I_\rho$ on $\cM_6$ itself
by setting (see also the discussion in section \ref{SU3}) 
\beq \label{complex_str}
   (I_\rho)^m_n = \frac{1}{\cH(\rho)} (e^m\wedge f_n \lrcorner \rho\wedge\rho)\ , 
\eeq
where $\cH(\rho)$ is defined in eqn.~\eqref{Hitchin-function}. 
With respect to $I_\rho$ one decomposes complex three-forms as
\beq
  \Lambda^3 T^*_{\bbC} = \Lambda^{(3,0)} \oplus \Lambda^{(2,1)} \oplus \Lambda^{(1,2)}  \oplus \Lambda^{(0,3)} \ .
\eeq
Using this decomposition the complex structure $\cI$ on $U^3$ is evaluated 
to be  $\cI = i$ on $\Lambda^{(3,0)}\oplus \Lambda^{(2,1)}$ and 
$\cI = -i$ on $\Lambda^{(1,2)}\oplus \Lambda^{(0,3)}$.
Furthermore, assuming that $\cM_6$ possesses a metric hermitian with respect to
$I_\rho$ the six-dimensional Hodge-star obeys $*_6 = i$ on $\Lambda^{(0,3)}\oplus \Lambda^{(2,1)}$, while 
$*_6 =- i$ on $\Lambda^{(3,0)}\oplus \Lambda^{(1,2)}$. 
This implies the identifications
\beq \label{Istar}
  \cI = *_6\ \text{on} \ \Lambda^{(2,1)}\oplus \Lambda^{(1,2)}\ , \qquad 
  \cI = -*_6\ \text{on} \ \Lambda^{(3,0)} \oplus \Lambda^{(0,3)} \ .
\eeq
The identity \eqref{Istar} is essential to show eqn.~\eqref{second_der} 
as we will see in appendix \ref{der_KQ}.
\subsection{Stable odd/even forms and the extended Hitchin functional}
Let us now briefly review the definition of general odd and even stable forms and 
their associated Hitchin functional. 
Many of the identities for stable three-forms naturally generalize to the 
more generic case. We consider real odd or even forms $\rho^{\rm ev/odd} \in S^{ev/odd}$,
where $S^{\rm ev/odd} = \Lambda^{\rm ev/odd} T^* \otimes |\det T|^{1/2}$ 
was already defined in equation \eqref{def-S}. In most of the 
discussion a distinction between the odd and even case 
is not needed and we simplify our notation by writing $\rho^\cdot \in  S^\cdot$, where $\cdot = {\rm ev}$ or 
$\cdot = {\rm odd}$.
As in section \ref{gen-half} the generalized tangent bundle is denoted by $E=T\oplus T^*$
(cf.~equation \eqref{def-E}).
A natural Clifford action of elements of $E$ on the forms $\rho^\cdot$
is defined in eqn.~\eqref{Cliff}. In this sense the elements $S^\cdot$
are spinors of the group $SO(6,6)$. In analogy to the definition 
\eqref{def-q} one introduces
\beq
  q(\rho^\cdot) = \big< e^m\wedge f_n \lrcorner \rho^\cdot , \rho^\cdot \big> 
           \big< e^n\wedge f_m \lrcorner \rho^\cdot,\rho^\cdot \big>\ ,
\eeq
where $e^m$ is a basis of $T^*$ and $f_m$ is a basis of $T$. The
anti-symmetric Mukai pairing $\big<\cdot ,\cdot \big>$ is defined in eqn.~\eqref{def-Mukai}. 
The map $q(\rho^\cdot)$ can be evaluated for elements $\rho^\cdot \in S^{\cdot}$
yielding a number. The set of stable spinors $\rho^\cdot$ is then defined as
\beq
    U^{\cdot} = \big\{\rho^\cdot \in S^{\cdot}: q(\rho^\cdot) < 0 \big\}
\eeq
All spinors in $U^\cdot$ define a reduction of the structure group $SO(6,6)$ of $E$ 
to $U(3,3)$. Furthermore, the elements of $U^\cdot$ can be decomposed 
as 
\beq
    \rho^\cdot = \tfrac{1}{2} \big(\Pi^\cdot + \bar \Pi^\cdot \big)\ ,
\eeq
where as above the spinor $\hat \rho^\cdot =\I(\Pi^\cdot)$ is unique up 
to ordering. It was shown that the complex spinors $\Pi^\cdot$ are eliminated by half
of the elements in $E$ via the Clifford action \eqref{Cliff}. Such spinors are called 
pure spinors.

In order to define the Hitchin functional we un-twist $S^\cdot \rightarrow \Lambda^\cdot T^*$ 
and consider $q(\rho^\cdot)$ on forms. In this case $\sqrt{-q(\rho^\cdot)}$ is a volume form
and we define the extended Hitchin functional \cite{Hitchin:2004ut}
\beq
   H[\rho^\cdot] = \int_{\cM_6}\sqrt{-\tfrac{1}{3} q(\rho^\cdot)} =  i \int_{\cM_6} \big<\Pi^\cdot, \bar \Pi^\cdot \big>\ .
\eeq
As in the three-form case the first (variational) derivative is precisely the form $\hat \rho^\cdot$ such that
\beq
   \partial_{\rho^\cdot} H = -4\hat \rho^\cdot\ , \qquad \qquad \partial_{\rho^\cdot} H(\alpha) =- 4\int_{\cM_6} \big<\hat \rho^\cdot, \alpha\big>\ ,
\eeq
where $\alpha \in\Lambda^\cdot T^*$. The second derivative of $H$ is shown to  define a complex structure on the 
space $U^\cdot$. Moreover, the space of stable spinors $U^\cdot$ naturally admits a rigid special K\"ahler structure.

\subsection{Derivation of the K\"ahler metric \label{der_KQ}}

In this appendix we give more details on the derivation of the expression \eqref{second_der}
and briefly discuss its generalizations. 
We first show that 
\beq \label{what_to_show}
   \big[ \partial_{\Pi^{\rm odd}_c }\, \partial_{\bar \Pi^{\rm odd}_c} K^{\rm Q} \big]  (\alpha', \alpha) = 2 e^{2D} \int_{\cM_6}  \big< \alpha', *_6\, \alpha \big>\ ,
\eeq 
where $\alpha,\alpha' \in \Lambda^3_+ T^*$ are real three-forms obeying an additional condition 
and $K^{\rm Q}$ is given in eqn.~\eqref{def-KQ_A}. To begin with, 
notice 
that the K\"ahler potential $K^{\rm Q}$ is independent of the real part of $\Pi^{\rm odd}_c$ and only depends on 
$\rho = \R(\Pi^{\rm odd}) = \I(\Pi^{\rm odd}_c)$. Hence, one infers 
$\partial_{\Pi^{\rm odd}_c }\, \partial_{\bar \Pi^{\rm odd}_c} K^{\rm Q} = \tfrac14  \partial_{\rho}\, \partial_{\rho} K^{\rm Q}  $.
Using the expressions \eqref{first-der}, \eqref{second-der} and \eqref{def-KQ_A} 
it is straight forward to derive
\beq \label{ddK}
   \tfrac14 \big[ \partial_{\rho}\, \partial_{\rho} K^{\rm Q} \big]  (\alpha', \alpha) = 2 e^{2D} \int_{\cM_6}  \big< \alpha', \cI\, \alpha \big>
   + 8 e^{4D} \int_{\cM_6} \big< \hat \rho , \alpha' \big> \int_{\cM_6}  \big< \hat \rho , \alpha \big>\ ,
\eeq
where $\hat \rho = \I(\Pi^{\rm odd})$. Since $\cM_6$ admits an almost complex structure \eqref{complex_str}
associated to $\rho$, each form $\alpha$ can be decomposed as
\beq
  \alpha = \alpha^{(3,0)+(0,3)} +  \alpha^{(2,1)+(1,2)}\ \quad \in \Lambda^3_+ T^*\ .
\eeq
Using the fact that $\hat \rho$ is a $(3,0)+(0,3)$-form one has $\big< \hat \rho, \alpha^{(2,1)+(1,2)}\big>=0$. 
Therefore it is an immediate consequence of the identifications \eqref{Istar} that 
\eqref{what_to_show} holds on $\alpha^{(2,1)+(1,2)}$. It remains to show that 
it is also true for $\alpha^{(3,0)+(0,3)}$. Since we demand 
$\alpha^{(3,0)+(0,3)} \in \Lambda^3_+ T^*$ one has $\alpha^{(3,0)+(0,3)} = f \rho$ 
for some function $f$ on $\cM_6/\sigma$. Note that to show eqn.~\eqref{what_to_show}
we need to pull $f$ through the integral and hence demand that $f$ is actually constant. 
With this restriction it is straight forward to use 
$\int_{\cM_6}\big<\hat \rho,\rho \big> = \frac{1}{2} e^{-2D}$
to show that equation \eqref{what_to_show} holds on general $(2,1)+(1,2)$-forms 
and forms $\alpha^{(3,0)+(0,3)} \propto \rho$.\footnote{%
An alternative derivation of the condition \eqref{what_to_show}
might be performed by using decompositions into $SU(3)$ representations, as done 
for the $G_2$ analog of \eqref{what_to_show}, for example, in ref.~\cite{Joyce2,BW}.}

Let us also briefly comment on the general case. As we have just argued, the identity 
\eqref{what_to_show} is easily shown on $(2,1)+(1,2)$ forms, while general $(3,0)+(0,3)$ forms are
problematic. To also include the generic $(3,0)+(0,3)$ case one defines the `K\"ahler potential'
\beq \label{def-genK}
   \check K = - 2 \ln\big[i \big<\Pi^{\rm odd},\bar \Pi^{\rm odd} \big> \big] \ ,
\eeq
where now $\Pi^{\rm odd} = (\rho + i \hat \rho)\otimes \epsilon^{-1/2}$ is an element of $S^{3} = \Lambda^3 T^* \otimes |\det T|^{1/2}$
and $\epsilon$ is the volume form
\beq
  i \big<\Pi^{\rm odd},\bar \Pi^{\rm odd} \big> \epsilon = 2 \big<\rho,\hat \rho \big>\ .  
\eeq 
Note that the product $\big<\Pi^{\rm odd},\bar \Pi^{\rm odd} \big>$
yields a number while $\big<\rho,\hat \rho \big>$ is a volume form. 
Both are depending on the coordinates of $\cM_6$. Furthermore, in contrast 
to the K\"ahler potential \eqref{def-KQ_A} there is no integration in the functional \eqref{def-genK} \cite{GLW}.
The first derivative of $ \check K$ is obtained from eqn.~\eqref{first-der} to be
\beq
  \tfrac{1}{2}\big[\partial_{\R(\Pi^{\rm odd})} \check K\big](\alpha) =  4 e^{ \check K/2}\, 
  \big<\hat \rho, \alpha\big> \otimes \epsilon^{-1/2}\ ,
\eeq
for a three-form $\alpha \in \Lambda^3 T^*$.
Using eqn.~\eqref{second-der} the second derivative reads 
\beq
  \tfrac{1}{4}\big[ \partial_{\R(\Pi^{\rm odd})} \partial_{\R(\Pi^{\rm odd})}\check K \big](\alpha,\alpha') 
  =  2 e^{\check  K/2}\big<\alpha,\cI \alpha'\big> + 
     8e^{\check K } \big<\hat \rho,\alpha\big> \big<\hat \rho,\alpha\big> \epsilon^{-1}\ ,
\eeq
for a three-forms $\alpha,\alpha' \in \Lambda^3 T^*$. This is precisely a volume 
form and integration over $\cM_6$ yields a metric on three-forms $\alpha,\alpha'$.
Following the same reasoning as above, it is now straight forward to show 
\beq \label{gen_metric}
  \int_{\cM_6} \big[ \partial_{\Pi^{\rm odd}_c} \partial_{\bar \Pi^{\rm odd}_c} \check K \big](\alpha,\alpha') =
  2\int_{\cM_6} e^{2D} \big<\alpha, *_6 \alpha'\big>\ ,
\eeq
on all elements $\alpha,\alpha' \in \Lambda^3 T^*_+$. We have used that 
$\Pi^{\rm odd}_c= C_3^{(0)}+ i \R(\Pi^{\rm odd})$ and that $\check K$
is independent of the R-R fields $C_3^{(0)}$. In this general case, the 
four-dimensional dilaton $D$ is defined as 
$e^{-2D} = i \big<\Pi^{\rm odd},\bar \Pi^{\rm odd} \big>$ and can vary along 
$\cM_6$. The equation \eqref{gen_metric} is the generalization of the identity 
\eqref{what_to_show}.



\begin{thebibliography}{99}


\bibitem{reviewPP}
For a review see, for example,
E.~Kiritsis,
``D-branes in standard model building, gravity and cosmology,''
Fortsch.\ Phys.\  {\bf 52} (2004) 200
[arXiv:hep-th/0310001];\\
A.~M.~Uranga,
``Chiral four-dimensional string compactifications with intersecting
D-branes,''
Class.\ Quant.\ Grav.\  {\bf 20}, S373 (2003)
[arXiv:hep-th/0301032];\\
D.~L\"ust,
``Intersecting brane worlds: A path to the standard model?,''
Class.\ Quant.\ Grav.\  {\bf 21} (2004) S1399
[arXiv:hep-th/0401156];\\
L.~E.~Ib\'a\~nez,
``The fluxed MSSM,''
Phys.\ Rev.\ D {\bf 71}, 055005 (2005)
  [arXiv:hep-ph/0408064]; \\
R.~Blumenhagen,
``Recent progress in intersecting D-brane models,''
arXiv:hep-th/0412025;
R.~Blumenhagen, M.~Cvetic, P.~Langacker and G.~Shiu,
``Toward realistic intersecting D-brane models,''
arXiv:hep-th/0502005,
and references therein.


\bibitem{reviewcosmo}
For a review see, for example,
A.~Linde,
``Prospects of inflation,''
arXiv:hep-th/0402051;\\
V.~Balasubramanian,
``Accelerating universes and string theory,''
Class.\ Quant.\ Grav.\  {\bf 21} (2004) S1337
[arXiv:hep-th/0404075];\\
C.~P.~Burgess,
``Inflationary String Theory?,''
arXiv:hep-th/0408037;
A.~D.~Linde, ``Particle Physics and Inflationary Cosmology,''
arXiv:hep-th/0503203,
and references therein.


\bibitem{JP}
A.~Sagnotti,
``Open Strings And Their Symmetry Groups,''
arXiv:hep-th/0208020; \\
J.~Dai, R.~G.~Leigh and J.~Polchinski,
``New Connections Between String Theories,''
Mod.\ Phys.\ Lett.\ A {\bf 4} (1989) 2073;\\
R.~G.~Leigh,
``Dirac-Born-Infeld Action From Dirichlet Sigma Model,''
Mod.\ Phys.\ Lett.\ A {\bf 4} (1989) 2767;\\
M.~Bianchi and A.~Sagnotti,
``On The Systematics Of Open String Theories,''
Phys.\ Lett.\ B {\bf 247} (1990) 517;
``Twist Symmetry And Open String Wilson Lines,''
Nucl.\ Phys.\ B {\bf 361} (1991) 519;\\
P.~Horava,
``Strings On World Sheet Orbifolds,''
Nucl.\ Phys.\ B {\bf 327} (1989) 461;\\
J.~Polchinski,
``Dirichlet-Branes and Ramond-Ramond Charges,''
Phys.\ Rev.\ Lett.\  {\bf 75} (1995) 4724
[arXiv:hep-th/9510017];\\
E.~G.~Gimon and J.~Polchinski,
``Consistency Conditions for Orientifolds and D-Manifolds,''
Phys.\ Rev.\ D {\bf 54} (1996) 1667
[arXiv:hep-th/9601038].


\bibitem{JPbook}
J.~Polchinski,
``String Theory'', Vol. 1\& 2,
Cambridge University Press, Cambridge, 1998.

\bibitem{AD}
For a review see, for example,
A.~Dabholkar,
``Lectures on orientifolds and duality,''
[arXiv:hep-th/9804208];\\
C.~Angelantonj and A.~Sagnotti,
``Open strings,''
Phys.\ Rept.\  {\bf 371} (2002) 1
[Erratum-ibid.\  {\bf 376} (2003) 339]
[arXiv:hep-th/0204089].


\bibitem{Ori}
C.~Angelantonj, M.~Bianchi, G.~Pradisi, A.~Sagnotti and Y.~S.~Stanev,
``Chiral asymmetry in four-dimensional open- string vacua,''
Phys.\ Lett.\ B {\bf 385} (1996) 96
[arXiv:hep-th/9606169];\\
M.~Berkooz and R.~G.~Leigh,
``A D = 4 N = 1 orbifold of type I strings,''
Nucl.\ Phys.\ B {\bf 483} (1997) 187
[arXiv:hep-th/9605049];\\
G.~Aldazabal, A.~Font, L.~E.~Ib\'a\~nez and G.~Violero,
``D = 4, N = 1, type IIB orientifolds,''
Nucl.\ Phys.\ B {\bf 536} (1998) 29
[arXiv:hep-th/9804026];\\
M.~Cvetic, G.~Shiu and A.~M.~Uranga,
``Chiral four-dimensional N = 1 supersymmetric type IIA orientifolds from
intersecting D6-branes,''
Nucl.\ Phys.\ B {\bf 615} (2001) 3
[arXiv:hep-th/0107166];


\bibitem{GKP}
S.~B.~Giddings, S.~Kachru and J.~Polchinski,
``Hierarchies from fluxes in string compactifications,''
Phys.\ Rev.\ D {\bf 66} (2002) 106006
[arXiv:hep-th/0105097].


\bibitem{AAHV}
B.~Acharya, M.~Aganagic, K.~Hori and C.~Vafa,
``Orientifolds, mirror symmetry and superpotentials,''
[arXiv:hep-th/0202208].


\bibitem{BBKL}
R.~Blumenhagen, V.~Braun, B.~K\"ors and D.~L\"ust,
``Orientifolds of K3 and Calabi-Yau manifolds with intersecting D-branes,''
JHEP {\bf 0207} (2002) 026
[arXiv:hep-th/0206038];
R.~Blumenhagen, V.~Braun, B.~K\"ors and D.~L\"ust,
``The standard model on the quintic,''
arXiv:hep-th/0210083.


\bibitem{BH}
I.~Brunner and K.~Hori,
``Orientifolds and mirror symmetry,''
JHEP {\bf 0411} (2004) 005
[arXiv:hep-th/0303135];\\
I.~Brunner, K.~Hori, K.~Hosomichi and J.~Walcher,
``Orientifolds of Gepner models,''
arXiv:hep-th/0401137.

\bibitem{Granarev}
For a review see, for example,
  M.~Gra\~na,
  ``Flux compactifications in string theory: A comprehensive review,''
  Phys.\ Rept.\  {\bf 423} (2006) 91
  [arXiv:hep-th/0509003], and references therein.



\bibitem{Curio:2000sc}
  G.~Curio, A.~Klemm, D.~Lust and S.~Theisen,
  ``On the vacuum structure of type II string compactifications on  Calabi-Yau
  spaces with H-fluxes,''
  Nucl.\ Phys.\ B {\bf 609} (2001) 3
  [arXiv:hep-th/0012213].

\bibitem{Kachru:2002he}
  S.~Kachru, M.~B.~Schulz and S.~Trivedi,
  ``Moduli stabilization from fluxes in a simple IIB orientifold,''
  JHEP {\bf 0310}, 007 (2003)
  [arXiv:hep-th/0201028].

\bibitem{BKL}
R.~Blumenhagen, D.~L{\"u}st and T.~R.~Taylor,
``Moduli stabilization in chiral type IIB orientifold models with fluxes,''
Nucl.\ Phys.\ B {\bf 663} (2003) 319
[arXiv:hep-th/0303016].

\bibitem{Curio:2005ew}
  G.~Curio, A.~Krause and D.~Lust,
  ``Moduli stabilization in the heterotic / IIB discretuum,''
  arXiv:hep-th/0502168.

\bibitem{Denef:2005mm}
  F.~Denef, M.~R.~Douglas, B.~Florea, A.~Grassi and S.~Kachru,
  ``Fixing all moduli in a simple F-theory compactification,''
  arXiv:hep-th/0503124.

\bibitem{Lust:2005dy}
  D.~Lust, S.~Reffert, W.~Schulgin and S.~Stieberger,
  ``Moduli stabilization in type IIB orientifolds. I: Orbifold limits,''
  arXiv:hep-th/0506090.



\bibitem{Derendinger:2004jn}
  J.~P.~Derendinger, C.~Kounnas, P.~M.~Petropoulos and F.~Zwirner,
  ``Superpotentials in IIA compactifications with general fluxes,''
  Nucl.\ Phys.\ B {\bf 715} (2005) 211
  [arXiv:hep-th/0411276].\\
  J.~P.~Derendinger, C.~Kounnas, P.~M.~Petropoulos and F.~Zwirner,
  ``Fluxes and gaugings: N = 1 effective superpotentials,''
  Fortsch.\ Phys.\  {\bf 53} (2005) 926
  [arXiv:hep-th/0503229].

\bibitem{VZ}
  G.~Villadoro and F.~Zwirner,
  ``N = 1 effective potential from dual type-IIA D6/O6 orientifolds with
  general fluxes,''
  JHEP {\bf 0506} (2005) 047
  [arXiv:hep-th/0503169].

\bibitem{DeWolfe:2005uu}
  O.~DeWolfe, A.~Giryavets, S.~Kachru and W.~Taylor,
  ``Type IIA moduli stabilization,''
  JHEP {\bf 0507} (2005) 066
  [arXiv:hep-th/0505160].


\bibitem{Camara:2005dc}
  P.~G.~Camara, A.~Font and L.~E.~Ibanez,
  ``Fluxes, moduli fixing and MSSM-like vacua in a simple IIA orientifold,''
  JHEP {\bf 0509} (2005) 013
  [arXiv:hep-th/0506066].



\bibitem{Becker:2002nn}
K.~Becker, M.~Becker, M.~Haack and J.~Louis,
``Supersymmetry breaking and alpha'-corrections to flux induced potentials,''
  JHEP {\bf 0206} (2002) 060
  [arXiv:hep-th/0204254].


\bibitem{TGL1}
  T.~W.~Grimm and J.~Louis,
  ``The effective action of N = 1 Calabi-Yau orientifolds,''
  Nucl.\ Phys.\ B {\bf 699} (2004) 387
  [arXiv:hep-th/0403067].


\bibitem{TGL2}
  T.~W.~Grimm and J.~Louis,
  ``The effective action of type IIA Calabi-Yau orientifolds,''
  Nucl.\ Phys.\ B {\bf 718} (2005) 153
  [arXiv:hep-th/0412277].


\bibitem{Soft} 
M.~Gra\~na,
``MSSM parameters from supergravity backgrounds,''
Phys.\ Rev.\ D {\bf 67}, 066006 (2003)
[arXiv:hep-th/0209200];\\
B.~K\"ors and P.~Nath,
``Effective action and soft supersymmetry breaking for intersecting D-brane
models,''
Nucl.\ Phys.\ B {\bf 681}, 77 (2004)
[arXiv:hep-th/0309167];\\
P.~G.~C\'amara, L.~E.~Ib\'a\~nez and A.~M.~Uranga,
``Flux-induced SUSY-breaking soft terms,''
Nucl.\ Phys.\ B {\bf 689} (2004) 195
[arXiv:hep-th/0311241];\\
M.~Gra\~na, T.~W.~Grimm, H.~Jockers and J.~Louis,
``Soft supersymmetry breaking in Calabi-Yau orientifolds with D-branes and fluxes,''
Nucl.\ Phys.\ B {\bf 690} (2004) 21
[arXiv:hep-th/0312232];\\
A.~Lawrence and J.~McGreevy,
``Local string models of soft supersymmetry breaking,''
arXiv:hep-th/0401034;\\
``Remarks on branes, fluxes, and soft SUSY breaking,''
arXiv:hep-th/0401233;
D.~L\"ust, P.~Mayr, R.~Richter and S.~Stieberger,
``Scattering of gauge, matter, and moduli fields from intersecting branes,''
Nucl.\ Phys.\ B {\bf 696}, 205 (2004)
[arXiv:hep-th/0404134];\\
D.~Lust, S.~Reffert and S.~Stieberger,
  ``Flux-induced soft supersymmetry breaking in chiral type IIb  orientifolds
  with D3/D7-branes,''
  Nucl.\ Phys.\ B {\bf 706}, 3 (2005)
  [arXiv:hep-th/0406092];\\
P.~G.~Camara, L.~E.~Ibanez and A.~M.~Uranga,
  ``Flux-induced SUSY-breaking soft terms on D7-D3 brane systems,''
  Nucl.\ Phys.\ B {\bf 708}, 268 (2005)
  [arXiv:hep-th/0408036];\\
D.~Lust, S.~Reffert and S.~Stieberger,
  ``MSSM with soft SUSY breaking terms from D7-branes with fluxes,''
  Nucl.\ Phys.\ B {\bf 727}, 264 (2005)
  [arXiv:hep-th/0410074];\\
A.~Font and L.~E.~Ibanez,
  ``SUSY-breaking soft terms in a MSSM magnetized D7-brane model,''
  JHEP {\bf 0503}, 040 (2005)
  [arXiv:hep-th/0412150];\\
D.~Lust, P.~Mayr, S.~Reffert and S.~Stieberger,
  ``F-theory flux, destabilization of orientifolds and soft terms on
  D7-branes,''
  Nucl.\ Phys.\ B {\bf 732}, 243 (2006)
  [arXiv:hep-th/0501139];\\
K.~Choi, A.~Falkowski, H.~P.~Nilles and M.~Olechowski,
  ``Soft supersymmetry breaking in KKLT flux compactification,''
  Nucl.\ Phys.\ B {\bf 718}, 113 (2005)
  [arXiv:hep-th/0503216];\\
M.~Bertolini, M.~Billo, A.~Lerda, J.~F.~Morales and R.~Russo,
  ``Brane world effective actions for D-branes with fluxes,''
  arXiv:hep-th/0512067;\\
C.~M.~Chen, T.~Li and D.~V.~Nanopoulos,
  ``Type IIA Pati-Salam flux vacua,''
  arXiv:hep-th/0601064.


\bibitem{Frey:2003tf}
  A.~R.~Frey,
  ``Warped strings: Self-dual flux and contemporary compactifications,''
  arXiv:hep-th/0308156.

\bibitem{Jockers:2005pn}
  H.~Jockers,
  ``The effective action of D-branes in Calabi-Yau orientifold
  compactifications,''
  Fortsch.\ Phys.\  {\bf 53} (2005) 1087
  [arXiv:hep-th/0507042].

\bibitem{TGthesis}
  T.~W.~Grimm,
  ``The effective action of type II Calabi-Yau orientifolds,''
  Fortsch.\ Phys.\  {\bf 53} (2005) 1179
  [arXiv:hep-th/0507153].


\bibitem{CS}
S.\ Chiossi and S.\ Salamon, ``The Intrinsic Torsion of $SU(3)$ and $G_2$
Structures,'' in \emph{Differential geometry, Valencia, 2001}, pp. 115,
arXiv: math.DG/0202282.

\bibitem{salamonb}
S. Salamon, {\it Riemannian Geometry and Holonomy Groups}, Vol.~201 of
{\it Pitman Research Notes in Mathematics}, Longman, Harlow, 1989.

\bibitem{joyce}
D.\ Joyce, ``Compact Manifolds with Special Holonomy'', Oxford
University Press, Oxford, 2000.




\bibitem{Rocek}
M.~Rocek,
``Modified Calabi--Yau manifolds with torsion,''
in: Essays on Mirror Manifolds, ed.\ S.T.\ Yau (International Press,
Hong Kong, 1992).\\
S.~J.~Gates, C.~M.~Hull, and M.~Rocek,
``Twisted Multiplets And New Supersymmetric Nonlinear Sigma Models,''
Nucl.\ Phys.\  {\bf B248} (1984) 157.

\bibitem{StromingerT}
A.~Strominger,
``Superstrings With Torsion,''
Nucl.\ Phys.\ B {\bf 274} (1986) 253.


\bibitem{hullT}
C.~M.~Hull,
``Superstring Compactifications With Torsion And Space-Time
Supersymmetry,''
in {\it Turin 1985, Proceedings, Superunification and Extra
 Dimensions}, 347-375; \\
``Compactifications Of The Heterotic Superstring,''
Phys.\ Lett.\ B {\bf 178} (1986) 357.


\bibitem{waldram}
J.~P.~Gauntlett, N.W.~Kim, D.~Martelli, and D.~Waldram,
``Fivebranes wrapped on SLAG three-cycles and related geometry,''
{\em JHEP} {\bf 0111} (2001) 018, {\tt hep-th/0110034};\\
J.~P.~Gauntlett, D.~Martelli, S.~Pakis, and D.~Waldram,
``G-structures and wrapped NS5-branes,'' {\tt hep-th/0205050}; \\
J.~P.~Gauntlett, D.~Martelli and D.~Waldram,
``Superstrings with intrinsic torsion,''
  Phys.\ Rev.\ D {\bf 69}, 086002 (2004)
  [arXiv:hep-th/0302158].

\bibitem{GLMW}
S.~Gurrieri, J.~Louis, A.~Micu and D.~Waldram,
``Mirror symmetry in generalized Calabi--Yau compactifications,''
Nucl.\ Phys.\ B {\bf 654} (2003) 61
[arXiv:hep-th/0211102];\\
S.~Gurrieri and A.~Micu,
``Type IIB theory on half-flat manifolds,''
Class.\ Quant.\ Grav.\  {\bf 20} (2003) 2181
[arXiv:hep-th/0212278].

\bibitem{Cardoso:2002hd}
  G.~L.~Cardoso, G.~Curio, G.~Dall'Agata, D.~Lust, P.~Manousselis and G.~Zoupanos,
  ``Non-Kaehler string backgrounds and their five torsion classes,''
  Nucl.\ Phys.\ B {\bf 652} (2003) 5
  [arXiv:hep-th/0211118].

\bibitem{KSTT}
  S.~Kachru, M.~B.~Schulz, P.~K.~Tripathy and S.~P.~Trivedi,
  ``New supersymmetric string compactifications,''
  JHEP {\bf 0303}, 061 (2003)
  [arXiv:hep-th/0211182].

\bibitem{KMPT}
P.~Kaste, R.~Minasian, M.~Petrini, and A.~Tomasiello,
``Kaluza--Klein bundles and manifolds of exceptional holonomy,''
JHEP {\bf 0209}, 033 (2002)
[arXiv:hep-th/0206213];
``Nontrivial RR two-form field strength and $\SU(3)$-structure,''
Fortsch.\ Phys.\  {\bf 51}, 764 (2003)
[arXiv:hep-th/0301063].

\bibitem{CCDL}
G.~L.~Cardoso, G.~Curio, G.~Dall'Agata and D.~L\"ust,
``BPS action and superpotential for heterotic string compactifications  with
fluxes,''
JHEP {\bf 0310} (2003) 004
[arXiv:hep-th/0306088].

\bibitem{Becker}
K.~Becker, M.~Becker, K.~Dasgupta and P.~S.~Green,
``Compactifications of heterotic theory on non-K\"ahler complex manifolds. I,''
JHEP {\bf 0304} (2003) 007
[arXiv:hep-th/0301161].\\
K.~Becker, M.~Becker, P.~S.~Green, K.~Dasgupta and E.~Sharpe,
``Compactifications of heterotic strings on non-K\"ahler complex manifolds. II,''
Nucl.\ Phys.\ B {\bf 678} (2004) 19
[arXiv:hep-th/0310058];\\
M.~Becker, K.~Dasgupta, A.~Knauf and R.~Tatar,
``Geometric transitions, flops and non-K\"ahler manifolds. I,''
Nucl.\ Phys.\ B {\bf 702}, 207 (2004)
[arXiv:hep-th/0403288];\\
S.~Alexander, K.~Becker, M.~Becker, K.~Dasgupta, A.~Knauf and R.~Tatar,
  ``In the realm of the geometric transitions,''
  Nucl.\ Phys.\ B {\bf 704} (2005) 231
  [arXiv:hep-th/0408192]\\
M.~Becker, K.~Dasgupta, S.~Katz, A.~Knauf and R.~Tatar,
  ``Geometric transitions, flops and non-Kaehler manifolds. II,''
Nucl.\ Phys.\ B {\bf 738} (2006) 124
[arXiv:hep-th/0511099].

\bibitem{BJ}
K.~Behrndt and C.~Jeschek,
``Fluxes in M-theory on 7-manifolds and G structures,''
JHEP {\bf 0304} (2003) 002
[arXiv:hep-th/0302047];\\
``Fluxes in M-theory on 7-manifolds: G-structures and superpotential,''
Nucl.\ Phys.\ B {\bf 694} (2004) 99
[arXiv:hep-th/0311119].


\bibitem{Fidanza:2003zi}
  S.~Fidanza, R.~Minasian and A.~Tomasiello,
  ``Mirror symmetric SU(3)-structure manifolds with NS fluxes,''
  Commun.\ Math.\ Phys.\  {\bf 254}, 401 (2005)
  [arXiv:hep-th/0311122].

\bibitem{Dalla}
  G.~Dall'Agata,
  ``On supersymmetric solutions of type IIB supergravity with general fluxes,''
  Nucl.\ Phys.\ B {\bf 695}, 243 (2004) [arXiv:hep-th/0403220].

\bibitem{Frey}
A.~R.~Frey,
  ``Notes on $\SU(3)$ structures in type IIB supergravity,''
  JHEP {\bf 0406}, 027 (2004)
  [arXiv:hep-th/0404107].

\bibitem{Jeschek}
C.~Jeschek,
``Generalized Calabi--Yau structures and mirror symmetry,''
arXiv:hep-th/0406046;\\
S.~Chiantese, F.~Gmeiner and C.~Jeschek,
``Mirror symmetry for topological sigma models with generalized Kaehler
geometry,''
arXiv:hep-th/0408169;


\bibitem{GMPT}
M.~Gra\~na, R.~Minasian, M.~Petrini and A.~Tomasiello,
``Supersymmetric backgrounds from generalized Calabi--Yau manifolds,''
JHEP {\bf 0408} (2004) 046
[arXiv:hep-th/0406137].


\bibitem{JW}
C.~Jeschek and F.~Witt,
``Generalised $G_2$-structures and type IIB superstrings,''
JHEP {\bf 0503} (2005) 053
[arXiv:hep-th/0412280].


\bibitem{KL}
  A.~Kapustin and Y.~Li,
  ``Topological sigma-models with H-flux and twisted generalized complex
  manifolds,''
  arXiv:hep-th/0407249.
 

\bibitem{Behrndt:2004mj}
  K.~Behrndt and M.~Cvetic,
  ``General N = 1 supersymmetric fluxes in massive type IIA string theory,''
  Nucl.\ Phys.\ B {\bf 708} (2005) 45
  [arXiv:hep-th/0407263];\\
  D.~L\"ust and D.~Tsimpis,
  ``Supersymmetric AdS(4) compactifications of IIA supergravity,''
  JHEP {\bf 0502} (2005) 027
  [arXiv:hep-th/0412250].


\bibitem{GuLuMi}
S.~Gurrieri, A.~Lukas and A.~Micu,
``Heterotic on half-flat,''
Phys.\ Rev.\ D {\bf 70} (2004) 126009
[arXiv:hep-th/0408121].



\bibitem{Zucchini}
  R.~Zucchini,
  ``A sigma model field theoretic realization of Hitchin's generalized complex
  geometry,''
  JHEP {\bf 0411}, 045 (2004)
  [arXiv:hep-th/0409181]; \\
  ``Generalized complex geometry, generalized branes and the Hitchin sigma
  model,''
  JHEP {\bf 0503}, 022 (2005)
  [arXiv:hep-th/0501062].

\bibitem{HMicu}
T.~House and A.~Micu,
``M-theory compactifications on manifolds with $G_2$ structure,''
Class.\ Quant.\ Grav.\  {\bf 22} (2005) 1709
[arXiv:hep-th/0412006].


\bibitem{Dall'Agata:2005ff}
G.~Dall'Agata and S.~Ferrara,
``Gauged supergravity algebras from twisted tori compactifications with fluxes,''
Nucl.\ Phys.\ B {\bf 717} (2005) 223
[arXiv:hep-th/0502066].
 
\bibitem{Tomasiello}
A.~Tomasiello, ``Topological mirror symmetry with fluxes,''
  arXiv:hep-th/0502148.

\bibitem{Behrndt:2005bv}
K.~Behrndt, M.~Cvetic and P.~Gao,
``General type IIB fluxes with $\SU(3)$ structures,''
arXiv:hep-th/0502154.

\bibitem{Pestun:2005rp}
  V.~Pestun and E.~Witten,
  ``The Hitchin functionals and the topological B-model at one loop,''
  Lett.\ Math.\ Phys.\  {\bf 74}, 21 (2005)
  [arXiv:hep-th/0503083];\\
  V.~Pestun,
  ``Black hole entropy and topological strings on generalized CY manifolds,''
  arXiv:hep-th/0512189.

\bibitem{Dall'Agata:2005mj}
G.~Dall'Agata, R.~D'Auria and S.~Ferrara,
``Compactifications on twisted tori with fluxes and free differential
algebras,''
arXiv:hep-th/0503122.

\bibitem{hull}
C.~M.~Hull and R.~A.~Reid-Edwards,
``Flux compactifications of string theory on twisted tori,''
J.\ Sci.\ Eng.\  {\bf 1} (2004) 411
[arXiv:hep-th/0503114].

\bibitem{DAuria}
R.~D'Auria, S.~Ferrara and M.~Trigiante,
``$E_{7(7)}$ symmetry and dual gauge algebra of M-theory on a twisted
seven-torus,''
arXiv:hep-th/0504108;\\
L.~Andrianopoli, M.~A.~Lledo and M.~Trigiante,
``The Scherk-Schwarz mechanism as a flux compactification with internal
torsion,''
JHEP {\bf 0505}, 051 (2005)
[arXiv:hep-th/0502083].


\bibitem{Berglund}
P.~Berglund and P.~Mayr,
``Non-Perturbative Superpotentials in F-theory and String Duality,''
arXiv:hep-th/0504058.

\bibitem{House:2005yc}
T.~House and E.~Palti,
``Effective action of (massive) IIA on manifolds with SU(3) structure,''
arXiv:hep-th/0505177.


\bibitem{Grana:2005sn}
  M.~Grana, R.~Minasian, M.~Petrini and A.~Tomasiello,
  ``Generalized structures of N = 1 vacua,''
  JHEP {\bf 0511}, 020 (2005)
  [arXiv:hep-th/0505212].


\bibitem{GLW}
  M.~Grana, J.~Louis and D.~Waldram,
  ``Hitchin functionals in N = 2 supergravity,''
  JHEP {\bf 0601}, 008 (2006)
  [arXiv:hep-th/0505264].



\bibitem{D-branecalib}
  P.~Koerber,
  ``Stable D-branes, calibrations and generalized Calabi-Yau geometry,''
  JHEP {\bf 0508} (2005) 099
  [arXiv:hep-th/0506154];\\
  L.~Martucci and P.~Smyth,
  ``Supersymmetric D-branes and calibrations on general N = 1 backgrounds,''
  JHEP {\bf 0511} (2005) 048
  [arXiv:hep-th/0507099];\\
  L.~Martucci,
  ``D-branes on general N = 1 backgrounds: Superpotentials and D-terms,''
  arXiv:hep-th/0602129.


\bibitem{Dall'Agata:2005fm}
  G.~Dall'Agata and N.~Prezas,
  ``Scherk-Schwarz reduction of M-theory on G2-manifolds with fluxes,''
  JHEP {\bf 0510} (2005) 103
  [arXiv:hep-th/0509052].

\bibitem{Chuang:2005qd}
  W.~y.~Chuang, S.~Kachru and A.~Tomasiello,
  ``Complex / symplectic mirrors,''
  arXiv:hep-th/0510042.


\bibitem{Jeschek:2005ek}
  C.~Jeschek and F.~Witt,
  ``Generalised geometries, constrained critical points and Ramond-Ramond
  fields,''
  arXiv:math.dg/0510131.


\bibitem{Tsimpis:2005kj}
  D.~Tsimpis,
  ``M-theory on eight-manifolds revisited: N = 1 supersymmetry and generalized
  Spin(7) structures,''
  arXiv:hep-th/0511047.

\bibitem{Behrndt:2005im}
  K.~Behrndt, M.~Cvetic and T.~Liu,
  ``Classification of supersymmetric flux vacua in M theory,''
  arXiv:hep-th/0512032.

\bibitem{Dall'Agata:2006nr}  
    G.~Dall'Agata,
  ``Non-Kaehler attracting manifolds,''
  arXiv:hep-th/0602045;\\
  J.~P.~Hsu, A.~Maloney and A.~Tomasiello,
  ``Black hole attractors and pure spinors,''
  arXiv:hep-th/0602142.  

\bibitem{Anguelova:2006qf}
  L.~Anguelova and K.~Zoubos,
  ``Flux superpotential in heterotic M-theory,''
  arXiv:hep-th/0602039.

\bibitem{Micu:2006ey}
  A.~Micu, E.~Palti and P.~M.~Saffin,
  ``M-theory on seven-dimensional manifolds with SU(3) structure,''
  arXiv:hep-th/0602163.



\bibitem{Shelton:2005cf}
  J.~Shelton, W.~Taylor and B.~Wecht,
  ``Nongeometric flux compactifications,''
  JHEP {\bf 0510}, 085 (2005)
  [arXiv:hep-th/0508133].

\bibitem{Aldazabal:2006up}
  G.~Aldazabal, P.~G.~Camara, A.~Font and L.~E.~Ibanez,
  ``More dual fluxes and moduli fixing,''
  arXiv:hep-th/0602089.


\bibitem{Villadoro:2006ia}
  G.~Villadoro and F.~Zwirner,
  ``D terms from D-branes, gauge invariance and moduli stabilization in flux
  compactifications,''
  arXiv:hep-th/0602120.



\bibitem{ADAF}
L.~Andrianopoli, R.~D'Auria and S.~Ferrara,
``Supersymmetry reduction of N-extended supergravities in four  dimensions,''
JHEP {\bf 0203} (2002) 025
[arXiv:hep-th/0110277];\\
``Consistent reduction of N = 2 $\to$ N = 1 four dimensional supergravity  coupled to matter,''
Nucl.\ Phys.\ B {\bf 628} (2002) 387
[arXiv:hep-th/0112192].

\bibitem{D'Auria:2005yg}
  R.~D'Auria, S.~Ferrara, M.~Trigiante and S.~Vaula,
  ``N = 1 reductions of N = 2 supergravity in the presence of tensor
  multiplets,''
  JHEP {\bf 0503} (2005) 052
  [arXiv:hep-th/0502219].



\bibitem{Hitchin:2000jd}
  N.~J.~Hitchin,
  ``The geometry of three-forms in six and seven dimensions,''
  arXiv:math.dg/0010054;\\
  N.~J.~Hitchin,
  ``The Geometry of three forms in six-dimensions,''
  J.\ Diff.\ Geom.\  {\bf 55} (2000) 547;\\
  N.~J.~Hitchin,
  ``Stable forms and special metrics,''
  arXiv:math.dg/0107101.

\bibitem{Hitchin:2004ut}
  N.~Hitchin,
  ``Generalized Calabi-Yau manifolds,''
  Quart.\ J.\ Math.\ Oxford Ser.\  {\bf 54} (2003) 281
  [arXiv:math.dg/0209099].

\bibitem{Hitchin:2005in}
  N.~Hitchin,
  ``Brackets, forms and invariant functionals,''
  arXiv:math.dg/0508618.




\bibitem{Mathai:2005fd}
  V.~Mathai and J.~M.~Rosenberg,
  ``T-duality for torus bundles via noncommutative topology,''
  Commun.\ Math.\ Phys.\  {\bf 253} (2004) 705
  [arXiv:hep-th/0401168];\\
  V.~Mathai and J.~M.~Rosenberg,
  ``On mysteriously missing T-duals, H-flux and the T-duality group,''
  arXiv:hep-th/0409073;\\
  P.~Bouwknegt, K.~Hannabuss and V.~Mathai,
  ``T-duality for principal torus bundles and dimensionally reduced Gysin
  sequences,''
  arXiv:hep-th/0412268;\\
  V.~Mathai and J.~Rosenberg,
  ``T-duality for torus bundles with H-fluxes via noncommutative topology. II:
  The high-dimensional case and the T-duality group,''
  arXiv:hep-th/0508084.

\bibitem{Dabholkar:2005ve}
  C.~M.~Hull,
  ``A geometry for non-geometric string backgrounds,''
  JHEP {\bf 0510} (2005) 065
  [arXiv:hep-th/0406102];\\
  A.~Dabholkar and C.~Hull,
  ``Generalised T-duality and non-geometric backgrounds,''
  arXiv:hep-th/0512005.
  
\bibitem{Gualtieri}
  M.~Gualtieri, ``Generalized complex geometry'', 
  Oxford University DPhil thesis, 107 pages
  math.DG/0401221.

\bibitem{Witt:2005sk}
  F.~Witt,
  ``Special metric structures and closed forms,''
  arXiv:math.dg/0502443.


\bibitem{TGinprep}
Work in preperation.

\bibitem{Strominger:1996it}
  A.~Strominger, S.~T.~Yau and E.~Zaslow,
  ``Mirror symmetry is T-duality,''
  Nucl.\ Phys.\ B {\bf 479} (1996) 243
  [arXiv:hep-th/9606040].


\bibitem{Giddings:2005ff}
  S.~B.~Giddings and A.~Maharana,
  ``Dynamics of warped compactifications and the shape of the warped
  landscape,''
  arXiv:hep-th/0507158.

\bibitem{Firouzjahi:2005qs}
  H.~Firouzjahi and S.~H.~Tye,
  ``The shape of gravity in a warped deformed conifold,''
  JHEP {\bf 0601} (2006) 136
  [arXiv:hep-th/0512076].


\bibitem{Sen}
  A.~Sen,
  ``F-theory and Orientifolds,''
  Nucl.\ Phys.\ B {\bf 475} (1996) 562
  [arXiv:hep-th/9605150].  
  
\bibitem{DP}
  A.~Dabholkar and J.~Park,
  ``Strings on Orientifolds,''
  Nucl.\ Phys.\ B {\bf 477} (1996) 701
  [arXiv:hep-th/9604178].


\bibitem{BBS}
K.~Becker, M.~Becker and A.~Strominger,
``Five-branes, membranes and nonperturbative string theory,''
Nucl.\ Phys.\ B {\bf 456} (1995) 130
[arXiv:hep-th/9507158].


\bibitem{MMMS}
M.~Marino, R.~Minasian, G.~W.~Moore and A.~Strominger,
``Nonlinear instantons from supersymmetric p-branes,''
JHEP {\bf 0001} (2000) 005
[arXiv:hep-th/9911206].




  
\bibitem{BKORvanP}
  E.~Bergshoeff, R.~Kallosh, T.~Ortin, D.~Roest and A.~Van Proeyen,
  ``New formulations of D = 10 supersymmetry and D8 - O8 domain walls,''
  Class.\ Quant.\ Grav.\  {\bf 18}, 3359 (2001)
  [arXiv:hep-th/0103233].

\bibitem{mass_tensors}
  R.~D'Auria and S.~Ferrara,
  ``Dyonic masses from conformal field strengths in D even dimensions,''
  Phys.\ Lett.\ B {\bf 606} (2005) 211
  [arXiv:hep-th/0410051];\\  
  J.~Louis and W.~Schulgin,
  ``Massive tensor multiplets in N = 1 supersymmetry,''
  Fortsch.\ Phys.\  {\bf 53} (2005) 235
  [arXiv:hep-th/0410149];\\
  U.~Theis,
  ``Masses and dualities in extended Freedman-Townsend models,''
  Phys.\ Lett.\ B {\bf 609} (2005) 402
  [arXiv:hep-th/0412177];\\
  S.~M.~Kuzenko,
  ``On massive tensor multiplets,''
  JHEP {\bf 0501} (2005) 041
  [arXiv:hep-th/0412190].

\bibitem{BGG}
P.~Binetruy, G.~Girardi and R.~Grimm,
``Supergravity couplings: A geometric formulation,''
Phys.\ Rept.\  {\bf 343} (2001) 255
[arXiv:hep-th/0005225].

\bibitem{WB}
J.~Wess and J.~Bagger,
``Supersymmetry And Supergravity,''
Princeton University Press, Princeton, 1992.

\bibitem{Candelas:1990pi}
  P.~Candelas and X.~de la Ossa,
  ``Moduli Space Of Calabi-Yau Manifolds,''
  Nucl.\ Phys.\ B {\bf 355} (1991) 455.


\bibitem{Romans}
L.~J.~Romans,
``Massive N=2a Supergravity In Ten-Dimensions,''
Phys.\ Lett.\ B {\bf 169} (1986) 374.

\bibitem{Jockers:2005zy}
  H.~Jockers and J.~Louis,
  ``D-terms and F-terms from D7-brane fluxes,''
  Nucl.\ Phys.\ B {\bf 718} (2005) 203
  [arXiv:hep-th/0502059].

\bibitem{GVW}
 S.~Gukov, C.~Vafa and E.~Witten,
``CFT's from Calabi-Yau four-folds,''
Nucl.\ Phys.\ B {\bf 584}, 69 (2000)
[Erratum-ibid.\ B {\bf 608}, 477 (2001)]
[arXiv:hep-th/9906070].

\bibitem{Taylor:1999ii}
  T.~R.~Taylor and C.~Vafa,
  ``RR flux on Calabi-Yau and partial supersymmetry breaking,''
  Phys.\ Lett.\ B {\bf 474} (2000) 130
  [arXiv:hep-th/9912152].


\bibitem{Klemm:2005tw}
  For a review see, for example, 
  A.~Klemm, ``Topological string theory on Calabi-Yau threefolds,''
  PoS {\bf RTN2005} (2005) 002, and references therein.

\bibitem{Gualtieri2}
    G.~R.~Cavalcanti and M.~Gualtieri, ``A surgery for generalized complex structures on 4-manifolds,''
     math.DG/0602333

\bibitem{VanProeyen:1999ni}
  A.~Van Proeyen,
  ``Tools for supersymmetry,''
  arXiv:hep-th/9910030.


\bibitem{Joyce2}
D.~D.~Joyce,
``Compact Riemannian 7-manifolds with Holonomy $G_2$, I./II.''
J. Diff. Geom. {\bf 43} (1996) 291-328

\bibitem{BW}
C.~Beasley and E.~Witten,
``A note on fluxes and superpotentials in M-theory compactifications on
manifolds of G(2) holonomy,''
JHEP {\bf 0207} (2002) 046
[arXiv:hep-th/0203061].

\end{thebibliography}
\end{document}